
\documentstyle{amsppt}
\magnification=1200
\baselineskip=18pt
\NoBlackBoxes
\TagsOnLeft
\document

\def\ts{\thinspace}
\def\tr{\text{{\rm tr}\ts}}
\def\tra{\text{{\rm tr}}}
\def\inv{\text{{\rm inv}\ts}}
\def\inva{\text{{\rm inv}}}
\def\ssl{{\frak {sl}}}
\def\oa{{\frak o}}
\def\Sym{{\frak S}}
\def\spa{{\frak {sp}}}
\def\U{{\operatorname {U}}}
\def\A{{\operatorname {A}}}
\def\Z{{\operatorname {Z}}}
\def\I{{\operatorname {I}}}
\def\Y{{\operatorname {Y}}}
\def\SY{{\operatorname {SY}}}
\def\SS{{\operatorname {S}}}
\def\C{{\Bbb C}}
\def\Left{\text{{\rm Left}}}
\def\Right{\text{{\rm Right}}}
\def\dega{\text{{\rm deg}\ts}}
\def\degaa{\text{{\rm deg}}}
\def\gr{\text{{\rm gr}}}
\def\sgn{\text{{\rm sgn}}}
\def\sign{\text{{\rm sign}}}
\def\qdet{\text{{\rm qdet}\ts}}
\def\qddet{\text{{\rm sdet}\ts}}
\def\End{\text{{\rm End}\ts}}
\def\Enda{\text{{\rm End}}}
\def\E{{\Cal E}}
\def\ra{\rightarrow}
\def\hra{\hookrightarrow}
\def\Proof{\noindent {\bf Proof. }}
\def\aa{{\frak a}}
\def\C{{\Bbb C}}
\def\EndCN{{\Enda(\C^N)}}
\def\g{{\frak g}}
\def\gl{{\frak{gl}}}
\def\glN{{\frak{gl}(N)}}
\def\o{{\frak o}}
\def\ot{\otimes}
\def\slN{{\frak{sl}(N)}}
\def\sp{{\frak{sp}}}
\def\Ya{{\operatorname{Y}(\aa)}}
\def\Yt{{\operatorname{Y}\bigl(\glN,\sigma\bigr)}}
\def\Ua{{\operatorname{U}(\aa)}}
\def\UglN{{\operatorname{U}\bigl(\glN\bigr)}}
\def\YglN{{\operatorname{Y}\bigl(\glN\bigr)}}
\def\td{\tilde}

\font\bigbf=cmbx10 scaled 1200

\
\bigskip
\bigskip
\bigskip
\bigskip
\bigskip
\heading{\bigbf YANGIANS AND CLASSICAL LIE ALGEBRAS} \endheading
\bigskip
\bigskip
\bigskip
\heading{Alexander Molev},
\endheading
\heading{\sl Centre for Mathematics and its Applications,
Australian National University,}
\endheading
\heading{\sl Canberra, ACT 0200, AUSTRALIA}
\endheading
\heading{\rm e-mail: Alexander.Molev\@maths.anu.edu.au}
\endheading
\bigskip
\heading{Maxim Nazarov},
\endheading
\heading{\sl Mathematics Department, University College of Swansea,}
\endheading
\heading{\sl Singleton Park, Swansea SA2 8PP, UNITED KINGDOM,}
\endheading
\heading{\rm e-mail: M.L.Nazarov\@swansea.ac.uk}
\endheading
\bigskip
\heading{\rm and}
\endheading
\bigskip
\heading{Grigori Olshanski}
\endheading
\heading{\sl Institute for Problems of Information Transmission,}
\endheading
\heading{\sl Ermolovoy 19, Moscow 101447, GSP-4, RUSSIA,}
\endheading
\heading{\rm e-mail:\ olsh\@ippi.ac.msk.su}
\endheading

\newpage
\heading
{\bf 0. Introduction}
\endheading

\noindent
The term \lq Yangian' was introduced by V. G. Drinfeld
to specify quantum groups related to rational solutions of the classical
Yang--Baxter equation; see Belavin--Drinfeld [BD1,BD2] for the description
of
these solutions. In Drinfeld [D1] for each simple finite-dimensional Lie
algebra $\aa$, a certain Hopf algebra $\Ya$ was constructed so that $\Ya$
is a
deformation of the universal enveloping algebra for the polynomial current
Lie
algebra $\aa[x]$. An alternative description of the algebra $\Ya$ was given
in
Drinfeld [D3]; see Theorem 1 therein.

Prior to the intruduction of the Hopf algebra $\Ya$ in Drinfeld [D1], the
algebra which may be called the Yangian for the reductive Lie algebra
$\glN$
and may be denoted by $\YglN$, was considered in the works of
mathematical physicists from St.-Petersburg; see for instance
Takhtajan--Faddeev [TF]. The latter algebra is a deformation of the
universal
enveloping algebra $\operatorname{U}\bigl(\glN[x]\bigr)$. Representations
of
the algebra $\YglN$ were studied in Kulish--Reshetikhin--Sklyanin [KRS] and
in Tarasov [T1,T2].

For any $\aa$ the Yangian $\Ya$ contains the universal enveloping
algebra $\Ua$ as a subalgebra. However, the case $\aa=\slN$ seems to be
exceptional since only for $\aa=\slN$ there exists a homomorphism
$\Ya\rightarrow\Ua$ identical on the subalgebra $\Ua$; see Drinfeld [D1],
Theorem 9. In the present article we concentrate on this distinguished
Yangian.
For each of the remaining classical Lie algebras
$\aa=\o(2n+1),\ts\sp(2n),\ts\o(2n)$ we introduce instead of the Yangian
$\Ya$ a
new algebra. This new algebra is a deformation of the universal enveloping
algebra for a certain twisted polynomial current Lie algebra.

Let $\aa$ be one of the latter three classical Lie algebras.
Consider $\aa$ as an involutive subalgebra in $\glN$ where
$N=2n+1,\ts2n,\ts2n$ respectively. Let $\sigma$ denote the corresponding
involution of $\glN$. The subalgebra
$$
\glN[x]^{\sigma}=\{\ts
A(x)\in\glN[x]\ts:\ts\sigma\bigl(A(x)\bigr)=A(-x)\ts\}
$$
in the Lie algebra $\glN[x]$ is called the twisted polynomial current Lie
algebra related to the symmetric Lie algebra $(\glN,\sigma)$;
it is an involutive
subalgebra in $\glN[x]$ also. In the present article we introduce an
algebra
$\Yt$ which is a deformation of the universal enveloping algebra
$\operatorname{U}(\glN[x]^{\sigma})$. The algebra $\Yt$ is a subalgebra in
$\YglN$ and we
call it \lq twisted Yangian'.

As well as the Yangian $\YglN$, the twisted Yangian $\Yt$ contains
the universal enveloping algebra $\Ua$ as a subalgebra and admits a
homomorphism $\Yt\rightarrow\Ua$ identical on $\Ua$. Contrary to the
Yangian
$\Ya$ defined in Drinfeld [D1], the twisted Yangian $\Yt$ has no natural
Hopf
algebra structure. However, it turns out to be a one-sided coideal in the
Hopf
algebra $\YglN$. Let us now describe the algebras $\YglN$ and $\Yt$ more
explicitly.

Let the indices $i,j$ run through the set
$\{\ts -n,\dots,-1,0,1,\dots,n\ts\}$ if $N=2n+1$ and through the set $\{\ts
-n,\dots,-1,1,\dots,n\ts\}$ if $N=2n$. Let $E_{ij}\in\EndCN$ be the
standard
matrix units. We will also regard them as generators of the algebra
$\UglN$.
The algebra $\YglN$ is generated by the elements $t_{ij}^{(k)}$ where
$k=1,2,\dots$\ts, subject to the following relations. Introduce the formal
power series in $u^{-1}$ with the coefficients in $\YglN\ot\EndCN$ $$
T(u)=\sum_{i,j}\ts t_{ij}(u)\ot E_{ij}, \quad
t_{ij}(u)=\delta_{ij}+\sum_k\ts
t_{ij}^{(k)}\ts u^{-k}. $$
Introduce also the formal power series in $u^{-1},\ts v^{-1}$ with the
coefficients in $\YglN\ot\EndCN\ot\EndCN$
$$ T_1=\sum_{i,j}\ts t_{ij}(u)\ot E_{ij}\ot1,
\quad T_2=\sum_{i,j}\ts t_{ij}(v)\ot1\ot E_{ij} $$ and put $$ R_{12}=1\ot
R(u,v), \quad
R(u,v)=1-(u-v)^{-1}\cdot\sum_{i,j}\ts E_{ij}\ot E_{ji}\ts. $$
Then the defining relations in $\YglN$ can be written as the
\lq ternary relation'
$$
R_{12}\ts T_1\ts T_2=T_2\ts T_1\ts R_{12}. \leqno (1)
$$
The exact meaning of relation (1) will be thoroughly explained
in Section 1.

The imbedding $\UglN\hookrightarrow\YglN$ and the homomorphism
$\YglN\rightarrow\UglN$ identical on $\UglN$ are defined respectively by
$E_{ij}\mapsto t_{ij}^{(1)}$ and $$ T(u)\ts\mapsto\ts
E(u)=1+u^{-1}\cdot\sum_{i,j}\ts E_{ij}\ot E_{ij}. \leqno (2) $$
Thus if we denote
$$
E_1=1+u^{-1}\cdot\sum_{i,j}\ts E_{ij}\ot E_{ij}\ot1, \quad
E_2=1+v^{-1}\cdot\sum_{i,j}\ts E_{ij}\ot1\ot E_{ij} $$
then the defining relations in $\UglN$ can be rewritten as $$
R_{12}\ts E_1\ts E_2=E_2\ts E_1\ts R_{12}. $$
It turns out that the defining relations for the generators
$F_{ij}=E_{ij}+\sigma(E_{ij})$ of the subalgebra $\Ua\subset\UglN$ can be
rewritten in an analogous way.

Let the superscript $t$ denote the transposition in $\EndCN$
corresponding to the bilinear (symmetric or alternating) form which is
preserved by the subalgebra $\aa\subset\glN$; then $F_{ij}=E_{ij}-E_{ij}^t$
in
$\EndCN$. The algebra $\Yt$ is generated by the elements $s_{ij}^{(k)}$
where
$k=1,2,\dots$\ts, subject to the following relations. As in the case of
$\YglN$, introduce the formal power series in $u^{-1}$ with the
coefficients in $\Yt\ot\EndCN$
$$
S(u)=\sum_{i,j}\ts s_{ij}(u)\ot E_{ij}, \quad
s_{ij}(u)=\delta_{ij}+\sum_k\ts s_{ij}^{(k)}\ts u^{-k}.
$$
Introduce also the
formal power series in $u^{-1},\ts v^{-1}$ with the coefficients
from\newline $\Yt\ot\EndCN\ot\EndCN$ $$ S_1=\sum_{i,j}\ts s_{ij}(u)\ot
E_{ij}\ot1, \quad S_2=\sum_{i,j}\ts s_{ij}(v)\ot1\ot E_{ij} $$
and put $$
R_{12}^\prime=1\ot R^\prime(-u,v), \quad
R^\prime(u,v)=1-(u-v)^{-1}\cdot\sum_{i,j}\ts E_{ij}^t\ot E_{ji}\ts. $$ Then
the
defining relations in $\Yt$ can be written as the \lq quaternary relation'
$$
R_{12}\ts S_1\ts R_{12}^\prime\ts S_2=
S_2\ts R_{12}^\prime\ts S_1\ts R_{12}
\leqno (3)
$$
along with the \lq symmetry relation'
$$
S(u)-S^t(-u)=\mp\frac1{2u}\ts\bigl(S(u)-S(-u)\bigr), \leqno (4)
$$
where
$$
S^t(u)=\sum_{i,j}\ts s_{ij}(u)\ot E_{ij}^t\ts. $$
Whenever the double sign $\pm$ or $\mp$ occurs, the upper sign corresponds
to the case $\aa=\o(N)$ while the lower sign corresponds to $\aa=\sp(N)$.

The imbedding $\Ua\hookrightarrow\Yt$ and the homomorphism
$\Yt\rightarrow\Ua$ identical on $\Ua$ are defined respectively by
$F_{ij}\mapsto s_{ij}^{(1)}$ and $$ S(u)\ts\mapsto\ts
F(u)=1+(u\pm\frac12)^{-1}\cdot \sum_{i,j}\ts F_{ij}\ot E_{ij}. \leqno (5)
$$
Thus if we denote
$$
F_1=1+(u\pm\frac12)^{-1}\cdot\sum_{i,j}\ts F_{ij}\ot E_{ij}\ot1, \quad
F_2=1+(v\pm\frac12)^{-1}\cdot\sum_{i,j}\ts F_{ij}\ot1\ot E_{ij} $$
then the defining relations in $\Ua$ can be rewritten as $$
\align
&
R_{12}\ts F_1\ts R_{12}^\prime\ts F_2
=F_2\ts R_{12}^\prime\ts F_1\ts R_{12},
\\
&
F(u)-F^t(-u)
=\mp\frac1{2u}\ts\bigl(F(u)-F(-u)\bigr),
\endalign
$$
where
$$
F^t(u)=1+(u\pm\frac12)^{-1}\sum_{i,j}\ts F_{ij}\ot E_{ij}^t\ts. $$
The imbedding $\Yt\hookrightarrow\YglN$ can be defined by $$
S(u)\ts\mapsto\ts T(u)\ts T^t(-u),
\quad\text{where}\ \ T^t(u)=\sum_{i,j}\ts t_{ij}(u)\ot E_{ij}^t. \leqno (6)
$$

The ternary relation (1) has a rich and extensive background;
see for instance Takhtajan--Faddeev [TF] and Drinfeld [D4]. This relation
originates from the quantum Yang--Baxter equation (see Kulish--Sklyanin
[KS1]),
and the Yangians themselves were primarily regarded as a vehicle for
producing
new solutions of that equation; cf. Drinfeld [D1]. Conversely, the ternary
relation (1) was used in Reshetikhin--Takhtajan--Faddeev [RTF] as a tool
for
studying quantum groups.

The quaternary relation (3) has its own history.
Relations of the type (3) appeared for the first time in Cherednik [C1] and
Sklyanin [S2], where integrable systems with boundary conditions were
studied.
Various versions of (3) were employed in
Reshetikhin--Semenov [RS] to extend the
approach of Reshetikhin--Takhtajan--Faddeev [RTF] from finite-dimensional
to
affine Lie algebras, and in Noumi [No] to obtain the $q$-analogues of
spherical functions on the classical symmetric spaces. Algebraic structures
related to (3) were discussed in Kulish--Sklyanin [KS3] and in
Kulish--Sasaki--Schwiebert [KSS]. In these two papers
a quaternary type relation
was called the `reflection equation'.

On the other hand, the Yangian $\YglN$ has proved to be useful
in the theory of finite-dimensional representations of the Lie algebra
$\glN$.
It gives rise to canonical generators of the center of the universal
enveloping
algebra $\UglN$ and to a variety of commutative subalgebras in $\UglN$; see
Cherednik [C2] and Kirillov--Reshetikhin [KR]. Applications of the algebra
$\YglN$ to constructing the Gelfand--Zetlin bases for irreducible
finite-dimensional representations of $\glN$ were considered in Cherednik
[C2] and in Nazarov--Tarasov [NT], Molev [M2].
All of these applications are based on the
existence of a homomorphism $\YglN\rightarrow\UglN$ identical on $\UglN$.
We
believe that the twisted Yangians will play the role of $\YglN$ for the
other
classical Lie algebras $\aa=\o(2n+1),\ts\sp(2n),\ts\o(2n)$.
The results of Molev [M3] come in support of this claim. In forthcoming
publications we will consider
further applications of the twisted Yangians to the theory of
finite-dimensional representations of the classical Lie algebras.

Our definition of the twisted Yangian $\Yt$ is motivated by
the results of Olshanski\u\i\ [O1] where a natural extension of the
universal
enveloping algebra $\operatorname{U}\bigl(\gl(\infty)\bigr)$ is
constructed. In
that paper the Yangian $\YglN$ arises in the following way. For each
$m=N+1,N+2,\dots$ consider the subalgebra
$$
\glN\oplus\gl(m-N)\subset\gl(m)
$$
and denote by $\A_N(m)$ the centralizer of $\gl(m-N)$
in the algebra $\operatorname{U}\bigl(\gl(m)\bigr)$.
In particular, $\A_0(m)$ coincides with the center
$\operatorname{Z}\bigl(\gl(m)\bigr)$ of
$\operatorname{U}\bigl(\gl(m)\bigr)$.
Then there is a canonical chain of homomorphisms $$
\A_N(N+1)\leftarrow \A_N(N+2)\leftarrow\ldots
\leftarrow \A_N(m)\leftarrow\ldots\ts.
$$
In particular, for $N=0$ we obtain a canonical chain $$
\operatorname{Z}\bigl(\gl(1)\bigr)
\leftarrow
\operatorname{Z}\bigl(\gl(2)\bigr)
\leftarrow\ldots\leftarrow
\operatorname{Z}\bigl(\gl(m)\bigr)
\leftarrow\ldots\ts.
$$
Then Theorem 2.1.5 from Olshanski\u\i\ [O1] establishes an isomorphism
$$
\underset{m\to\infty}\to{\operatorname{lim}\ts\operatorname{proj}}\ts\ts
\A_N(m) \cong\YglN\ot
\underset{m\to\infty}\to{\operatorname{lim}\ts\operatorname{proj}}\ts\ts
\operatorname{Z}\bigl(\gl(m)\bigr). $$
It was shown in Olshanski\u\i\ [O2] that by applying an analogous
construction
to the Lie algebra $\aa$ instead of $\glN$, one obtains
the twisted Yangian $\Yt$ in place of $\YglN$.

A systematic study of finite-dimensional representations
of Yangians was commenced in Drinfeld [D3]. The case of the Yangian
$\operatorname{Y}\bigl({\frak{sl}}(2)\bigr)$ was primarily investigated in
Tarasov [T1,T2]; see also Chari--Pressley [CP1]. Finite-dimensional
representations of the Yangian $\Ya$ were studied in Reshetikhin [R], and
the
paper Chari--Pressley [CP2] is concerned with general Yangians. However,
the
distinguished case of the Yangian $\YglN$ is the most fully studied; for
this
case an analogue of the classical Schur--Weyl duality is established in
Cherednik [C2] and Drinfeld [D2]. We are convinced that finite-dimensional
representations of the twisted Yangians also deserve a thorough study.

In Molev [M1] a general theory of finite-dimensional re\-pre\-sent\-ations
of the twist\-ed Yangian $\Yt$ was addressed. In that paper an analogue of
the
classification theorem from Drinfeld [D3] was obtained and the simplest
cases
$\aa=\sp(2),\ts\o(2)$ were thoroughly examined.

In the present paper we study in
detail the algebraic structure of the Yangian $\YglN$ and that of the
twisted
Yangian $\Yt$. Most of the results about the structure of $\YglN$ are
known but there is no exposition of them available; the results concerning
$\Yt$ were announced in Olshanski\u\i\ [O2] without proofs. Let us now give
an
overview of the contents of the present paper.

In Section 1 we begin with the definition of the Yangian $\YglN$ and
introduce certain useful automorphisms of this algebra. The main result of
Section 1 is an analogue of the Poincar\'e--Birkhoff--Witt theorem (Theorem
1.22). We also introduce a filtration on the algebra $\YglN$ such that the
corresponding graded algebra coincides with the universal enveloping
algebra
$\operatorname{U}\bigl(\glN[x]\bigr)$; see Theorem 1.26.

In Section 2 we give a complete description (Theorem 2.13) of
the center of the algebra $\YglN$.
Here we use an important notion of
\lq quantum determinant' from Kulish--Sklyanin [KS2]. For $N=2$ this notion
appeared earlier in Izergin--Korepin [IK]. Consider
the formal series in $u^{-1}$ with the coefficients in $\YglN$ $$
\sum_{p\in \Sym_N}\ts \sgn(p)\ts
t_{p(1),1}(u)\ts t_{p(2),2}(u-1)\dots t_{p(N),N}(u-N+1)=
1+ \sum_k\ts d_k\ts u^{-k},
\leqno (7)
$$
where $\Sym_N$ is the symmetric group. This series is called
the quantum determinant of the $N\times N$-matrix formed by $t_{ij}(u)$. We
prove that the coefficients $d_1,\ts d_2,\ts\ldots$ generate the center of
$\YglN$. The images of the elements $d_1,\ts d_2,\ts\ldots\ts,\ts d_N$
under
the homomorphism $\YglN\rightarrow\UglN$ defined by (2) turn out to be the
generators of the center $\operatorname{Z}\bigl(\gl(N)\bigr)$, introduced
in
Capelli [Ca1,Ca2] and also considered in Carter--Lusztig [CL] and Howe [H].
Furthermore, the algebra $\operatorname{Y}\bigl(\slN\bigr)$ can be defined
as
the quotient of $\YglN$ by the relations $d_1=d_2=\ldots=0$ (Corollary
2.18).

In Section 3 we give two alternative definitions of the algebra $\Yt$
and establish their equivalence.
By one of the definitions, $\Yt$ is the algebra
with the generators $s_{ij}^{(k)}$ and relations (3,4). We prove (Theorem
3.5)
that the mapping (6) extends to a homomorhism of algebras
$\Yt\rightarrow\YglN$ and that this homomoprhism is injective (Theorem
3.8).
Thus $\Yt$ can be also defined as a certain subalgebra in
$\YglN$. Moreover, this subalgebra turns out to be a left coideal in the
Hopf
algebra $\YglN$; see Theorem 4.17. We point out (Remark 3.14) an analogue
of
the Poincar\'e--Birkhoff--Witt theorem for the algebra $\Yt$. As well as in
Section 1, we introduce a filtration on the algebra $\Yt$ such that the
corresponding graded algebra coincides with
$\operatorname{U}(\glN[x]^{\sigma})$, see Theorem 3.15.

In Section 4 we construct generators $c_1,\ts c_2,\ts\ldots$ of
the center of $\Yt$ analogous to $d_1,\ts
d_2,\ts\ldots\ts\in\YglN$; see Proposition 4.4, Theorem 4.7 and Theorem
4.11.
Here we generalize one particular construction from Sklyanin [S2].
Through the homomorphism
$\Yt\rightarrow\Ua$ defined by (5), we then obtain generators of the center
of
$\Ua$ which seem to be new; cf. Howe--Umeda [HU], Appendix 2. The quotient
of
the algebra $\Yt$ by the relations $c_1=c_2=\dots=0$ is a deformation of
the
universal enveloping algebra of the Lie algebra $\slN[x]^{\sigma}$;
see Corollary 4.16.

In Section 5 we construct generators $z_1,\ts z_2,\ts\ldots$ of
the center of the
algebra $\YglN$ different from those considered in Section 2, cf. [N1].
The images of these generators under the homomorphism
$\YglN\rightarrow\UglN$
defined by (2) essentially coincide with the elements of
$\operatorname{Z}\bigl(\gl(N)\bigr)$ found in Perelomov--Popov [PP].
We describe
explicitly (Theorem 5.11) the automorphism ${\operatorname{S}}^2$ of the
algebra $\YglN$, where $\operatorname{S}$ stands for the antipode; this
description also involves the elements $z_1,\ts z_2,\ts\ldots$. We provide
a
formula which links these elements with $d_1,\ts d_2,\ts\ldots$ (Theorem
5.7).

In Section 6 we construct generators of the center of the algebra $\Yt$
analogous to $z_1,\ts z_2,\ts\ldots\in\YglN$. Their images with respect to
the
homomorphism $\Yt\allowmathbreak\rightarrow\Ua$ defined by (5) again
essentially coincide with the elements $\operatorname{Z}(\aa)$ from
Perelomov--Popov [PP]. The results of this section also allow us to
reformulate the symmetry relation (4) in a rather elegant way (Theorem
6.4).

Finally, in Section 7 we resume consideration of the
quantum determinant (7).
In that section we provide a quantum analogue (Theorem 7.3) of the
well-known
expansion of the determinant of a block matrix $$ \operatorname{det}
\left(\matrix A&B\\C&D\endmatrix\right)=
\operatorname{det}A\cdot\operatorname{det}(D-C\ts A^{-1}B) $$
where the block $A$ is assumed to be invertible.
We also provide an analogue of the last expansion for the algebra $\Yt$;
see Theorem 7.6.

A word of explanation is necessary in regard to the scheme of referring to
formulae adopted in the present paper.
We number the formulae
in every subsection independently. When we refer back to these formulae
in later subsections, triple numbering is employed. For example, formula
(1) in Subsection 3.2 is referred to as (1) in that subsection, and as
(3.2.1)
later on.

At various stages of this work we benefited from discussions with
I. V. Cherednik,
V. G. Drinfeld,
B. L. Feigin,
P. P. Kulish,
N. Yu. Reshetikhin,
E. K. Sklyanin
and
V. O. Tarasov.
It is a pleasure to express our gratitude to all of them.
We are also very grateful to D. W. Holtby for his kind
help in preparing the manuscript.

\newpage
\heading {\bf 1. The Yangian $\Y(N)$}\endheading

\noindent
In this section, we keep $N \in\{ 1,2, \ldots \}$ fixed.
The Yangian $\Y(N)$ is introduced as an associative
algebra with the defining relations (1.1.1).
Then we describe the basic tool to work with $\Y(N)$ --- the $R$-matrix
formalism. Further we define some automorphisms of the algebra $\Y(N)$
which will be extensively used later. In Subsections 1.16\ts--\ts1.19 we
discuss
fundamental links between $\Y(N)$ and the universal enveloping algebra
of $\gl(N)$. Then we define two filtrations in $\Y(N)$ and prove the main
result of this section --- the Poincar\'e--Birkhoff--Witt theorem (Theorem
1.22
and Corollary 1.23). We also show that $\Y(N)$
is a flat
deformation of the universal enveloping algebra of the polynomial current
Lie
algebra $\gl(N)\ot\C[x]$. Finally we discuss the Hopf algebra structure of
$\Y(N)$.
\bigskip
\noindent {\bf 1.1. Definition.} The {\it Yangian} $\Y(N) = \Y(\gl(N))$ is
defined as the complex associative
unital algebra with countably many generators
$t^{(1)}_{ij}$, $t^{(2)}_{ij} , \ldots$ where $1 \leq i,j \leq N$,
and defining relations
$$
[t_{ij}^{(M+1)},t_{kl}^{(L)}]-
[t_{ij}^{(M)},t_{kl}^{(L+1)}]=
t_{kj}^{(M)}t_{il}^{(L)}-t_{kj}^{(L)}t_{il}^{(M)},
\leqno(1)
$$
where $M,L=0,1,2,\ldots\;$ and $t_{ij}^{(0)}:=\delta_{ij}\cdot1$.

\proclaim
{\bf 1.2. Proposition}
The system {\rm (1.1.1)} is equivalent to the system of
commutation relations: $$
[ t^{(M)}_{ij} , t^{(L)}_{kl} ] =  \sum^{\min (M,L)-1}_{r=0}
(t^{(r)}_{kj} t^{(M+L-1-r)}_{il} - t^{(M+L-1-r)}_{kj}
t^{(r)}_{il}),
\leqno(1)
$$
where $M,L = 1,2, \ldots  \;$ and \  $ 1 \leq i,j,k,l \leq N$.
\endproclaim

\noindent{\bf Proof.}
To simplify the notation, let us denote by $\Left(M,L)$
and $\Right(M,L)$ the left and right hand sides of (1.1.1), respectively.
The system of relations (1.1.1) is clearly equivalent to the
following system of relations:
$$
\displaylines{\quad
\Left(M,L)+\Left(M-1,L+1)+\ldots+\Left(0,M+L)\hfill\cr
\hfill=\Right(M,L)+\Right(M-1,L+1)+\ldots+\Right(0,M+L),  \quad (2)}
$$
where $M,L=0,1,2,\ldots$\ . Observe that
$$
\Left(0,M+L)=[t^{(1)}_{ij}, t^{(M+L)}_{kl}]-[t^{(0)}_{ij},t^{(M+L+1)}_{kl}]
=[t^{(1)}_{ij},t^{(M+L)}_{kl}],
$$
since $t^{(0)}_{ij}=\delta_{ij}$. Thus (2) may be rewritten as
$$
[t^{(M+1)}_{ij}, t^{(L)}_{kl}]=\sum^M_{r=0}\Right(r,M+L-r). \leqno(3)
$$
Furthermore, observe that
$$
\Right(r,s)=-\Right(s,r)\qquad\hbox{for}\quad r,s=0,1,\ldots,
\leqno(4)
$$
so that
$$
\sum^M_{r=L}\Right(r,M+L-r)=0\qquad\hbox{if}\quad M\ge L,
$$
and (3) becomes
$$
[t^{(M+1)}_{ij}, t^{(L)}_{kl}]=\sum^{\min(M,L-1)}_{r=0}\Right(r,M+L-r).
$$
Now by replacing $M$ by $M-1$ we obtain the relation (1).

\bigskip
\noindent {\bf 1.3.} The next few subsections contain preliminaries on
the
$R$-matrix formalism. In this formalism, one deals with the multiple tensor
products $\C^N\otimes \cdots \otimes\ts\C^N$ and operators therein. Let us
set  $\E=\C^N$. For an operator
$A\in \End\E$ and a number $m=1,2,\ldots$ we set
$$
A_k:=1^{\otimes (k-1)}\otimes A\otimes 1^{\otimes (m-k)}
\in\End\E^{\otimes m}, \quad 1\leq k\leq m. \leqno(1)
$$
If $A\in\End\E^{\otimes 2}$ then for any $k, l$\  such that
$1\leq k, l\leq m$\ and $k\neq l$, we denote by $A_{kl}$ the operator in
$\E^{\otimes m}$
 which acts as $A$ in the product of $k$-th and $l$-th copies and as
1 in all other copies. That is,
$$
A=\sum_{r,s,t,u} a_{rstu} E_{rs}\otimes E_{tu}, \quad
a_{rstu}\in\C\qquad \Rightarrow
\quad A_{kl}=\sum_{r,s,t,u} a_{rstu} (E_{rs})_k\ (E_{tu})_l
\leqno(2)
$$
where, in accordance with (1),
$$
(E_{rs})_k=1^{\otimes (k-1)}\otimes E_{rs}\otimes 1^{\otimes (m-k)}.
\leqno(3)
$$

\bigskip
\noindent {\bf 1.4}. We will denote by $P$ the permutation operator
in $\E\otimes\E$:
$$
P:=\sum_{i,j=1}^N E_{ij}\otimes E_{ji}. \leqno(1)
$$
Set
$$
R(u):= 1-{P\over u} \in \Enda\left(\E\otimes\E\right)\otimes\C(u),
\leqno(2)
$$
where $u$ is a formal variable; this object is called the
{\it Yang $R$-matrix}.

Let $m=2,3,\ldots$ and $u_1,\ldots,u_m$ be formal variables. For
$1\leq k,l\leq m$,\  $k\neq l$  consider the operator $P_{kl}$ obtained
from $P$ by using the general rule (1.3.2); this is simply the permutation
of
$k$-th and $l$-th factors in $\E^{\otimes m}$. Now set
$$
R_{kl}(u_k-u_l):=1-{P_{kl}\over u_k-u_l}\in
\End\E^{\otimes m}\otimes\C(u_1,\ldots, u_m). \leqno(3)
$$
When $m=2$, we will write $R(u-v)$ instead of $R_{12}(u_1-u_2)$.

\proclaim
{\bf 1.5. Proposition} If $i,j,k$ are pairwise distinct, then the
following identity holds:
$$
R_{ij}(u)R_{ik}(u+v)R_{jk}(v)=R_{jk}(v)R_{ik}(u+v)R_{ij}(u). \leqno(1)
$$
\endproclaim

This identity is called the {\it Yang--Baxter equation}. Sometimes it is
convenient  to write (1) in a slightly different form
$$
R_{12}(u_1-u_2)R_{13}(u_1-u_3)R_{23}(u_2-u_3)
=R_{23}(u_2-u_3)R_{13}(u_1-u_3)R_{12}(u_1-u_2). \leqno(2)
$$

\bigskip
\noindent
{\bf Proof.} One may assume that $i=1$, $j=2$, $k=3$. Using (1.4.2) and
multiplying (1) by $uv(u+v)$, we obtain, after obvious transformations,
$$
P_{12}P_{13}v+P_{12}P_{23}(u+v)+P_{13}P_{23}u-P_{12}P_{13}P_{23}=
$$
$$
P_{13}P_{12}v+P_{23}P_{12}(u+v)+P_{23}P_{13}u-P_{23}P_{13}P_{12}.
$$
Since $P_{12}P_{13}P_{23}=P_{23}P_{13}P_{12}$ (this is essentially an
identity
in the symmetric group $\Sym_3$),  we have to verify the following two
identities:
$$
\displaylines{
P_{12}P_{23}+P_{13}P_{23}=P_{23}P_{12}+P_{23}P_{13} \cr
P_{12}P_{13}+P_{12}P_{23}=P_{13}P_{12}+P_{23}P_{12}. \cr}
$$
However, these identities are indeed true since
$$
P_{12}P_{23}=P_{23}P_{13}=P_{13}P_{12},\qquad
P_{13}P_{23}=P_{23}P_{12}=P_{12}P_{13}.
$$

\bigskip
\noindent
{\bf 1.6.} Now we introduce the so-called $T$-matrix which is a
matrix-valued formal generating series for the generators $t_{ij}^{(M)}$
of the Yangian $\Y(N)$. For certain reasons (see, e.g. Remark 2.2), it is
convenient to deal with series in the {\it negative} powers of a formal
variable.

Firstly, for any $i,j=1,\ldots, N$ define the  generating
series for the sequence {$t_{ij}^{(M)}, M=1,2,\ldots$} as follows:
$$
t_{ij} (u) = \delta_{ij} + t^{(1)}_{ij} u^{-1} + t^{(2)}_{ij}u^{-2} +
\cdots   \in \Y (N)[[ u^{-1} ]]. \leqno(1)
$$
Then combine all these series into a single ``$T$-matrix'':
$$
T(u) :=\sum^N_{i,j=1} t_{ij}(u)\otimes E_{ij}
\in \Y(N)[[u^{-1}]]\otimes \End\E. \leqno(2)
$$
More generally, given a number $m=2,3\ldots$ and formal variables
$u_1,\ldots,u_m$,
we set for any $k=1,\ldots, m$
$$
T_k (u_k) :=\sum_{i,j=1}^N t_{ij} (u_k) \otimes (E_{ij})_k
\in \Y(N)[[u_1^{-1},\ldots,u_m^{-1}]]\otimes\End\E^{\otimes m}.
\leqno(3)
$$
If $m=2$, we will prefer to write $u,v$ instead of $u_1,u_2$.

\bigskip
\noindent
{\bf 1.7.} We will often have to deal with the operators
$R_{kl}(u_k-u_l)$ and $T_k(u_k)$ simultaneously. Then the algebra
$\Y(N)[[u_1^{-1},\ldots,u_m^{-1}]]$
should be replaced by an appropriate extension
$\Y(N)[[u_1^{-1},\ldots,u_m^{-1}]]_{ext}$ containing the elements
$(u_k-u_l)^{-1}$. It is easy to construct such an extension. For example,
we write
$$
(u_k-u_l)^{-1}=-{u_k^{-1}u_l^{-1}\over u_k^{-1}-u_l^{-1}}
$$
and then localize $\Y(N)[[u_1^{-1},\ldots, u_m^{-1}]]$ with respect to the
multiplicative family generated by the elements $u_k^{-1}-u_l^{-1}$,
$k\neq l$. The localization is well-defined since the algebra
$\Y(N)[[u_1^{-1},\ldots,u_m^{-1}]]$ has no divisors of zero. This is the
minimal possible extension, and sometimes we will need a larger one
(see, e.g., Subsection 1.10).

\bigskip
\proclaim {\bf 1.8. Proposition}
The system {\rm (1.1.1)} of the defining relations of $\Y(N)$ is equivalent
to the following single relation on the $T$-matrix:
$$
R (u-v) T_1 (u) T_2 (v) = T_2 (v) T_1 (u) R (u-v).   \leqno(1)
$$
\endproclaim

We will refer to (1) as the {\it ternary relation}.

\bigskip
\noindent {\bf Proof.}
It is easily seen that, in terms of the generating series (1.6.1),
the initial system (1.1.1) may be rewritten as follows:
$$
[t_{ij} (u), t_{kl} (v) ] =  {1\over u-v} (t_{kj} (u) t_{il} (v) -
t_{kj} (v) t_{il} (u)) \leqno(2)
$$
where $1 \leq i,j,k,l \leq N$. Indeed, if we multiply both sides of (2) by
$u-v$ and compare the terms having the same degrees in $u$ and $v$, then
we will return to (1.1.1).

On the other hand, by definitions (1.3.1) and (1.4.2), formula (1)
reads as follows:
$$
\displaylines{
\qquad (1-{P\over u-v})\sum_{i,j,k,l}t_{ij}(u)t_{kl}(v)\,
(E_{ij}\otimes E_{kl})= \hfill\cr
\hfill=\sum_{i,j,k,l} t_{kl}(v)t_{ij}(u)\, (E_{ij}\otimes E_{kl})\,
(1-{P\over u-v}).\qquad (3)\cr}
$$
This may be rewritten as
$$
\gather
\sum_{i,j,k,l}\ts [t_{ij}(u),t_{kl}(v)]\,(E_{ij}\otimes E_{kl})=
\\
={1\over u-v}\sum_{i,j,k,l}t_{ij}(u)t_{kl}(v)\,
P\,(E_{ij}\otimes
E_{kl})
-{1\over u-v}\sum_{i,j,k,l}t_{kl}(v)t_{ij}(u)\,(E_{ij}\otimes
E_{kl})\, P.
\ \ \tag{4}
\endgather
$$
Observe now that, by definition of $P$,
$$
P\,(E_{ij}\otimes E_{kl})=E_{kj}\otimes E_{il},\qquad
(E_{ij}\otimes E_{kl})\, P=E_{il}\otimes E_{kj}.
$$
Substituting this in (4) and changing the notation of
the indices in an obvious manner one finally obtains that
$$
\displaylines{
\qquad\qquad \sum_{i,j,k,l}[t_{ij}(u), t_{kl}(v)]\,
(E_{ij}\otimes E_{kl})=\hfill\cr
\hfill = {1\over u-v}\sum_{i,j,k,l}(t_{kj}(u)t_{il}(v)-
t_{kj}(v)t_{il}(u))\,(E_{ij}\otimes E_{kl}),\qquad\qquad\cr}
$$
which is equivalent to the system (2).

\bigskip
\noindent
{\bf 1.9. Remark.} One could propose the following informal interpretation
of identity (1.8.1). Let us suppose
that the generators $t^{(1)}_{ij} , t^{(2)}_{ij} , \ldots\;$ operate in a
certain vector space $W$ (the nature of $W$ is irrelevant; e.g., one
may take the left regular representation of the algebra $\Y(N)$).
Then $T(u)$ may be regarded as an operator in
$W\otimes\E$ depending on a (formal) parameter $u$, so that (1.8.1)
may be regarded as a relation in $\End W\otimes\E\otimes\E$.

\bigskip
\noindent
{\bf 1.10. Remark.} The commutation relations (1.2.1) can be also written
as
follows:
$$
[T_1^{(M)}, T_2^{(L)}]=\sum_{r=0}^{\min(M,L)-1} (PT_1^{(r)}
T_2^{(M+L-1-r)}-T_2^{(M+L-1-r)}T_1^{(r)}P) \leqno(1)
$$
where
$$
T^{(M)}:=\sum_{i,j=1}^N t_{ij}^{(M)}\otimes E_{ij}\in \Y(N)\otimes
\End\E
$$
and
$T_1^{(M)}$ and $T_2^{(M)}$ are built following the general prescription
(1.3.1).

Let us show how (1) can be derived from the ternary relation (1.8.1).
First, we write (1.8.1) as
$$
[T_1(u), T_2(v)]={1\over u-v}(PT_1(u)T_2(v)-T_2(v)T_1(u)P). \leqno(2)
$$
Next we write
$$
{1\over u-v}={u^{-1}\over {1-vu^{-1}}}=\sum_{s=0}^{\infty} v^su^{-s-1}
\leqno(3)
$$
and regard both sides of (2) as the elements of the extended algebra
$$
\Y(N)((v^{-1}))[[u^{-1}]]\otimes\Enda(\E\otimes\E) \leqno(4)
$$
(note that (3) does belong to this algebra). Then (2) is rewritten as
the system of the following relations for $M,L=0,1,2,\ldots$
$$
[T_1^{(M)}, T_2^{(L)}]=\sum_{s=0}^{\infty} (PT_1^{(M-s-1)}
T_2^{(L+s)}-T_2^{(L+s)}T_1^{(M-s-1)}P). \leqno(5)
$$
The sum in the right hand side of (5) is actually taken over
$s=0,1,\ldots, M-1$ so that we may replace $s$ by $r:=M-1-s$.
Then (5) takes the form
$$
[T_1^{(M)}, T_2^{(L)}]=\sum_{r=0}^{M-1} (PT_1^{(r)}
T_2^{(M+L-1-r)}-T_2^{(M+L-1-r)}T_1^{(r)}P) \leqno(6)
$$
which is simply another form of (1.2.3).

It remains to note that for any $r,s$ the conjugation by $P$
sends $T_1^{(r)}$ into
$T_2^{(r)}$ and $T_2^{(s)}$ into $T_1^{(s)}$, so that the expression
$$
PT_1^{(r)}T_2^{(s)}-T_2^{(s)}T_1^{(r)}P
$$
is antisymmetric in $(r, s)\ $ (compare with (1.2.4)). This shows that
the summation in (6)
can be actually made over $r=0,1,\ldots,\  \min(M,L)-1$.

\bigskip
\proclaim {\bf 1.11. Proposition} There exists an involutive
antiautomorphism
of the algebra $\Y(N)$ defined by
$$
\sign : T(u)\mapsto T(-u). \leqno (1)
$$
\endproclaim

\Proof  This is almost trivial. We have to check that
$$
R(u-v)T_2(-v)T_1(-u)=T_1(-u)T_2(-v)R(u-v)\leqno(2)
$$
but this follows from the ternary relation (1.8.1), if we conjugate both
of its sides
by $P$ and then replace $(u,v)$ by $(-v,-u)$.

\bigskip
\proclaim {\bf 1.12. Proposition} The following mappings define
automorphisms of the algebra $\Y(N)$.
\indent {\rm (i)} The shift in u:
$$
\sigma_a : T(u)\mapsto T(u+a), \quad a\in\C. \leqno(1)
$$
\indent {\rm (ii)} The multiplication by a formal power series:
$$
\mu_f: T(u)\mapsto f(u)T(u) \leqno(2)
$$
where
$$
f(u):=1+f_1u^{-1}+f_2u^{-2}+\ldots \in \C[[u^{-1}]] \leqno(3)
$$
or, more explicitly,
$$
t_{ij}^{(1)}\mapsto t_{ij}^{(1)}+f_1\delta_{ij},\quad
t_{ij}^{(2)}\mapsto t_{ij}^{(2)}+f_1t_{ij}^{(1)}+f_2\delta_{ij}, \quad
\hbox{etc.} \leqno(4)
$$
\indent {\rm (iii)} Inversion:
$$
\inv: T(u)\mapsto T^{-1}(-u). \leqno(5)
$$
\indent {\rm (iv)} Transposition:
$$
T(u)\mapsto T^t(-u) \leqno(6)
$$
where $t:\End\E \rightarrow \End\E $ is an arbitrary antiautomorphism
of the algebra $\End\E$ (e.g., $E_{ij}\mapsto E_{ji}$) and
$$
T^t(u):=\sum t_{ij}(u)\otimes (E_{ij})^t. \leqno(7)
$$
\endproclaim

\Proof We will verify first that each of the mappings (i) -- (iv) preserves
the defining relations (1.8.1) of $\Y(N)$.

(i) This seems to be trivial since the ternary relation is
clearly invariant under the shift of the parameter $u$. There is, however,
an important detail: it should be stressed that a shift of the (formal)
parameter $u$ is a well defined operation in $\Y(N)[[u^{-1}]]$. Note
that in  the case of $\Y(N)[[u]]$ this is no longer true.

(ii) It suffices to multiply both sides of the ternary relation
by $f(u)f(v)$.

(iii) We have to verify the relation
$$
R (u-v)T_1^{-1}(-u)T_2^{-1}(-v) = T_2^{-1}(-v)T_1^{-1}(-u)R(u-v). \leqno(8)
$$
This can be done as follows. First, one multiplies both sides of the
ternary relation by $T_1^{-1}(u)T_2^{-1}(v)$ on the left and by
$T_2^{-1}(v)T_1^{-1}(u)$ on the right; the result looks like
$$
T_1^{-1}(u)T_2^{-1}(v)R(u-v)=R(u-v)T_2^{-1}(v)T_1^{-1}(u). \leqno(9)
$$
Next one interchanges both sides of (9) and conjugates them by the
permutation operator $P$; the result then looks like
$$
R(u-v)T_1^{-1}(v)T_2^{-1}(u)=T_2^{-1}(u)T_1^{-1}(v)R(u-v). \leqno(10)
$$
Finally one replaces $(u,v)$ by $(-v,-u)$; this clearly transforms (10)
to (8).

(iv) First of all, observe that any antiautomorphism
of $\End\E$ can be written
as the composition of  the ``standard'' transposition
$E_{ij}\mapsto E_{ji}$ and an interior automorphism (i.e., conjugation
by an invertible operator). It implies that $P\in\Enda(\E\otimes\E)$ is
invariant with respect to
$t\otimes t$, so that $R(u-v)$ is invariant too.

Introduce the partial transpositions
$$
t_1:=t\otimes 1,\qquad t_2:=1\otimes t. \leqno(11)
$$
Our claim is equivalent to the validity of the relation
$$
R (u-v)T_1^{t_1}(-u)T_2^{t_2}(-v) = T_2^{t_2}(-v)T_1^{t_1}(-u)R(u-v).
\leqno(12)
$$
We will deduce (12) from the ternary relation by means of the following
transformations.

Firstly, apply $t_1$ to both sides of the ternary relation.
It is easy to see that the result is
$$
(R(u-v)T_1(u))^{t_1}T_2(v)=T_2(v)(T_1(u)R(u-v))^{t_1}.\leqno(13)
$$

Secondly, observe that one may regard $R(u-v)$ and $T_1(u)$ as $N\times N$
matrices, say $A$ and $B$, such that each coefficient  of $A$ commutes
with any coefficient of $B$. (In fact, the coefficients of $A$ are
essentially  operators in the second copy of $\E$ while the coefficients of
$B$ are essentially elements of the Yangian.) In such a situation, we
have $(AB)^t=B^tA^t$, therefore we may write
$$
\gather
(R(u-v)T_1(u))^{t_1}=T_1^{t_1}(u)R^{t_1}(u-v),
\\
(T_1(u)R(u-v))^{t_1}=R^{t_1}(u-v)T_1^{t_1}(u).
\endgather
$$
Substituting this in (13), we obtain
$$
T_1^{t_1}(u)R^{t_1}(u-v)T_2(v)=T_2(v)R^{t_1}(u-v)T_1^{t_1}(u).\leqno(14)
$$

Thirdly, applying the partial transposition $t_2$ to (14) and using the
invariance of the $R$-matrix under $t_2\circ t_1=t\otimes t$, we obtain
that
$$
T_1^{t_1}(u)T_2^{t_2}(v)R(u-v)=R(u-v)T_2^{t_2}(v)T_1^{t_1}(u).\leqno(15)
$$

Finally, conjugating both sides of (15) by $P$, then interchanging them and
replacing $(u,v)$ by $(-v,-u)$, we arrive at (12).

Thus each of the mappings (i)\ts--\ts(iv) preserves the defining relations
of $\Y(N)$. To prove Proposition 1.12 it remains to verify that all these
mappings are invertible. In cases (i), (ii) and (iv) this is clear.
For (iii) this needs a bit of work. We start with the equality
$$
(\inv T(u))T(-u)=1
$$
and apply $\inva$ to both sides. Then we get
$$
\gather
(\inva\circ\inva)(T(u))\ts\inva(T(-u))=1,
\\
(\inva\circ\inva)(T(u))(T^{-1}(u))=1,
\\
(\inva\circ\inva)(T(u))=T(u)
\endgather
$$
so that $\inva\circ\inva=\text{{\rm id}}$.

\bigskip
\proclaim {\bf 1.13. Corollary to Propositions 1.11 and 1.12}
The mappings
$$
\SS:=\inva\circ\sign : T(u)\mapsto T^{-1}(u),\leqno(1)
$$
$$
t\circ\sign : T(u)\mapsto T^t(u)\leqno(2)
$$
define antiautomorphisms of the algebra $\Y(N)$.
\endproclaim

\bigskip
\noindent
{\bf 1.14. Remark.} Note that the antiautomorphisms
$\inva$ and $\sign$ do not commute, so that
$\SS^{-1}=\sign\circ\inva\neq \SS$ and the automorphism $\SS$
is not involutive! In
fact,
$$
(\inva\circ\sign)(t_{ij}^{(M)})=\inva(t_{ij}^{(M)})(-1)^M.
$$
On the other hand, $\inva(t_{ij}^{(M)})$ is a rather complicated expression
on the generators $t_{kl}^{(L)}$, e.g.,
$$\align
\inva(t_{ij}^{(1)}) &=t_{ij}^{(1)}, \\
\inva(t_{ij}^{(2)}) &=-t_{ij}^{(2)}+\sum_{a=1}^N t_{ia}^{(1)}t_{aj}^{(1)},
\\
\inva(t_{ij}^{(3)}) &=t_{ij}^{(3)}-\sum_{a=1}^N (t_{ia}^{(1)}t_{aj}^{(2)}+
t_{ia}^{(2)}t_{aj}^{(1)})+\sum_{a,b=1}^N
t_{ia}^{(1)}t_{ab}^{(1)}t_{bj}^{(1)},
\endalign$$
so that its behavior under the antiautomorphism which sends
$t_{kl}^{(L)}$ into $t_{kl}^{(L)}(-1)^L$, is not easy to describe.
We will
calculate the square of $\SS$ later on, see Subsection 5.11.

\bigskip
\noindent
{\bf 1.15. Remark.} In what follows, we shall often use an
assertion which is a natural generalization of that used in the proof of
Proposition 1.12: if $A$ and $B$ are matrices whose coefficients
belong to an associative algebra and each
coefficient  of $A$ commutes  with any coefficient of $B$, then
$
(AB)^t=B^tA^t
$
for any antiautomorphism $t$ of the matrix algebra.

\bigskip\noindent
Now we need some preparations to prove an important result, Theorem 1.22.

\bigskip
\noindent
{\bf 1.16.} Let $E_{ij}$ be the natural basis of the Lie algebra $\gl(N)$
formed by matrix units.

\bigskip
\proclaim
{\bf Proposition} The mapping
$$
\xi : t_{ij}(u)\mapsto \delta_{ij}+E_{ij}u^{-1} \leqno(1)
$$
defines the homomorphism of algebras
$
\xi : \Y(N)\ra \U(\gl(N)).
$
\endproclaim

\noindent {\bf Proof.} By using the automorphism (1.12.6) of $\Y(N)$
we obtain that the required property of $\xi$ is equivalent to that of
the mapping
$$
\xi': t_{ij}(u)\mapsto \delta_{ji}-E_{ji}u^{-1}.
$$
To prove that $\xi'$ also defines an algebra homomorphism
$
\xi':\Y(N)\ra\U(\gl(N)),
$
we have to verify that the ternary relation holds for
$$
T(u)=1-u^{-1}\sum_{i,j}E_{ij}\otimes E_{ji}. \leqno(2)
$$
However, due to (1.4.1 and 1.4.2), the expression (2) coincides with
$R(u)$.
Hence the required statement is equivalent to the formula
$$
R_{23}(u-v)R_{12}(u)R_{13}(v)=R_{13}(v)R_{12}(u)R_{23}(u-v).
$$
After conjugation by $P_{23}$ it turns into the Yang-Baxter equation
(1.5.2)
written in a slightly different form.

\bigskip
\proclaim
{\bf 1.17. Proposition} The mapping
$
\eta: E_{ij}\mapsto t_{ij}^{(1)}
$
defines the inclusion of the algebra $\U(\gl(N))$ into $\Y(N)$.
\endproclaim

\noindent {\bf Proof.} It follows from the commutation relations (1.2.1)
that $\eta$ is extended to an algebra homomorphism. It is clear that
$\xi\circ\eta=\text{id}$, so the kernel of $\eta$ is trivial.

\bigskip
\noindent
{\bf 1.18. Remark.} Denote by $E$ the $N\times N$-matrix whose entries are
$E_{ij}$, i.e.,
$$
E:=\sum_{i,j}E_{ij}\otimes E_{ij}\in\Enda(\E\otimes \E)
$$
and set
$
\Cal T(u):=1+Eu^{-1}.
$
We can summarize the previous results as follows: the fact that $\Cal T(u)$
satisfies the ternary relation is equivalent
to the fact that the basis elements
$E_{ij}$ satisfy the commutation relations
$$
[E_{ij}, E_{kl}]=\delta_{kj}E_{il}-\delta_{il}E_{kj}.
$$

\bigskip
\noindent
{\bf 1.19.} We shall also need the composition
$
\xi\circ\inv: \Y(N)\ra \U(\gl(N))
$
of the homomorphism $\xi$ and the inversion (1.12.5):
$$
\xi\circ\inv: T(u)\mapsto \Cal T(-u)^{-1},
$$
that is
$$
\xi\circ\inva(t_{ij}^{(M)})=\sum_{a_1,\cdots,a_{M-1}=1}^N E_{ia_1}
E_{a_1a_2}\dots E_{a_{M-1}j}.\leqno (1)
$$

\bigskip
\noindent
{\bf 1.20. Definition.} The algebra $\Y(N)$ is equipped with two different
ascending filtrations which are obtained by defining the degree of a
generator
in two different ways:
$$
\degaa_1(t^{(M)}_{ij})=M \qquad\hbox{and}\qquad
\degaa_2(t^{(M)}_{ij})=M-1, $$
respectively. Let $\gr_1\Y(N)$ and $\gr_2\Y(N)$ denote the corresponding
graded algebras.

\bigskip
\proclaim
{\bf 1.21. Corollary} The algebra $\gr_1\Y(N)$ is commutative.
\endproclaim

\noindent {\bf Proof.} This follows directly from (1.2.1) since the degree
$\degaa_1(\cdot)$
of each term in the right hand side of (1.2.1) is less than that of the
left hand side.

\proclaim
{\bf 1.22. Theorem} Let $\bar t^{\,(M)}_{ij}$ stand for the image of the
generator $t^{(M)}_{ij}$ in the $M$-th component of $\gr_1\Y(N)$. The
elements
$\bar t^{\,(M)}_{ij}$ are algebraically independent, so that $\gr_1\Y(N)$
is the algebra of polynomials in countably many variables
$\bar t^{\,(M)}_{ij}$.
\endproclaim

\noindent {\bf Proof.} It follows from the defining relations (1.1.1) that
for any $N'\geq N$ there is a natural homomorphism
$
\iota:\Y(N)\ra \Y(N').
$
Taking the composition of $\iota$ and the homomorphism
$
\xi\circ\inva: \Y(N')\ra\U(\gl(N')),
$
we get another homomorphism
$\zeta:=\xi\circ\inva\circ\iota$,
$$
\zeta: \Y(N)\to \U(\gl(N'))
$$
such that
$$
\zeta(t_{ij}^{(M)})=\sum_{a_1,\cdots,a_{M-1}=1}^{N'} E_{ia_1}
E_{a_1a_2}\dots E_{a_{M-1}j}.
$$

It prerserves the filtration of $\Y(N)$ defined by $\degaa_1$ and the
canonical
filtration of $\U(\gl(N'))$. Therefore it determines a homomorphism of
graded
algebras
$$
\bar\zeta: \gr_1\Y(N)\ra\SS(\gl(N')).
$$
We shall consider elements of the symmetric algebra $\SS(\gl(N'))$
as polynomial
functions on  $\gl(N')$. Thus the image of $\bar t^{\,(M)}_{ij}$ under
$\bar\zeta$ is the polynomial $p_{ij}^{(M)}$ such that
$$
p_{ij}^{(M)}(x)=(x^M)_{ij},\qquad x\in\gl(N').
$$
Now it suffices to prove that for each fixed positive integer $M'$ all the
polynomials $p_{ij}^{(M)},\  1\leq i,j\leq N,\   1\leq M\leq M'$,
are algebraically independent for sufficiently large $N'$.

For any triple $(i,j,M)$ satisfying the conditions
$1\leq i,j\leq N,\  1\leq M\leq M'$
we can choose a subset
$$
\Omega_{ij}^{(M)}\subset\{N+1, N+2,\dots\}
$$
of cardinality $M-1$ in such a way that all these subsets are disjoint.
Let $N'$ be so large that all of them belong to $\{N+1, N+2,\dots,N'\}$.
Let $y_{ij}^{(M)}$ be complex parameters. Define a linear
operator $x_{ij}^{(M)}$ in $\C^{N'}$ depending on $y_{ij}^{(M)}$ as
follows.
Let $e_1,\dots,e_{N'}$ be the canonical basis in $\C^{N'}$ and
$a_1<\dots<a_{M-1}$ be all the  elements of $\Omega_{ij}^{(M)}$. Then put
$$
\gather
x_{ij}^{(M)}: e_j\mapsto y_{ij}^{(M)}e_{a_{M-1}},\ \  e_{a_{M-1}}\mapsto
e_{a_{M-2}},\ \dots, e_{a_1}\mapsto e_i,
\\
x_{ij}^{(M)}: e_k\mapsto 0\quad \text{for}\quad  k\notin \{j\}\cup
\Omega_{ij}^{(M)}
\endgather
$$
and set
$$
x=\sum_{i,j,M}  x_{ij}^{(M)}. \leqno(1)
$$
Then for any matrix $x$ of the form (1) we have
$$p_{ij}^{(M)}(x)=y_{ij}^{(M)}+\psi$$
for certain polynomial $\psi$ in $y_{kl}^{(L)}$ where $L<M$.
Thus the polynomials $p_{ij}^{(M)}$ are algebraically independent even if
they are restricted to the affine subspace
of matrices of the form (1). Theorem 1.22 is proved.

\proclaim
{\bf 1.23. Corollary} Given an arbitrary linear order on the set of
the generators
$t^{(M)}_{ij}$, any element of the algebra $\Y(N)$ is uniquely written
as a linear combination of ordered monomials in the generators.
\endproclaim

\bigskip
\noindent {\bf 1.24. Remark.} Theorem 1.22 (or the equivalent statement
given in Corollary 1.23) is a fundamental fact which may be called {\it the
Poincar\'e--Birkhoff--Witt theorem for the Yangian} $\Y(N)$.

\bigskip
\noindent {\bf 1.25. Remark.} Theorem 1.22
implies that $\Y(N)$ can be viewed as
a flat deformation of the algebra of polynomials in countably many
variables. To see this, for each $h\in {\Bbb C}\setminus\{0\}$ consider the
algebra $\Y(N,h)$ with the generators $t_{ij}^{(M)}$ and the relations
obtained from (1.2.1) by multiplying the right hand side by $h$:
$$
[\ts t_{ij}^{(M)}, t_{kl}^{(L)}\ts]=h\cdot \sum_{r=0}^{\min(M,L)-1}
(\ts t_{kj}^{(r)}\ts t_{il}^{(M+L-1-r)}-t_{kj}^{(M+L-1-r)}\ts
t_{il}^{(r)}\ts).
\leqno (1)
$$
The algebras $\Y(N,h)$ are all isomorphic to each other; an isomorphism
$\Y(N,h)\to \Y(N)$ can be defined by $t_{ij}^{(M)}\mapsto t_{ij}^{(M)}h^M$.
On the other hand, in the limit $h\to 0$ we obtain from $\Y(N,h)$ the
algebra of polynomials in the generators $t_{ij}^{(M)}$.

\bigskip
\noindent {\bf 1.26.} Now let us turn to the second filtration. Consider
the
{\it polynomial current Lie algebra}
$$
\gl(N)[x]:=\gl(N) \otimes_{\C}{\C}[x] \leqno(1)
$$
and its universal enveloping algebra $\U(\gl(N)[x])$.
There is a natural basis in $\gl(N)[x]$ formed by the elements
$E_{ij}x^{M-1}$ where $1\le i,j\le N$ and $M=1,2,\ldots$\ .

\proclaim
{\bf Theorem} The algebra $\gr_2\Y(N)$ is isomorphic to the algebra
$\U(\gl(N)[x])$.
\endproclaim

\noindent {\bf Proof.} Let us examine the degree
$\degaa_2(\cdot)$ of the different terms in (1.2.1).
The degree of the left hand side
equals $M+L-2$. The degree of each of the terms in the right hand side
equals $M+L-3$ except the term with $r=0$. The latter one may be written as
$$
t^{(0)}_{kj}t^{(M+L-1)}_{il}-t^{(M+L-1)}_{kj}t^{(0)}_{il}=
\delta_{kj}t^{(M+L-1)}_{il}-\delta_{il}t^{(M+L-1)}_{kj} \leqno(2)
$$
and has the same degree $M+L-2$ as the left hand side. This implies that
in the algebra $\gr_2\Y(N)$, the commutation relations take the form
$$
[\tilde t^{\,(M)}_{ij}, \tilde t^{\,(L)}_{kl}]=
\delta_{kj}\tilde t^{\,(M+L-1)}_{il}-
\delta_{il}\tilde t^{\,(M+L-1)}_{kj},
\leqno(3)
$$
where $\tilde t^{\,(M)}_{ij}$ stands for the image of $t^{(M)}_{ij}$
in the $(M-1)$-th component of $\gr_2\Y(N)$. Observe now that
relations (3) are exactly
the commutation relations of the Lie algebra $\gl(N)[x]$ in its basis
$\{E_{ij}x^{M-1}\}$. Thus there exists an algebra morphism
$E_{ij}x^{M-1} \mapsto \tilde t^{\,(M)}_{ij}$ from $\U(\gl(N)[x])$ onto
$\gr_2\Y(N)$. The kernel of this morphism is trivial by Theorem 1.22.

\bigskip
\noindent {\bf 1.27. Remark} (cf. Remark 1.25). The algebra $\Y(N)$
may be also regarded as a flat deformation of the algebra $\U(\gl(N)[x])$.
Indeed, let us renormalize the generators of $\Y(N)$ by multiplying
$t^{(M)}_{ij}$ by $h^{M-1}$ (instead of $h^M$ as before). This results in
the
following modification of relations (1.2.1): the numerical factor $h$
will appear in all the terms of the right hand side
of (1.2.1) except for the term (1.2.1) corresponding to $r=0$.
Let $\Y_h(N)$ denote the algebra defined by these modified relations.
Then $\Y_1(N)=\Y(N)$ and $\Y_0(N)=\U(\gl(N)[x])$. The flatness
of the deformation $\{\Y_h(N): h\in\C\}$ is again guaranteed by Theorem
1.22.

\bigskip
\proclaim {\bf 1.28. Theorem } The Yangian $\Y(N)$ is a Hopf algebra
with respect to the coproduct $\Delta : \Y(N)\rightarrow
\Y(N)^{\otimes 2}$ defined by
$$
\Delta (t_{ij}(u)):=\sum_{a=1}^N t_{ia}(u)\otimes
t_{aj}(u),
$$
the antipode $\SS$ defined by {\rm (1.13.1)} and the counit
$\epsilon$ 
defined by
$
\epsilon (T(u)):=1.
$
\endproclaim

\Proof  To work with the coproduct $\Delta$ it is more convenient to
rewrite
its definition in terms of the $T$-matrix. To do this we will
generalize the notation adopted in Subsection 1.6 a little bit.

Suppose that we are dealing with the tensor product of $m$ copies of
$\Y(N)[[u^{-1}]]$ and $n$ copies of $\End\E$. Then for any numbers $k$,$l$
such
that $1\le k\le m$ and $1\le l\le n$, set
$$
\eqalignno{
T_{[k]l}(u)&:=\sum_{i,j=1}^N (1^{\otimes k-1}\otimes t_{ij}(u)
\otimes 1^{\otimes m-k})\otimes(1^{\otimes l-1}\otimes E_{ij}
\otimes 1^{\otimes n-l})\cr
&\in \Y(N)[[u^{-1}]]^{\otimes m}\otimes (\End\E)^{\otimes n}.\cr}
$$
Using the informal language of Remark 1.9 one could say that
$T_{[k]l}(u)$ is the operator in
$W^{\otimes m}\otimes \E^{\otimes n}$ which acts as $T(u)$ in the
product of the $k$-th copy of $W$ and the $l$-th copy of $\E$ and
as 1 in all other copies of these spaces.

When $m=1$, we prefer to abbreviate $T_{l}(u):=T_{[1]l}(u)$ according
to our usual convention, and when $n=1$, we abbreviate
$T_{[k]}(u):=T_{[k]1}(u)$.
Now we may rewrite the definition of $\Delta$ in the form
$$
\Delta (T(u)):= T_{[1]}(u)T_{[2]}(u),
$$
which is most suitable for our purposes.

Let us verify the `main' axiom, the compatibility of the product
and the coproduct. This means that $\Delta$ is an algebra morphism
of $\Y(N)$ to $\Y(N)\otimes \Y(N)$ or, by the definition of the Yangian,
that $\Delta (T(u))$ satisfies the ternary relation (1.8.1), or else that
$$
R_{12}(u-v)T_{[1]1}(u)T_{[2]1}(u)T_{[1]2}(v)T_{[2]2}(v)=
T_{[1]2}(v)T_{[2]2}(v)T_{[1]1}(u)T_{[2]1}(u)R_{12}(u-v).\leqno(1)
$$

The key observation here is the fact that $T_{[2]1}(u)$ and
$T_{[1]2}(v)$, as well as $T_{[1]1}(u)$ and $T_{[2]2}(v)$, commute.
Using  this, we transform the left hand side of (1) to the right one
as follows.
We interchange first the commuting $T$-matrices
$T_{[2]1}(u)$ and $T_{[1]2}(v)$, then
$T_{[1]1}(u)$ with $T_{[1]2}(v)$ using the ternary relation
$$
R_{12}(u-v)T_{[1]1}(u)T_{[1]2}(v)=
T_{[1]2}(v)T_{[1]1}(u)R_{12}(u-v),
$$
then $T_{[2]1}(u)$ with $T_{[2]2}(v)$ using
the ternary relation again, and finally we interchange the commuting
$T$-matrices $T_{[1]1}(u)$ and $T_{[2]2}(v)$. The result of these
transformations is the right hand side of (1).

Other axioms follow directly from the definitions of $\Delta$,
$\SS$ and $\epsilon$.

\bigskip
\noindent
{\bf 1.29. Remark.} It is easily verified that the Yangian $\Y(N)$
is a deformation of the universal enveloping algebra
$\U(\gl(N)[x])$ not only as an algebra (Remark 1.27) but as a Hopf
algebra too.

\bigskip
\noindent
{\bf 1.30. Remark.} The coproduct $\Delta$ is not cocommutative.

\bigskip
\noindent
{\bf 1.31. Comments.} The main information about the structure
 of
(general) Yangians is contained in Drinfeld's works [D1, D3, D4].

Concerning the
$R$-matrix formalism see for instance the papers Takhtajan--Faddeev [TF],
Kulish--Sklyanin [KS2], Reshetikhin--Takhtajan--Faddeev [RTF].

The fact that the commutation relations of $\gl(N)$ can be written in an
$R$-matrix form shows that the Yangian $\Y(N)$ is a natural
`superstructure'
over $\U(\gl(N))$. The existence of a projection $\Y(N)\to \U(\gl(N))$
gives rise to a connection between the Yangians and conventional
representation theory (see Cherednik [C3] and Nazarov--Tarasov [NT]
for some applications).

Prior to the appearance of the Yangians, the idea of
combining generators of a classical Lie algebra into a matrix was used in
the
work of Perelomov--Popov [PP] and in the works of
Bracken--Green [BG] and Green [Gr] on the so-called
characteristic identities.

The Poincar\'e--Birkhoff--Witt
theorem (PBW) for general Yangians is due to V.G.Drinfeld. As he
has communicated to one of the authors, he derived  this
theorem from the PBW for quantized loop algebras. Another proof of PBW has
recently been given in Levendorski\u\i's note [L1]. Our proof of PBW,
presented in Subsection 1.22, follows the approach of Olshanski\u\i's paper
[O1]; see especially Lemma 2.1.11 in [O1].

One of Drinfeld's results ([D1,
Theorem 2]) shows that the Yangians admit a characterization as the
canonical
deformations of the current Lie algebras, where `canonical' means
`satisfying
certain natural conditions'.

The coproduct $\Delta$, the antipodal
map $\SS$ and the shift automorphisms of
$\Y(N)$ play a key role in constructing
the finite-dimensional representations of $\Y(N)$.

\newpage

\heading {\bf 2. The quantum determinant $\qdet T(u)$
and the center of $\Y(N)$}\endheading

\noindent
Here we introduce the quantum determinant of the matrix $T(u)$,
which is a formal power series in $u^{-1}$ with coefficients
from the Yangian $\Y(N)$. We prove that all the coefficients
belong to the center of $\Y(N)$, that they are algebraically independent
and generate the whole center. We introduce the Yangian for the Lie
algebra $\ssl(N)$ and prove that the algebra
$\Y(N)$ is isomorphic to the tensor product
of its center and the Yangian for $\ssl(N)$.
We will keep to the notation of Section 1.

\bigskip
\noindent
{\bf 2.1.} Let $u_1,\ldots,u_m$ be formal variables. Set
$$
R(u_1,\ldots , u_m) :=(R_{m-1,m})(R_{m-2,m}R_{m-2,m-1}) \cdots (R_{1m}
\cdots
R_{12}),\leqno(1)
$$
where  we abbreviate $R_{ij}:=R_{ij}(u_i-u_j)$.

\proclaim
{\bf Proposition}
We have the following fundamental identity:
$$
R(u_1,\ldots, u_m)\, T_1 (u_1) \cdots T_m (u_m) = T_m (u_m) \cdots T_1
(u_1)\, R(u_1,\dots, u_m).\leqno(2)
$$
\endproclaim

\Proof To simplify the notation, set $T_i := T_i(u_i)$.
First, let us check the identity
$$
(R_{1m}\cdots R_{12})T_1(T_2\cdots T_m)=
(T_2\cdots T_m)T_1(R_{1m}\cdots R_{12}). \leqno(3)
$$
Indeed, the left hand side of (3) equals
$$
\eqalign{
(R_{1m}\cdots R_{12})T_1(T_2\cdots T_m)&=
R_{1m}\cdots R_{13}(R_{12}T_1T_2)T_3\cdots T_m \cr
&=R_{1m}\cdots R_{13}(T_2T_1R_{12})T_3\cdots T_m  \cr
&=T_2(R_{1m}\cdots R_{13})T_1(T_3\cdots T_m)R_{12}, \cr}
$$
where the passage from the first to the second line is based on the ternary
relation, and  the last transformation is justified by the fact that
the matrices $R_{ij}$ and $T_k$ with disjoint indices are
pairwise permutable.
Repeating the same procedure we can interchange
$T_1$ with $T_3,\ldots ,T_m$. This proves (3).

Next, observe that
$$
R(u_1,\ldots ,u_m)=R(u_2,\ldots ,u_m)\,(R_{1m}\ldots R_{12}).\leqno(4)
$$
Using this and (3), we can interchange $T_1$ with $(T_2\cdots T_m)$ as
follows:
$$
\eqalign{
R(u_1,\ldots ,u_m)T_1T_2\cdots T_m&=
R(u_2,\ldots ,u_m)\,(R_{1m}\cdots R_{12})T_1T_2\cdots T_m\cr
&=R(u_2,\ldots ,u_m)\,(T_2\cdots T_m)T_1(R_{1m}\cdots R_{12}).\cr}
$$
Similarly we interchange $T_2$ with ${(T_3\cdots T_m)}$ etc. Finally we
arrive
at the right hand side of (2).

\bigskip
\noindent
{\bf 2.2. Remark.} Let $u$ and $v$ be formal variables and $c\in \C$ a
constant. In contrast to the case of the algebra $\Y(N)[[u,v]]$, in the
algebra
$\Y(N)[[u^{-1}, v^{-1}]]$ it is possible to perform the specialization
$v=u-c$. This means that there exists a natural algebra morphism
$$
\Y(N)[[u^{-1},v^{-1}]] \rightarrow \Y(N)[[u^{-1}]] \leqno(1)
$$
such that
$$
\sum_{k,l=0}^{\infty}a_{kl}u^{-k}v^{-l}\mapsto
\sum_{k,l=0}^{\infty}a_{kl}u^{-k}(u-c)^{-l}
=\sum_{k,l=0}^{\infty}a_{kl}u^{-k-l}(1+\sum_{r=1}^{\infty}
c^ru^{-r}).
$$
Also note that this specialization is compatible with the localization
relative to $u^{-1}-v^{-1}$ provided $c\neq 0$.
This remark will allow us to use the fundamental identity (2.1.2) when
$u_1,\ldots, u_m$ are not independent but subject to certain relations
with each other.

\bigskip
\noindent
{\bf 2.3.} Let $\Sym_m$ denote the symmetric group realized as the group of
permutations of the set $\{1,\ldots,m\}$ and let
$$
a_m=\sum_{p\in \Sym_m}\sgn(p)\cdot p\in\C[\Sym_m] \leqno(1)
$$
denote the antisymmetrizer in the group ring. Consider the natural action
of
$\Sym_m$ in the tensor space $\E^{\otimes m}$ and denote by $A_m$ the
image of the normalized antisymmetrizer $(m!)^{-1}a_m$.

\proclaim
{\bf Proposition}  If $u_i-u_{i+1}=1$ for $i=1,\ldots,m-1$ then
$$
R(u_1,\ldots,u_m)=m!\,A_m. \leqno(2)
$$
\endproclaim

\noindent
{\bf Proof.} Let $p_{ij} \in \Sym_m$ denote the transposition
$(i,j)$. Then, in the notation of Proposition 2.1,
$$
R_{ij}=\ \ \hbox{the image of}\quad 1-{p_{ij}\over u_i-u_j}.
$$
Hence (2) is provided by the following `multiplicative formula' for
the antisymmetrizer $a_m\in \C[\Sym_m]$ :
$$
a_m=\prod^{m-1}_{k=1}{}^{\leftarrow}
\prod^{m}_{l=k+1}{}^{\leftarrow}(1-{p_{kl}\over l-k}).
\leqno(3)
$$
Here and below the symbol $\prod{}^{\leftarrow}$ means that the factors
in the product are written from right to left.

Denoting the right hand side of (3) by $b_m$,
let us prove $a_m=b_m$ by induction on $m$. For $m=2$ this is obvious:
$$
b_2=1-{p_{12}\over 2-1}=1-p_{12}=a_2.
$$
Now, assuming that $m>2$ and $a_{m-1}=b_{m-1}$, let us check that
$a_m=b_m$.

Let $\Sym_{m-1}$ be identified with the stabilizer of $m$ in $\Sym_m$ so
that
$\C[\Sym_{m-1}]$ is contained in $\C[\Sym_m]$. Observe that
$$
a_m=(1-p_{1m}-\cdots-p_{m-1,m})a_{m-1}. \leqno(4)
$$

On the other hand, observe that in the double product (3) all
the factors with $l=m$ may be moved to the left, so that we obtain
$$
b_m=\{\prod_{k=1}^{m-1}{}^{\leftarrow}(1-{p_{km}\over m-k})\}\, b_{m-1}
=\{\prod_{k=1}^{m-1}{}^{\leftarrow}(1-{p_{km}\over m-k})\}\, a_{m-1}\,1;
\leqno(5)
$$
the assumption $a_{m-1}=b_{m-1}$ has been used here.

Now we will consecutively open the brackets in the right hand side of (5).
First do this in the factor with $k=1$ which is the extreme right:
$$
b_m=\{\prod^{m-1}_{k=2}{}^{\leftarrow}(1-{p_{km}\over m-k})\}\,
a_{m-1}\,-\,{1\over m-1}\{\prod^{m-1}_{k=2}{}^{\leftarrow}
(1-{p_{km}\over m-k})\}\,p_{1m} a_{m-1}. \leqno(6)
$$
However, $p_{km}p_{1m}=p_{1m}p_{k1}$ for $k=2,\ldots,m-1$, so that
$$
{1\over m-1}\{\prod^{m-1}_{k=2}{}^{\leftarrow}(1-{p_{km}\over m-k})\}\,
p_{1m}
={p_{1m}\over m-1}\,(1-{p_{m-1,1}\over 1})\,\cdots\,(1-{p_{21}\over m-2})\,
a_{m-1}. \leqno(7)
$$
Since $p_{k1}a_{m-1}=-a_{m-1}$ for $k=2,\ldots,m-1$, the right hand side
of (7) equals
$$
{p_{1m}\over m-1}\,(1+{1\over 1})\, \cdots \,(1+{1\over m-2})\, a_{m-1}=
{p_{1m}\over m-1}\;{2\over 1}\;{3\over 2}\;\cdots\;{m-1\over m-2}\, a_{m-1}
=p_{1m}a_{m-1}.
$$
Substituting this into (6), we obtain (compare with (5))
$$
b_m=\{\prod^{m-1}_{k=2}{}^{\leftarrow}(1-{p_{km}\over m-k})\}\, a_{m-1}\,
-\,p_{1m}a_{m-1}. \leqno(8)
$$

Next we open the brackets in the factor with $k=2$, repeat the same
transformations etc. Finally we arrive at the last factor
$(1-{p_{m-1,m}/1})$
for which no transformations are required, and
obtain
$$
b_m=(1-p_{m-1,m}-\cdots -p_{1m})\,a_{m-1}. \leqno(9)
$$
Combining (9) with (4) we conclude that $a_m=b_m$.

\proclaim
{\bf 2.4. Proposition} The following identity holds:
$$
A_N\, T_1(u)\cdots T_N(u-N+1)=T_N(u-N+1)\cdots T_1(u)\, A_N. \leqno(1)
$$
Moreover, we have
$$
\eqalignno{
(1)&=A_N\, T_1(u)\cdots T_N(u-N+1)\, A_N, &(2) \cr
(1)&=A_N\, T_N(u-N+1)\cdots T_1(u)\, A_N. &(3) \cr}
$$
\endproclaim

\noindent
{\bf Proof.} Applying Propositions 2.1 and 2.3 to $m=N$, we obtain (1).
To prove (2), we have to multiply
both sides of (1) by $A_N$ on the right. Then the right hand side
will not change since $A_N^2=A_N$, and the left hand side will turn
into the right hand side of (2). Similarly, to prove (2), it suffices to
multiply (1) by $A_N$ on the left.

\bigskip
\proclaim
{\bf 2.5. Proposition} There exists a formal series
$$
\qdet T(u):=1+d_1u^{-1}+d_2u^{-2}+\cdots \in \Y(N)[[u^{-1}]] \leqno(1)
$$
such that {\rm (2.4.1)} equals $\qdet T(u) A_N$.
\endproclaim

\Proof Observe that $A_N$ is a one-dimensional projection: it projects
$\E^{\otimes N}$ onto $\C \xi$, where
$$
\xi :=\sum_{p\in \Sym_N} \sgn (p)e_{p(1)}\otimes\cdots\otimes e_{p(N)}
\leqno(2)
$$
and $e_1,\ldots, e_N$ is the canonical basis of $\C^N$. Hence (2.4.1)
equals $A_N$ times a formal series in $u^{-1}$ with coefficients in
$\Y(N)$. It remains to check that this series begins with 1; but this
follows from the fact that each of the series $T_i(u-i+1)$,
$i=1,\ldots, N$, begins with 1.

\bigskip\noindent
{\bf 2.6. Definition.} $\qdet T(u)$ is called the {\it quantum determinant}
of the matrix $T(u)$.

\bigskip
\proclaim
{\bf 2.7. Proposition} We have
$$
\eqalignno{
\qdet\, T(u)&=\sum_{p\in \Sym_N} \sgn(p)\, t_{p(1),1}(u)\cdots
t_{p(N),N}(u-N+1)
&(1)\cr
&=\sum_{p\in \Sym_N} \sgn(p)\, t_{1,p(1)}(u-N+1)\cdots
t_{N,p(N)}(u).&(2)\cr}
$$
\endproclaim

For example, if $N=1$, then $\qdet T(u)=t_{11}(u)$;
if $N=2$, then
$$
\eqalignno{
\qdet T(u)&=t_{11}(u)t_{22}(u-1)-t_{21}(u)t_{12}(u-1) \cr
           &=t_{11}(u-1)t_{22}(u)-t_{12}(u-1)t_{21}(u). &(3)\cr}
$$

\Proof To prove (1), we start with the identity
$$
\qdet T(u)\, A_N=A_N\,T_1(u)\cdots T_N(u-N+1). \leqno(4)
$$
Let us apply both sides of (4) to the vector
$e_1\otimes\cdots\otimes e_N$.
Then on the left we obtain
$$
\qdet T(u)\, A_N(e_1\otimes\cdots\otimes e_N)=
(N!)^{-1}\,\qdet T(u)\xi, \leqno(5)
$$
while on the right we get
$$
\displaylines{
\quad A_N\,\sum_{i_1,\ldots,i_N,j_1,\ldots,j_N=1}^N t_{i_1j_1}(u)\cdots
t_{i_Nj_N}(u-N+1)\,(E_{i_1j_1}\otimes\cdots\otimes E_{i_Nj_N})
(e_1\otimes\cdots\otimes e_N)\hfill\cr
\hfill {}=\sum_{i_1,\ldots,i_N=1}^N t_{i_1,1}(u)\cdots t_{i_N,N}(u-N+1)\,
A_N
\,(e_{i_1}\otimes\cdots \otimes e_{i_N}).\qquad (6)\cr}
$$
If the indices $i_1,\ldots,i_N$ are pairwise distinct, then the vector
$A_N(e_{i_1}\otimes\cdots\otimes e_{i_N})$ is equal to
$(N!)^{-1}{\sgn\,((i_1,\ldots,i_N))\,\xi}$; otherwise it equals 0.
Hence (6) equals
$(N!)^{-1}\,c\xi$ where $c$ stands for the right hand side of (1). Thus
$\qdet T(u)\xi=c\xi$, and (1) is proved.

To prove (2), we start with the identity
$$
\qdet\, T(u)\, A_N=T_N(u-N+1)\cdots T_1(u)\, A_N \leqno(7)
$$
and apply both of its sides to $\xi$. Since $A_N\xi=\xi$, we obtain
$$
\qdet T(u)\, \xi=T_N(u-N+1)\cdots T_1(u)\xi. \leqno(8)
$$
We may decompose the right hand side of (8) relative to the canonical basis
of $\E^{\otimes N}$, and
a similar calculation shows that the basis vector
$e_1\otimes\cdots \otimes e_N$ enters into this decomposition with
coefficient equal to the right hand side of (2).
This concludes the proof.

\bigskip\noindent
{\bf 2.8. Remark.} Taking the basis vector
$e_{q(1)}\otimes\cdots\otimes e_{q(N)}, \  q\in \Sym_N$,\ instead of
$e_1\otimes\cdots\otimes e_N$ in the above proof, one
could obtain two other
expressions for $\qdet T(u)$, namely
$$
\eqalignno{
\qdet T(u)&=\sgn(q)\,\sum_{p\in \Sym_N}\sgn(p)\, t_{p(1),q(1)}(u)
\cdots t_{p(N),q(N)}(u-N+1)&(1)\cr
&=\sgn(q)\, \sum_{p\in \Sym_N}\sgn(p)\, t_{q(1),p(1)}(u-N+1)\cdots
t_{q(N),p(N)}(u).&(2)\cr}
$$

\bigskip\noindent
{\bf 2.9. Remark.} Let $X(u)=(x_{ij}(u))_{i,j=1}^N$ be an arbitrary
matrix whose
entries are formal power series in $u^{-1}$ with coefficients from
$\Y(N)$. Then one can define the {\it quantum determinant}
of the matrix $X(u)$ as follows (cf. (2.7.1)):
$$
\qdet\, X(u)=\sum_{p\in \Sym_N} \sgn(p)\, x_{p(1),1}(u)\cdots
x_{p(N),N}(u-N+1).
\leqno(1)
$$
In order to avoid an ambiguity we shall sometimes enclose the matrix
$X(u)$ in brackets. For example, if $N\geq 2$, then the series
$$
\qdet T(-u)=1+d_1(-u)^{-1}+d_2(-u)^{-2}+\dots
$$
(see (2.5.1)) does not coincide with the quantum determinant
$\qdet (T(-u))$ of the matrix $X(u)=T(-u)$.

\bigskip
\proclaim
{\bf 2.10. Theorem} $\qdet T(u)$ lies in the center of $\Y(N)$. That is,
all of its coefficients are central elements.
\endproclaim

\Proof Consider the auxiliary tensor space
$\E^{\otimes (N+1)}$ where the copies of $\E$ are enumerated
by the indices $0,\ldots,N$, and consider the $N+1$ operators
$$
T_0:=T_0(v),\quad T_1:=T_1(u),\,\ldots,\, T_N:=T_N(u-N+1). \leqno(1)
$$
We shall prove the identity
$$
T_0(v)\,\qdet\,T(u)\, A_N=\qdet\,T(u)\,T_0(v)\, A_N, \leqno(2)
$$
where $\qdet\,T(u)$ is built from $T_1,\ldots, T_N$ as above and $A_N$
corresponds to the antisymmetrization relative to the indices $1,\ldots,N$.
It is easy to see that (2) implies the theorem. We shall derive (2)
from the fundamental identity (2.1.2) in several steps.

{\sl Step} 1. Applying (2.1.2) to the operators (1), we obtain
$$
R(v,u,u-1,\ldots,u-N+1)\, T_0T_1\cdots T_N=T_N\cdots T_1T_0\,
R(v,u,u-1,\ldots,u-N+1). \leqno(3)
$$
Since
$$
R(v,u,u-1,\ldots,u-N+1)=R(u,u-1,\ldots,u-N+1)\,\prod_{i=1}^N{}^{\leftarrow}
R_{0i}=N!\,A_N\,\prod_{i=1}^N{}^{\leftarrow}R_{0i},
$$
(3) may be rewritten as follows
$$
A_N\, (\prod_{i=1}^N{}^{\leftarrow}R_{0i})\, T_0T_1\cdots T_N=
T_N\cdots T_1T_0A_N\,(\prod_{i=1}^N{}^{\leftarrow}R_{0i}). \leqno(4)
$$

{\sl Step} 2. Let us prove that
$$
A_N\,R_{0N}\cdots R_{01}=R_{01}\cdots R_{0N}\, A_N. \leqno(5)
$$
To do this, rewrite (5) as
$$
A_N\, R_{01}^{-1}\cdots R_{0N}^{-1}=R_{0N}^{-1}\cdots R_{01}^{-1}\, A_N
\leqno(6)
$$
and observe that the structure of this formula is quite similar to that of
(2.4.1).

Now, if we examine the proof of (2.4.1), then we will see that it is
based entirely on the identity
$$
R_{ij}T_iT_j=T_jT_iR_{ij}.
$$
But the same identity holds when $T_i$, $T_j$ are replaced by
$R_{0i}^{-1}$, $R_{0j}^{-1}$ respectively. Indeed,
$$
R_{ij}R_{0i}^{-1}R_{0j}^{-1}=R_{0j}^{-1}R_{0i}^{-1}R_{ij}
$$
is simply equivalent to the Yang--Baxter equation (see (1.5.1))
$$
R_{0i}R_{0j}R_{ij}=R_{ij}R_{0j}R_{0i}.
$$
Hence the proof of (2.4.1) works for (6) as well.

{\sl Step} 3. Using (5), we transform the left hand side of (4)
as follows:
$$
\eqalign{
&A_N\,(R_{0N}\cdots R_{01})\, T_0T_1\cdots T_N \cr
&\quad =A_N^2\,(R_{0N}\cdots R_{01})\, T_0T_1\cdots T_N \cr
&\quad =A_N\,(R_{01}\cdots R_{0N})\, A_NT_0T_1\cdots T_N
\quad\hbox{by (5)}\cr
&\quad =A_N\,(R_{01}\cdots R_{0N})\, A_N^2T_0T_1\cdots T_N.\cr}
$$
Since $T_0$ and $A_N$ commute, this equals
$$
\eqalign{
&{}A_N\,(R_{01}\cdots R_{0N})\, A_NT_0A_NT_1\cdots T_N \cr
&\quad =A_N\,(R_{01}\cdots R_{0N})\, A_NT_0\,\qdet T(u)\, A_N. \cr}
$$
By applying similar transformations to the right hand side of (4) we
arrive at the following identity:
$$
\eqalignno{
&{}A_N\,(R_{01}\cdots R_{0N})\, A_NT_0(v)\,\qdet T(u)\, A_N \cr
&\quad =\qdet T(u)\, A_NT_0(v)A_N\,(R_{01}\cdots R_{0N})\, A_N. &(7)\cr}
$$

{\sl Step} 4. Let us prove the identity
$$
A_N\,(R_{01}\ldots R_{0N})\,A_N=f(u,v)\,A_N, \leqno(8)
$$
where $f(u,v)$ is a {\sl non zero} element of an appropriate extension
of $\C$ containing
$$
(v-u_i)^{-1}=(v-u+i-1)^{-1},\qquad\hbox{where}\quad i=1,\ldots, N.
$$
Indeed, write $\E^{\otimes(N+1)}$ as ${\E\otimes\E^{\otimes N}}$
and recall that $A_N$ is a one--dimensional projection in $\E^{\otimes N}$.
It follows that the left hand side of (8) may be written as
${X\otimes A_N}$, where $X$ is an operator in $\E$ (more correctly, an
element
of $\End\E$ tensored with our extension of $\C$).

On the other hand, the whole picture is clearly equivariant relative to the
action of the group ${\text{{\rm Aut}}\,\E=\text{{\rm GL}}(N,\C)}$, so that
$X$
is a scalar operator, i.e., ${X=f(u,v)\cdot 1}$.

It remains to check that $f(u,v)\neq 0$. To do this, take as
the above-mentioned  extension
the algebra ${\C[u][[v^{-1}]]}$ and observe that
$$
\eqalign{
R_{0i}&=1-{P_{0i}\over v-u+i-1} \cr
      &=1-P_{0i}v^{-1}(1+(u-i+1)v^{-1}+(u-i+1)^2v^{-2}+\ldots). \cr}
$$
This implies that $f(u,v)$, as a power series in $v^{-1}$ with coefficients
in $\C[u]$, begins with 1. Thus $f(u,v)\neq 0$.

{\sl Step} 5. Now, by (8) identity (7) reads as follows:
$$
T_0(v)\qdet T(u)\, A_N=\qdet T(u)\,T_0(v)\, A_N.
$$
This clearly means that $\qdet T(u)$ is central.

\bigskip\noindent
{\bf 2.11. Remark.} Theorem 2.10 may be applied to obtain
the (well-known) description of the center of the universal
enveloping algebra $\U(\gl(N))$. To do this one uses the homomorphism
$\xi$ (see (1.16.1)). It is easy to see that
$$
u(u-1)\dots(u-N+1)\,\xi(\qdet T(u))=
$$
\
$$
\det\left( \matrix
E_{11}+u&E_{12}&\hdots&E_{1N}\\
E_{21}&E_{22}+u-1&\hdots&E_{2N}\\
\vdots&\vdots&&\vdots\\
E_{N1}&E_{N2}&\hdots&E_{NN}+u-N+1
\endmatrix \right), \leqno(1)
$$
where the `determinant' $\det A$ of a
noncommutative matrix $A=(a_{ij})_{i,j=1}^N$ is defined as
$$
\det A:=\sum_{p\in \Sym_N}\sgn(p)a_{p(1),1}\dots a_{p(N),N}.
$$
Let us denote the right hand side of (1) by $Q(u)$. Then
$$
Q(u)=u^N+z_1u^{N-1}+\dots+z_N, \qquad z_i\in \U(\gl(N)).
$$
Theorem 2.10 implies that all the coefficients $z_i$ belong to the
center of $\U(\gl(N))$. By using the Harish-Chandra homomorphism
one can show that the elements $z_1,\dots,z_N$ are algebraically
independent and hence generate the whole center of the algebra
$\U(\gl(N))$ (cf. Theorem 2.13). Moreover, the polynomial
$\tilde Q(u)=Q(-u+N-1)$ may be
considered as the `characteristic polynomial' for the matrix
$E=(E_{ij})$ (see Remark 1.18), and the following analogue of
the Cayley--Hamilton theorem holds:
$$
\tilde Q(E)=0;
\tag{2}
$$
see Subsection 2.24 for more references and comments.

\bigskip
\noindent
{\bf 2.12.} Let $x$ be a formal variable.
The following auxiliary assertion will be used in the proof
of Theorem 2.13.

\proclaim
{\bf Proposition} Let $\frak a$ be a Lie algebra whose center is trivial.
Then
the center of the universal enveloping algebra $\U(\frak a[x])$ is also
trivial.
\endproclaim

\Proof Let us use the fact that for any Lie algebra $\g$
the symmetrization map $\SS(\g)\ra\U(\g)$ yields an
isomorphism of $\g$-modules $\U(\g)$ and $\SS(\g)$.
Then we have to prove that the symmetric algebra $\SS(\frak a[x])$ regarded
as the adjoint $\frak a[x]$-module, has no nontrivial invariant
elements.
Let $\{e_1,\dots,e_n\}$ be a basis of $\frak a$ and
$$
[e_i,\ e_j]=\sum_{k=1}^n c_{ij}^k\,e_k,
$$
where $c_{ij}^k$ are structure constants. The monomials
$$
\prod_{i,r}(e_i\,x^r), \qquad \text {(finite product)}
$$
then form a basis of $\SS(\frak a[x])$. Now let $A\in \SS(\frak a[x])$
be an $\frak a[x]$-invariant element and $m$ be the maximal
integer such that the element $e_i\,x^m$ occurs in $A$ for some
$i\in\{1,\dots,n\}$.
Then $A$ has the form
$$
A=\sum_d A_d (e_1\,x^m)^{d_1}\dots(e_n\,x^m)^{d_n},
$$
where $d=(d_1,\dots,d_n),\ \ d_1\geq 0,\dots,d_n\geq 0$, and $A_d$ is a
polynomial in the variables $e_i\,x^r$ with $r<m$. By the definition of $A$
the
following relation holds:
$$
\text{ad}(e_i\,x)(A)=0 \quad \text{for} \quad i=1,\dots,n. \leqno(1)
$$
The component of the left hand side of (1) that contains
the elements of the form $e_k\,x^{m+1}$
must be zero, i.e.,
$$
\sum_d A_d\,\sum_{j=1}^n d_j(e_1\,x^m)^{d_1}\dots(e_j\,x^m)^{d_j-1}
\dots(e_n\,x^m)^{d_n}\sum_{k=1}^n c_{ij}^k\,e_k x^{m+1}=0. \leqno(2)
$$
Taking the coefficient of $e_k\,x^{m+1}$ in this equality, we obtain:
$$
\sum_d A_d\,\sum_{j=1}^n d_jc_{ij}^k(e_1\,x^m)^{d_1}\dots(e_j\,x^m)^{d_j-1}
\dots(e_n\,x^m)^{d_n}=0.
$$
Thus, for any multi-index
$d^{\,\prime}=(d_1^{\,\prime},\dots,d_n^{\,\prime})$
with nonnegative components we have
$$
\sum_{j=1}^n A_{d^{\,\prime}+\delta_j}(d_j^{\,\prime}+1)c_{ij}^k=0,
\qquad 1\leq i,k\leq n,\leqno(3)
$$
where $d^{\,\prime}+\delta_j$ denotes the multi-index
$(d_1^{\,\prime},\dots,d_j^{\,\prime}+1,\dots,d_n^{\,\prime})$.
Fix $d^{\,\prime}$ and observe
that the elements
$$
e_j'=(d_j^{\,\prime}+1)e_j,\qquad j=1,\dots,n,
$$
also form a basis of $\frak a$. Since the center of $\aa$
is trivial, the system
of linear equations
$$
[e_i,\ \sum_{j=1}^n x_je_j']=0, \qquad i=1,\dots,n
$$
for the variables $x_j$, has only trivial solution. This system
can be rewritten as the system of $n^2$ equations
$$
\sum_{j=1}^n x_j(d_j^{\,\prime}+1)c_{ij}^k=0, \qquad i,k=1,\dots,n.
$$
Comparing this with (3) we see that
$A_{d^{\,\prime}+\delta_j}=0$. Thus we obtain that
$A_d=0$ for
all $d\ne0$, which proves the proposition.

\bigskip
\proclaim
{\bf 2.13. Theorem} The coefficients $d_1, d_2,\ldots$ of
$\qdet T(u)$ are algebraically independent and
generate the whole center of $\Y(N)$.
\endproclaim

\Proof The key idea is to reduce this assertion to an analogous one for the
algebra $\gr_2\Y(N)$. Recall that $\gr_2\Y(N)$ is isomorphic to
$\U(\gl(N)[x])$; see Theorem 1.26.

{\sl Step} 1. Set
$$
Z:=E_{11}+\cdots+E_{NN}, \leqno(1)
$$
so that $\gl(N)=\C Z\oplus \ssl(N)$. Then for any $M=1,2,\ldots $, the
$M$-th
coefficient $d_M$ of $\qdet\, T(u)$ has degree $M-1$ relative to
$\degaa_2(\cdot)$, and its image in the $(M-1)$-th component of
$\gr_2\Y(N)$
coincides with $Zx^{M-1}$.

Indeed, by formula (2.7.1) for $\qdet T(u)$, $d_M$ is a linear
combination of monomials of the form
$$
t_{p(1),1}^{(M_1)}\cdots t_{p(N),N}^{(M_N)},\quad\hbox{where}\quad
M_1+\cdots+M_N\le M. \leqno(2)
$$
By Definition 1.20 of $\degaa_2(\cdot)$ it is clear that the degree of
(2) is strictly less than $M-1$ with the exception of the case when, for
some $i$, we have $M_j=\delta_{ij}M$, $1\le j\le N$. Assume this is exactly
the case. Then, since ${t_{kl}^{(0)}=\delta_{kl}}$, the permutation $p$
has to be trivial, otherwise the monomial (2) vanishes. Hence,
$$
d_M=t_{11}^{(M)}+\cdots+t_{NN}^{(M)}+(\hbox{terms of degree
}<M-1),\leqno(3)
$$
which proves the assertion. This implies that the elements
$d_1,d_2,\dots$ are algebraically independent.

{\sl Step} 2. It remains to prove the
following claim: the center of the algebra $\U(\gl(N)[x])$ is generated by
$Z, Zx, Zx^2,\ldots$\ . Since
$$
\U(\gl(N)[x])=\C[Z, Zx, Zx^2,\ldots]\otimes \U(\ssl(N)[x]),
$$
this claim is equivalent to the triviality of the center of
$\U(\ssl(N)[x])$. But this follows from Proposition 2.12, because
the center of the Lie algebra $\ssl(N)$ is trivial.
This concludes the proof of  the theorem.

\bigskip
\noindent
{\bf 2.14. Definition.} Consider the subalgebra in $\Y(N)$
$$
\SY(N):=\{y\in \Y(N)\vert\ts \mu_f(y)=y\ \  \text{for every}\ \ f\}
\leqno(1)
$$
(see (1.12.2)).
This subalgebra is called the {\it Yangian of the Lie algebra} $\ssl(N)$.

\bigskip
\noindent
{\bf 2.15.} The following statement will be frequently used later on.

\proclaim {\bf Proposition} Let $\Cal A$ be an arbitrary commutative
associative algebra and $u$ be a formal variable. Then for any series
$$
a(u)=1+a_1u^{-1}+a_2u^{-2}+\dots \in \Cal A[[u^{-1}]]
$$
and any positive integer $N$ there exists a unique series
$$
\tilde a(u)=1+\tilde a_1u^{-1}+\tilde a_2u^{-2}+\dots \in \Cal A[[u^{-1}]]
$$
such that
$$
a(u)=\tilde a(u)\tilde a(u-1)\dots\tilde a(u-N+1). \leqno(1)
$$
\endproclaim

\Proof Write (1) in terms of the coefficients of the series $a(u)$
and $\tilde a(u)$:
$$
a_k=N\tilde a_k+(\dots), \qquad k=1,2,\dots,
$$
where $(\dots)$ stands for a certain polynomial in the variables
$\tilde a_1,\dots,\tilde a_{k-1}$. Hence, each element $\tilde a_k$
may be uniquely expressed as a polynomial in $a_1,\dots,a_k$,
which proves the proposition.

\bigskip
\noindent
{\bf 2.16.} Denote by $\Z(N)$ the center of the algebra $\Y(N)$. We have

\proclaim {\bf Proposition} The algebra $\Y(N)$ is isomorphic to the tensor
product of its subalgebras $\Z(N)$ and $\SY(N)$:
$$
\Y(N)=\Z(N)\otimes \SY(N).
$$
In particular, the center of $\SY(N)$ is trivial.
\endproclaim

\Proof Let us apply Proposition 2.15 to $\Cal A=\Z(N)$ and $a(u)=\qdet
T(u)$.
The corresponding element $\tilde a(u)$ will be denoted by $\tilde d(u)$.

{\sl Step} 1. Consider the automorphism $\mu_f$ (see (1.12.2))
and prove that
$$
\mu_f(\tilde d(u))=\tilde d(u)f(u). \leqno(1)
$$
It follows from  Proposition 2.7 that
$$
\mu_f(\qdet T(u))=\qdet T(u)f(u)f(u-1)\dots f(u-N+1). \leqno(2)
$$
On the other hand, by the definition of $\tilde d(u)$,
$$
\mu_f(\qdet T(u))=\mu_f(\tilde d(u))\mu_f(\tilde d(u-1))\dots
\mu_f(\tilde d(u-N+1)).
$$
Comparing this with (2) and applying Proposition 2.15 to
$a(u)=\mu_f(\qdet T(u))$, we obtain (1).

{\sl Step} 2. Let us prove that
$$
\Y(N)=\Z(N)\,\SY(N). \leqno(3)
$$
Set
$$
\tau_{ij}(u)=\tilde d(u)^{-1}\,t_{ij}(u), \qquad 1\leq i,j\leq N.
$$
Then by (1) we have
$
\mu_f(\tau_{ij}(u))=\tau_{ij}(u)\quad \text{for any series}\ f.
$
Hence, all the coefficients $\tau_{ij}^{(M)}$ lie in $\SY(N)$,
and (3) follows from the decomposition
$
t_{ij}(u)=\tilde d(u)\tau_{ij}(u).
$

{\sl Step} 3. Let $n$ be the minimum positive integer such that
there exists a nonzero polynomial $P\in \SY(N)[x_1,\dots,x_n]$
for which
$
P(\tilde d_1,\dots, \tilde d_n)=0.
$
Set $f(u)=1+au^{-n},\ \ a\in \C$. Then
$$
\mu_f: P(\tilde d_1,\dots, \tilde d_n)\mapsto
P(\tilde d_1,\dots, \tilde d_{n-1},\tilde d_n+a).
$$
Thus, $P(\tilde d_1,\dots, \tilde d_{n-1},\tilde d_n+a)=0$ for any
$a\in \C$. This means that the polynomial $P$ does not depend on $x_n$,
which contradicts the choice of $n$. The proposition is proved.

\bigskip
\proclaim
{\bf 2.17. Corollary} The elements $\tau_{ij}^{(M)},\ 1\leq i,j\leq N,\
M=1,2,\dots$, introduced in the proof of Proposition {\rm 2.16}, are
generators
of the algebra $\SY(N)$.
\endproclaim

\Proof It follows from the proof of Proposition 2.16 that any element
$y\in \Y(N)$ can be uniquely written as
$$
y=\sum_a z_a\otimes S_a, \leqno(1)
$$
where $\{z_a\}$ is the basis of $\Z(N)$, formed by the monomials
$\tilde d_{i_1}\dots \tilde d_{i_k},\ 1\leq i_1\leq \dots\leq i_k$,
and $S_a$ are (non-commutative) polynomials in the variables
$\tau_{ij}^{(M)}$. On the other hand, if $y\in \SY(N)$, then (1)
obviously must have the form
$
y=1\otimes y.
$
Hence, $y$ is a polynomial in $\tau_{ij}^{(M)}$.

\bigskip
\proclaim
{\bf 2.18. Corollary} The algebra $\SY(N)$ is isomorphic to the
factor-algebra
$$
\Y(N)/(\qdet T(u)=1)=\Y(N)/(d_1=d_2=\dots=0).
$$
\endproclaim

\Proof Let $\I$ be the ideal of $\Y(N)$, generated by all the
elements $d_1,d_2,\dots$ (or equivalently, by all the elements $\tilde
d_1,\tilde d_2, \dots$).
Proposition 2.16 immediately proves that
$$
\Y(N)=\I\oplus \SY(N).
$$

\bigskip
\proclaim
{\bf 2.19. Proposition} We have the equality
$$
\Delta (\qdet T(u))= \qdet T(u)\otimes \qdet T(u). \leqno(1)
$$
\endproclaim

\Proof We will regard $\Delta$ as a homomorphism of algebras
$$
\Delta: \Y(N)\otimes \End\E^{\otimes N}\ra
\Y(N)\otimes \Y(N)\otimes \End\E^{\otimes N},
$$
which is identical on $\End\E^{\otimes N}$. Using the notation of
Subsection 1.28, we obtain
$$
\Delta(\qdet T(u)A_N)=\Delta(A_NT_1\dots T_N)=
A_NT_{[1]1}T_{[2]1}\dots T_{[1]N}T_{[2]N}, \leqno(2)
$$
where $T_i=T_i(u-i+1)$. Note that the elements $T_{[i]j}$ and
$T_{[k]l}$ commute if $i\ne k$ and $j\ne l$. Hence, (2) may be
rewritten as
$$
A_NT_{[1]1}\dots T_{[1]N}T_{[2]1}\dots T_{[2]N}=
$$
$$
\qdet T_{[1]}(u)A_NT_{[2]1}\dots T_{[2]N}=
\qdet T_{[1]}(u)\qdet T_{[2]}(u)A_N.
$$
This implies (1).

\bigskip
\proclaim
{\bf 2.20. Corollary} $\Delta(\tilde d(u))=\tilde d(u)\otimes \tilde d(u)$.
\endproclaim

\Proof Recall that $\tilde d(u)$ is defined by the equality
$$
\qdet T(u)=\tilde d(u)\tilde d(u-1)\dots\tilde d(u-N+1)
$$
(see Subsection 2.16). Hence,
$$
\Delta(\qdet T(u))=
\Delta(\tilde d(u))\Delta(\tilde d(u-1))\dots\Delta(\tilde d(u-N+1)).
$$
On the other hand, by Proposition 2.19,
$$
\Delta(\qdet T(u))=\tilde d(u)\dots\tilde d(u-N+1)\otimes
\tilde d(u)\dots\tilde d(u-N+1)=
$$
$$
(\tilde d(u)\otimes\tilde d(u))(\tilde d(u-1)\otimes\tilde d(u-1))
\dots(\tilde d(u-N+1)\otimes\tilde d(u-N+1)).
$$
Applying Proposition 2.15 to the algebra $\Cal A=\Z(N)\otimes \Z(N)$
and the element $a(u)=\Delta(\qdet T(u))\in\Cal A[[u^{-1}]]$, we
complete the proof.

\bigskip
\proclaim
{\bf 2.21. Proposition} The subalgebra $\SY(N)\subset \Y(N)$ is
a Hopf algebra, that is, the coproduct, antipode and counit on
$\SY(N)$ can be obtained by restricting those of the Hopf
algebra $\Y(N)$ to $\SY(N)$.
\endproclaim

\Proof It follows from Corollary 2.20 that
$$
\Delta(\tilde d(u)^{-1})=\tilde d(u)^{-1}\otimes\tilde d(u)^{-1}.
$$
Hence,
$$\align
\Delta(\tau_{ij}(u))=&\Delta(\tilde d(u)^{-1}t_{ij}(u))\\
&=(\tilde d(u)^{-1}\otimes\tilde d(u)^{-1})
\sum_{a=1}^N t_{ia}(u)\otimes t_{aj}(u)=
\sum_{a=1}^N \tau_{ia}(u)\otimes \tau_{aj}(u).
\endalign$$
By Corollary 2.17 the last equality proves that
$
\Delta(\SY(N))\subset \SY(N)\otimes \SY(N).
$
We omit the verification of the axioms for the antipode and counit,
which are obvious.

\bigskip
\noindent
{\bf 2.22.} Note that $\SY(N)$ inherits both the filtrations of $\Y(N)$
defined
in Subsection 1.20. Now we will describe the associated graded algebras
$\gr_1\SY(N)$ and $\gr_2\SY(N)$;
the result will be analogous to Theorems 1.22 and 1.26.

\proclaim {\bf Proposition} The algebra $\gr_1\SY(N)$ is commutative
and isomorphic to the algebra of polynomials in the generators
$$
\bar t_{ij}^{\,(M)}\quad (i\neq j),\quad \bar t_{kk}^{\,(M)}-
\bar t_{k+1,k+1}^{\,(M)};\qquad k=1,\ldots, N-1;\  M=1,2,\ldots\,,
\leqno(1)
$$
where the bar has the same meaning as in Subsection {\rm 1.22.}
\endproclaim

\Proof The commutativity of $\gr_1\SY(N)$ follows from that of
$\gr_1\Y(N)$. Furthermore, according to (2.7.1),
$$
\bar d_M=\bar t_{11}^{\,(M)}+\cdots +\bar t_{NN}^{\,(M)}+ (\ldots),
\leqno(2)
$$
where $(\ldots)$ stands for a sum of (commutative) monomials in  letters
$\bar t_{ij}^{\,(L)}$ with degrees $L<M$. Together with the
Poincar\'e--Birkhoff--Witt theorem (1.22), this implies that the elements
$\bar d_M$ and (1) form a system of algebraically independent generators of

$\gr_1\Y(N)$. Since the elements $d_M$ are central in $\Y(N)$ it easily
follows
that the factorization of $\Y(N)$ by them results simply in eliminating the

elements $\bar d_M$ from $\gr_1\Y(N)$.

\bigskip
\proclaim
{\bf 2.23. Proposition} The algebra $\gr_2\SY(N)$ is isomorphic to
$\U(\ssl(N)[x])$.
\endproclaim

\Proof Consider the ideal $\I$ of $\Y(N)$, introduced in the proof of
Corollary
2.18. By using Step 1 of the proof of Theorem 2.13, we obtain that the
image of $\gr_2\I$ under the isomorphism
$$
\gr_2\Y(N)\ra \U(\gl(N)[x])
$$
(see Theorem 1.26) coincides with the ideal $\tilde\I$ of $\U(\gl(N)[x])$
generated by the elements $Z,Zx,Zx^2,\dots\ts$. It is clear that
the factor-algebra $\U(\gl(N)[x])/\tilde\I$ is isomorphic to
the algebra $\U(\ssl(N)[x])$.
This completes the proof.

\bigskip
\noindent
{\bf 2.24. Comments.}
The definition of the quantum determinant $\qdet T(u)$ primarily
appeared in Izergin--Korepin [IK] in the case $N=2$.
The basic ideas and formulae associated with the
quantum
determinant for an arbitrary $N$
are contained in Kulish--Sklyanin's survey paper
[KS2]. The quantum determinant has been used in many papers: see Cherednik
[C2, C3], Drinfeld [D3], Molev [M2], Nazarov [N1],
Nazarov--Tarasov [NT], Tarasov [T1, T2]. However, to our
knowledge, no detailed exposition of the properties of the quantum
determinant $\qdet T(u)$ has been published. In this section we have
attempted to fill this gap.

Formula (2.11.1) has a long history. It is closely related
to the celebrated Capelli identity [Ca1, Ca2], which is discussed in
Weyl's book on classical groups [W, Chapter II, Section
4]. The fact that the determinant (2.11.1) lies in the center of the
enveloping algebra $\U(\gl(N))$ is by no means trivial. A proof of this
result proposed in Howe [H] contained an error which was corrected in
Howe--Umeda [HU]. The approach to this result based on the Yangian seems
to be fruitful, and we hope to return to this subject later. Note that
Nazarov [N1] constructed a superanalogue of $\qdet T(u)$ (the quantum
Berezinian) and applied it to the derivation of a `super' Capelli
identity; in that case the construction is much more complicated
because the Berezinian is not a polynomial function of matrix coefficients
[B].

Identity (2.11.2) is one of the examples of the polynomial
identities satisfied by the generators of a semi-simple Lie algebra.
They were investigated in the papers
Bracken--Green [BG], Green [Gr], Kostant [K],
O'Brien--Cant--Carey [BCC], Gould [G] and others.
Nazarov and Tarasov [NT] have proved
identity (2.11.2) and its $q$-analogue by using the properties of the
quantum
determinant.

The idea of using the reduction to the algebra $\gr_2\Y(N)$ in the
proof of Theorem 2.13 was communicated to one of us by V.G.Drinfeld.

Following
the general philosophy of Drinfeld [D4], the Yangian of $\ssl(N)$ must be
defined as a factor-algebra of $\Y(N)$. The fact that it can also be
realized as
a subalgebra of $\Y(N)$ was observed by the third author and communicated
to
V.G.Drinfeld, who proposed in reply an elegant characterization of
$\Y(\ssl(N))\subset \Y(N)$ in terms of the automorphisms $\mu_f$.

\newpage

\heading {\bf 3. The twisted Yangian $\Y^{\pm}(N)$}\endheading

\noindent
In this section we introduce the twisted Yangians $\Y^+(N)$ and $\Y^-(N)$.
They are defined as certain subalgebras of the Yangian $\Y(N)$ (Definition
3.5).
We also find a realization of $\Y^{\pm}(N)$ via
generators and defining relations. Finally we prove an analogue of the
Poincar\'e--Birkhoff--Witt theorem for the algebras $\Y^{\pm}(N)$.

\bigskip
\noindent {\bf 3.1.}
As before, we will denote by $\E$
the vector space $\C^N$ and by $\{e_i\}$
its canonical basis. But from now on it will be convenient to parametrize
the basis vectors by the numbers $i=-n,-n+1,\ldots,n-1,n$, where $n:=[N/2]$
and $i=0$ is skipped when $N$ is even.

Let us equip $\E$ with a nondegenerate bilinear form $<\cdot,\cdot>=
<\cdot,\cdot>_{\pm}$ which may be either symmetric or alternating:
$$
<e_i,e_j>_+=\delta_{i,-j},\qquad <e_i,e_j>_-=\sgn(i)\delta_{i,-j}.
$$
Here the symbol $\sgn(i)$ equals $1$ for $i>0$ and $-1$ for $i<0$. Of
course,
the alternating case may occur only when $N$ is even.

Both of the cases, symmetric and alternating, will be considered
simultaneously unless stated otherwise. It will be convenient to use
the symbol $\theta_{ij}$ which is defined as follows:
\medskip
$$
\theta_{ij}:=\cases 1,\quad&\text{in the symmetric case};\\
\sgn(i)\sgn(j),\quad&\text{in the alternating case}.\endcases
$$
\bigskip

Whenever the double sign $\pm{}$ or $\mp{}$ occurs,
the upper sign corresponds to the symmetric case and the lower sign to
the alternating one.

By $A\mapsto A^t$ we will denote the transposition relative to the form
$<\cdot,\cdot>$. This is an antihomomorphism of the algebra $\End\E$;
on the matrix units the transposition acts as follows:
$$
(E_{ij})^t=\theta_{ij}E_{-j,-i}.
\tag{1}
$$

We will often use partial transpositions $t_1,t_2,\ldots$ in multiple
tensor
products of the form $\Cal A\otimes(\End\E)^{\otimes m}$, where $\Cal A$ is
a
certain algebra. By definition, $t_k$ denotes the transposition
corresponding
to the $k$-th copy of $\E$.

\bigskip
\noindent {\bf 3.2.} Let $P$ stand for the permutation in
$\E^{\otimes 2}$ as usual.

\proclaim {\bf Proposition} We have:
$$
Q:=P^{t_1}=P^{t_2}=\sum_{i,j} \theta_{ij}E_{-j,-i}\otimes E_{ji}, \leqno(1)
$$
\
$$
Q^2=NQ, \leqno(2)
$$
\
$$
Q\E^{\otimes 2}=\C\xi, \quad\text{where}\quad \xi:=\cases
\sum e_{-j}\otimes e_j,\quad&\text{in the symmetric case},\\
\sum
\sgn(j)e_{-j}\otimes e_j,\quad&\text{in the alternating
case},\endcases\leqno(3)
$$
\
$$
PQ=QP=\pm Q. \leqno(4)
$$
\endproclaim

\noindent {\bf Proof.}
We have
$$
Q:=P^{t_1}=\sum E_{ij}^t\otimes E_{ji}=
\sum \theta_{ij}E_{-j,-i}\otimes E_{ji},
\leqno(5) $$
$$
P^{t_2}=\sum E_{kl}\otimes E_{lk}^t=\sum \theta_{kl}E_{kl}\otimes
E_{-k,-l}.
\leqno(6)
$$
Replacing $(k,l)$ by $(-j,-i)$ in (6) and using the obvious equality
$\theta_{ij}=\theta_{-j,-i}$, we see that (5) equals (6). This proves (1).

Further, taking into account (5), we obtain:
$$
\eqalign{
Q^2 &=(\sum_{i,j} \theta_{ij}E_{-j,-i}\otimes E_{ji})
(\sum_{k,l} \theta_{kl}E_{-l,-k}\otimes E_{lk}) \cr
&=\sum_{i,j,k,l} \theta_{ij}\theta_{kl}(E_{-j,-i}E_{-l,-k}\otimes
E_{ji}E_{lk})\cr
&=\sum_{i,j,k,l}\theta_{ij}\theta_{kl}\delta_{il}(E_{-j,-k}\otimes
E_{jk})
=N\sum_{k,j}\theta_{kj}(E_{-j,-k}\otimes E_{jk})
=NQ\cr}
$$
and
$$
\eqalign{
Qe_k\otimes e_l&=
\sum_{i,j}\theta_{ij}(E_{-j,-i}\otimes E_{ji})(e_k\otimes e_l) \cr
&=\sum_{i,j} \theta_{ij}\delta_{-i,k}\delta_{il}\cdot e_{-j}\otimes e_j
=\delta_{-k,l}\sum_j \theta_{lj}\cdot e_{-j}\otimes e_j
=\delta_{-k,l}\sgn(l)\cdot\xi. \cr}
$$
This proves (2) and (3).
Finally,
$$
\eqalign{
PQ&=(\sum_{i,j} E_{ij}\otimes E_{ji})(\sum_{k,l} \theta_{kl}
E_{-l,-k}\otimes E_{lk}) \cr
&=\sum_{i,j,k,l} \theta_{kl}(E_{ij}E_{-l,-k}\otimes E_{ji}E_{lk})
=\sum_{i,k}\theta_{ki}E_{i,-k}\otimes E_{-i,k}. \cr}
$$
In the symmetric case $\theta_{ki}\equiv 1$, so that $PQ=Q$.
In the alternating case $\theta_{ki}=-\theta_{k,-i}$, so that
$PQ=-Q$. Thus we have verified (4).

\bigskip
\noindent {\bf 3.3.}
{}From (3.2.1) we obtain
$$
R'(u):=R^{t_1}(u)=R^{t_2}(u)=1-{Q\over u}.  \leqno(1)
$$
Note that
$$
R(u)R'(v)=R'(v)R(u), \leqno(2)
$$
since $PQ=QP$.

\proclaim {\bf Proposition}
The $T$-matrix satisfies the following relations:
$$
\eqalignno{
T^{t_1}_1(u)R'(u-v)T_2(v) & =T_2(v)R'(u-v)T^{t_1}_1(u), &(3)\cr
T_1(u)R'(-u+v)T^{t_2}_2(v)= & T^{t_2}_2(v)R'(-u+v)T_1(u), &(4)\cr
R(u-v)T^{t_1}_1(-u)T_2^{t_2}(-v) & =T_2^{t_2}(-v)T^{t_1}_1(-u)R(u-v).
&(5)\cr}
$$
\endproclaim

\noindent{\bf Proof.}
Each of these relations is equivalent to the basic ternary relation
(1.8.1).
Indeed, applying $t_1$ to both sides of (1.8.1)
and repeating the same arguments as in the proof of
Proposition 1.12(iv), we obtain (3).

Further, multiply both sides of (3) by $P$ from the left and from the
right.
Then we obtain
$$
T_2^{t_2}(u)R'(u-v)T_1(v)=T_1(v)R'(u-v)T_2^{t_2}(u),\leqno(6)
$$
or, which is the same,
$$
T_1(v)R'(u-v)T_2^{t_2}(u)=T_2^{t_2}(u)R'(u-v)T_1(v).
$$
Replacing $u,v$ by $v,u$ we obtain (4).

Finally, applying $t_1$ to both sides of (4)
and replacing $u,v$ by $-u,-v$ we
arrive at (5).

\bigskip
\noindent {\bf 3.4.} Relation (3.3.5) implies the following
analogue of Proposition 1.12(iv) for the transposition (3.1.1).
\proclaim {\bf Corollary}
The mapping $T(u)\mapsto T^t(-u)$
defines an involutive automorphism of the algebra $\Y(N)$.
\endproclaim

\bigskip
\noindent {\bf 3.5.}
Set
$$
S(u):=T(u)T^t(-u).\leqno(1)
$$
Then
$$
S(u)=\sum_{i,j} s_{ij}(u)\otimes E_{ij}, \quad\hbox{where}\quad
s_{ij}(u):=\sum_a \theta_{aj}t_{ia}(u)t_{-j,-a}(-u).\leqno(2)
$$
Further,
$$
s_{ij}(u)=\delta_{ij}+s_{ij}^{(1)}u^{-1}+s_{ij}^{(2)}u^{-2}+\cdots
\leqno(3)
$$
with
$$
s_{ij}^{(M)}=\sum_a\sum_{r=0}^M \theta_{aj}(-1)^rt_{ia}^{(M-r)}
t_{-j,-a}^{(r)}. \leqno(4)
$$
For example,
$$
s_{ij}^{(1)}=t_{ij}^{(1)}-\theta_{ij}t_{-j,-i}^{(1)},\ \
s_{ij}^{(2)}=t_{ij}^{(2)}+\theta_{ij}t_{-j,-i}^{(2)}-\sum_a
\theta_{aj}t_{ia}^{(1)}t_{-j,-a}^{(1)}.\leqno(5)
$$
\bigskip
\noindent
{\bf Definition.} The {\it twisted Yangian} $\Y^{\pm}(N)$ is the subalgebra
of
the Yangian $\Y(N)$
generated by the entries of the $S$-matrix, i.e. by the elements
$s_{ij}^{(M)}$, where $M=1,2,\ldots$ and $-n\leq i,j\leq n$.

\bigskip
\noindent {\bf 3.6.}
Set
$$
S_1(u):=\sum_{ij}s_{ij}(u)\otimes E_{ij}\otimes 1, \qquad
S_2(v):=\sum_{kl}s_{kl}(v)\otimes 1\otimes E_{kl}.\leqno(1)
$$
These are elements of the algebra
$\Y^{\pm}(N)[[u^{-1}, v^{-1}]]\otimes \End\E^{\otimes 2}$.

\proclaim {\bf Theorem} The $S$-matrix satisfies the following relations:
$$
R(u-v)S_1(u)R'(-u-v)S_2(v)=S_2(v)R'(-u-v)S_1(u)R(u-v), \leqno(2)
$$
$$
S^t(-u)=S(u)\pm {S(u)-S(-u)\over 2u}.\leqno(3)
$$
\endproclaim
\bigskip

We will refer to (2) and (3) as the {\it quaternary relation} and
the {\it symmetry relation} respectively.

\
\Proof The quaternary relation on the $S$-matrix
is derived from the ternary relation on the
$T$-matrix as follows:
$$
\displaylines{
R(u-v)S_1(u)R'(-u-v)S_2(v)=\cr
=R(u-v)T_1(u)T_1^{t_1}(-u)R'(-u-v)T_2(v)T_2^{t_2}(-v)\cr
=R(u-v)T_1(u)T_2(v)R'(-u-v)T_1^{t_1}(-u)T_2^{t_2}(-v)
\quad\hbox{by}\quad (3.3.3)\cr
=T_2(v)T_1(u)R(u-v)R'(-u-v)T_1^{t_1}(-u)T_2^{t_2}(-v)\cr
\hbox{(by the ternary relation)}\cr
=T_2(v)T_1(u)R'(-u-v)R(u-v)T_1^{t_1}(-u)T_2^{t_2}(-v)
\quad\hbox{by}\quad (3.3.2)\cr
=T_2(v)T_1(u)R'(-u-v)T_2^{t_2}(-v)T_1^{t_1}(-u)R(u-v)
\quad\hbox{by}\quad (3.3.5)\cr
=T_2(v)T_2^{t_2}(-v)R'(-u-v)T_1(u)T_1^{t_1}(-u)R(u-v)
\quad\hbox{by}\quad (3.3.4)\cr
=S_2(v)R'(-u-v)S_1(u)R(u-v).\cr}
$$

To establish the symmetry relation, we will use (3.5.2) and
the commutation relations
(1.8.2):
$$
\displaylines{
(S^t(-u))_{ij}=\theta_{ij} s_{-j,-i}(-u)\cr
=\sum_a\theta_{ij}\theta_{-i,a}t_{-j,a}(-u)t_{i,-a}(u)
\quad\hbox{by}\quad (3.5.2)\cr
=\sum_a\theta_{ja}t_{-j,-a}(-u)t_{i,a}(u),
\quad\hbox{replacing $a$ by $-a$}.\cr}
$$
By (1.8.2), this can be written as
$$
\displaylines{
\sum_a\theta_{ja}t_{ia}(u)t_{-j,-a}(-u)\cr
-{1\over 2u}\sum_a\theta_{ja}t_{i,-a}(-u)t_{-j,a}(u)
+{1\over 2u}\sum_a\theta_{ja}t_{i,-a}(u)t_{-j,a}(-u).\cr}
$$
Observe now that $\theta_{ja}=\pm \theta_{j,-a}$. Substituting this into
the
second and the third sum, we obtain finally
$$
\theta_{ij}s_{-j,-i}(-u)=s_{ij}(u)\pm {s_{ij}(u)-s_{ij}(-u)\over 2u}
\quad\hbox{for all}\quad i,j,\leqno(4)
$$
which is just the symmetry relation.

\bigskip
\proclaim {\bf 3.7. Proposition}
The quaternary relation {\rm (3.6.2)} may be written in the form
$$
\align
[S_1(u),S_2(v)]&={1\over u-v}(PS_1(u)S_2(v)-S_2(v)S_1(u)P)
\\
&-{1\over u+v}(S_1(u)QS_2(v)-S_2(v)QS_1(u))
\\
&+{1\over u^2-v^2}(PS_1(u)QS_2(v)-S_2(v)QS_1(u)P)
\tag{1}
\endalign
$$
or else as the following system of relations: for all $i,j,k,l$
$$
\align
[s_{ij}(u),s_{kl}(v)]&={1\over u-v}(s_{kj}(u)s_{il}(v)-s_{kj}(v)s_{il}(u))
\\
&-{1\over u+v}(\theta_{k,-j}s_{i,-k}(u)s_{-j,l}(v)-
\theta_{i,-l}s_{k,-i}(v)s_{-l,j}(u))
\\
&+{1\over u^2-v^2}(\theta_{i,-j}s_{k,-i}(u)s_{-j,l}(v)-
\theta_{i,-j}s_{k,-i}(v)s_{-j,l}(u)).
\tag{2}
\endalign
$$
\endproclaim
\

Note that relations (2) are analogous to  the relations (1.8.2) for the
Yangian
$\Y(N)$.

\Proof To derive (1), it suffices to substitute in (3.6.2)
$$
R(u-v)=1-{P\over u-v}, \qquad R'(-u-v)=1+{Q\over u+v}.
$$
To derive (2), one simply rewrites (1) in terms of $s_{ij}(u)$ using
(3.6.1)
and the explicit forms of $P$ and $Q$:
$$
P=\sum E_{ij}\otimes E_{ji}, \quad Q=\sum \theta_{kl}E_{-l,-k}\otimes
E_{lk}.
$$

\bigskip
\proclaim {\bf 3.8. Theorem} The quaternary relation
{\rm (3.6.2)} and the symmetry
relation {\rm (3.6.3)} are precisely the defining relations for the
twisted Yangian $\Y^{\pm}(N)$.
\endproclaim

\Proof Let us consider two algebras ${\Cal A}_1$ and ${\Cal A}_2$  where
${\Cal A}_1:=\Y^{\pm}(N)$ and ${\Cal A}_2$ is generated by {\sl arbitrary}
generators $s_{ij}^{(M)}$ (where $M=1,2,\ldots$ and $-n\leq i,j\leq n$)
subject to relations (3.6.2) and (3.6.3). Then there is an obvious
surjective
morphism ${\Cal A}_2\rightarrow {\Cal A}_1$, and we have to verify that it
is
injective.

To do this, we endow both of the algebras with  filtrations: that of
${\Cal A}_1=\Y^{\pm}(N)$ is induced by the {\sl first} filtration of the
Yangian   (see Definition 1.20) and that of ${\Cal A}_2$ is defined by
setting
$\dega(s_{ij}^{(M)})=M$. The mapping  ${\Cal A}_2\rightarrow {\Cal A}_1$
preserves filtration, so that it defines a surjective morphism
$\gr{\Cal A}_2\rightarrow \gr{\Cal A}_1$ of the corresponding graded
algebras.

It suffices to show that the latter morphism is injective. To do this we
will
describe both graded algebras more explicitly.

First consider  $\gr{\Cal A}_1$ which is a subalgebra of $\gr_1\Y(N)$.
Recall that
$\gr_1\Y(N)$ is commutative by Corollary 1.21, hence $\gr{\Cal A}_1$
is commutative too. Denote by $\bar t_{ij}^{\,(M)}$ and $\bar s_{ij}^{(M)}$
the images of $t_{ij}^{(M)}$ and $s_{ij}^{(M)}$ in the $M$-th homogeneous
component of $\gr_1\Y(N)$ respectively. Then (3.5.4) implies
$$
\bar s_{ij}^{(M)}=\bar t_{ij}^{\,(M)}+(-1)^M\theta_{ij}\bar
t_{-j,-i}^{\,(M)}+
\sum_a\sum_{r=1}^{M-1}(-1)^r\theta_{ja}
\bar t_{ia}^{\,(M-r)}\bar t_{-j,-a}^{\,(r)}.
\leqno(1)
$$
Note also that
$$
\theta_{ij}\bar s_{-j,-i}^{(M)}=(-1)^M\bar s_{ij}^{(M)}. \leqno(2)
$$
Indeed, this follows from (1) since we are dealing with a commutative
algebra
(this also follows from the symmetry relation).

Recall that the generators $\bar t_{ij}^{\,(M)}$ are algebraically
independent by
Theorem 1.22. Since the sum in (1) involves only generators of degree
strictly
less than $M$, we see that the algebra $\gr{\Cal A}_1$ is isomorphic to the
algebra of polynomials in the letters $\bar s_{ij}^{(M)}$ subject to the
symmetry condition (2).

Now let us turn to the algebra $\gr{\Cal A}_2$. Here the crucial
observation is
that this algebra is also commutative. Indeed, to show this it suffices to
verify that
$$
\dega [s_{ij}^{(M)},s_{kl}^{(L)}] <M+L.
$$

Let us examine the commutation relation (3.7.2) and regard its right hand
side
as an element of the algebra
$
\gr{\Cal A}_2\otimes \C((v^{-1}))[[u^{-1}]].
$
In $\C((v^{-1}))[[u^{-1}]]$ we may write
$$
{1\over u-v}=u^{-1}(1-vu^{-1})^{-1}=\sum_{r=0}^{\infty} v^ru^{-r-1},
$$
$$
{1\over u+v}=u^{-1}(1+vu^{-1})^{-1}=\sum_{r=0}^{\infty} (-1)^rv^ru^{-r-1},
\leqno(3)
$$
$$
{1\over u^2-v^2}=u^{-2}(1-v^2u^{-2})^{-1}
=\sum_{r=0}^{\infty} v^{2r}u^{-2r-2}.
$$
Substituting these in (3.7.2) and comparing the coefficients of
$u^{-M}v^{-L}$ in both the sides, we see that $[s_{ij}^{(M)},s_{kl}^{(L)}]$
is
a finite sum of expressions of degree $M+L-1$ or $M+L-2$.
(Note that this reasoning is quite similar to that used in the passage
from (1.1.1) to (1.2.1).)

Thus we have verified the commutativity of $\gr{\Cal A}_2$. Now let
$\bar s_{ij}^{(M)}$ have the same meaning as before: the image of the
(abstract) generators $s_{ij}^{(M)}$ in the $M$-th component of $\gr{\Cal
A}_2$.
By the symmetry relation (3.6.3), the  (abstract) generators $\bar
s_{ij}^{(M)}$
satisfy the symmetry relation (2). Since the generators of $\gr{\Cal A}_1$
are
algebraically independent, we conclude that the morphism
$\gr{\Cal A}_2\rightarrow \gr{\Cal A}_1$ is injective.

\bigskip
\proclaim {\bf 3.9. Proposition} The mapping
$$
S(u)\mapsto S^t(u) \leqno(1)
$$
defines an involutive antiautomorphism of the algebra $\Y^{\pm}(N)$.
\endproclaim

\Proof Let us apply the antiautomorphism sign (see (1.11.1)) to
$s_{ij}(u)$.
Due to (3.5.2) we have
$$\align
\sign(s_{ij}(u))&=\sum_a\theta_{aj}t_{-j,-a}(u)t_{ia}(-u)\\
&=\sum_a\theta_{-a,j}t_{-j,a}(u)t_{i,-a}(-u)\ \ (\text{we replaced}\  a\
\text{by}\  -a)\\
&=\sum_a\theta_{ij}\theta_{-a,i}t_{-j,a}(u)t_{i,-a}(-u)=\theta_{ij}s_{-j,-i}
(u).
\endalign$$
Thus, the subalgebra $\Y^{\pm}(N)$ is invariant under the antiautomorphism
$\sign$, and its restriction to $\Y^{\pm}(N)$ gives the antiautomorphism
(1).

\bigskip
\proclaim {\bf 3.10. Proposition} Let
$
g(u)=1+g_1u^{-2}+g_2u^{-4}+\dots\in\C[[u^{-2}]]
$
be a formal power series. Then the mapping
$$
\nu_g: S(u)\mapsto g(u)S(u) \leqno(1)
$$
defines an automorphism of the algebra $\Y^{\pm}(N)$.
\endproclaim

\Proof The series $g(u)$ may be written in the form
$
g(u)=f(u)f(-u),
$
where $f(u)=1+f_1u^{-1}+f_2u^{-2}+\dots\in\C[[u^{-1}]]$. Let us consider
the
automorphism $\mu_f$ of the algebra $\Y(N)$ (see (1.12.2)). By (3.5.1) we
have
$$
\mu_f(S(u))=f(u)T(u)f(-u)T^t(-u)=f(u)f(-u)S(u)=g(u)S(u).
$$
That is, the subalgebra $\Y^{\pm}(N)$ is invariant under
the automorphism $\mu_f$,
and its restriction to $\Y^{\pm}(N)$ gives the automorphism $\nu_g$.

\bigskip
\noindent
{\bf 3.11.} For $-n\leq i,j\leq n$ set
$$
F_{ij}:=E_{ij}-\theta_{ij}E_{-j,-i} \leqno(1)
$$
and denote by $\frak g(n)$ the Lie subalgebra of $\gl(N)$ spanned
by the elements (1). Then $\frak g(n)$ is isomorphic to $\oa(2n)$ or
$\spa(2n)$ (resp. $\oa(2n+1)$), if $N=2n$ (resp. $2n+1$). Note that the
generators (1) satisfy the following symmetry relations:
$$
F_{-j,-i}=-\theta_{ij}F_{ij}. \leqno(2)
$$
\proclaim
{\bf Proposition} The mapping
$$
\xi: s_{ij}(u)\mapsto \delta_{ij}+(u\pm {1\over 2})^{-1}\cdot F_{ij}
\leqno(3)
$$
defines the homomorphism of algebras
$
\xi: \Y^{\pm}(N)\ra \U(\frak g(n)).
$
\endproclaim

\Proof We have to verify that relations (3.7.2) and (3.6.4) hold for
$$
s_{ij}(u)=\delta_{ij}+(u\pm {1\over 2})^{-1}\cdot F_{ij}. \leqno(4)
$$
Let us set $u'=u\pm {1\over 2},\ v'=v\pm {1\over 2}$
and substitute (4) into (3.7.2). Multiplying both sides by
$u'v'$, we obtain:
$$
[F_{ij},F_{kl}]={1\over
u'-v'}((\delta_{kj}u'+F_{kj})(\delta_{il}v'+F_{il})-
(\delta_{kj}v'+F_{kj})(\delta_{il}u'+F_{il}))
$$
$$
-{1\over u'+v'\mp
1}(\theta_{k,-j}(\delta_{i,-k}u'+F_{i,-k})(\delta_{-j,l}v'+F_{-j,l})
-\theta_{i,-l}(\delta_{k,-i}v'+F_{k,-i})(\delta_{-l,j}u'+F_{-l,j}))
$$
$$\align
+{1\over (u'-v')(u'+v'\mp 1)}
\theta_{i,-j}((\delta_{k,-i}u'+F_{k,-i})&(\delta_{-j,l}v'+F_{-j,l})\\
&-(\delta_{k,-i}v'+F_{k,-i})(\delta_{-j,l}u'+F_{-j,l})),
\endalign$$
which is equal to
$$
\delta_{kj}F_{il}-\delta_{il}F_{kj}
-{1\over u'+v'\mp
1}((\theta_{k,-j}\delta_{i,-k}F_{-j,l}-\theta_{i,-l}\delta_{-l,j}F_{k,-i})u'
$$
$$
+(\theta_{k,-j}\delta_{-j,l}F_{i,-k}-\theta_{i,-l}\delta_{k,-i}F_{-l,j})v'
-\theta_{i,-j}(\delta_{k,-i}F_{-j,l}-\delta_{-j,l}F_{k,-i})).
$$
Here we used the relations
$$
\theta_{k,-j}\delta_{i,-k}\delta_{-j,l}-\theta_{i,-l}\delta_{k,-i}\delta_{-l
,j}=0
\quad\text{and}\quad
\theta_{k,-j}F_{i,-k}F_{-j,l}-\theta_{i,-l}F_{k,-i}F_{-l,j}=0.
$$
The former is obvious, while the latter follows from (2).
Relations (2) also imply that
$$
\theta_{k,-j}\delta_{i,-k}F_{-j,l}-\theta_{i,-l}\delta_{-l,j}F_{k,-i}
=\theta_{k,-j}\delta_{-j,l}F_{i,-k}-\theta_{i,-l}\delta_{k,-i}F_{-l,j}
$$
$$
=\pm \theta_{i,-j}(\delta_{k,-i}F_{-j,l}-\delta_{-j,l}F_{k,-i}).
$$
Thus, we get the equality
$$
[F_{ij},F_{kl}]=\delta_{kj}F_{il}-\delta_{il}F_{kj}-
\theta_{k,-j}\delta_{i,-k}F_{-j,l}+\theta_{i,-l}\delta_{-l,j}F_{k,-i},
\leqno(5)
$$
which coincides with the commutation relations of the Lie algebra $\frak
g(n)$.

Now we substitute (4) into (3.6.4). By using (2), we obtain for
the left hand side:
$$
\theta_{ij}(\delta_{-j,-i}+(-u\pm {1\over 2})^{-1}\cdot F_{-j,-i})=
\delta_{ij}+(u\mp {1\over 2})^{-1}\cdot F_{ij},
$$
and for the right hand side:
$$
\delta_{ij}+(u\pm {1\over 2})^{-1}\cdot F_{ij}\pm
{(u\pm {1\over 2})^{-1}-
(-u\pm {1\over 2})^{-1}\over 2u}
\cdot
F_{ij}=\delta_{ij}+(u\mp {1\over 2})^{-1}\cdot F_{ij},
$$
and the proof is complete.

\bigskip
\proclaim {\bf 3.12. Proposition} The mapping
$$
\eta: F_{ij}\mapsto s_{ij}^{(1)} \leqno(1)
$$
defines the inclusion of the algebra $\U(\frak g(n))$ into $\Y^{\pm}(N)$.
\endproclaim

\Proof Using the decompositions (3.8.3), we derive from relations (3.7.2)
that
$$
[s_{ij}^{(1)},s_{kl}^{(1)}]=\delta_{kj}s_{il}^{(1)}-\delta_{il}s_{kj}^{(1)}-
\theta_{k,-j}\delta_{i,-k}s_{-j,l}^{(1)}+\theta_{i,-l}\delta_{-l,j}s_{k,-i}^
{(1)}.
$$
Further, (3.6.4) implies the equality
$$
-\theta_{ij}s_{-j,-i}^{(1)}=s_{ij}^{(1)}.
$$
Comparing this with (3.11.5) and (3.11.2) we conclude that (1)
is a  homomorphism of algebras. It is clear that
$\xi\circ\eta=\text{id}$. Therefore the kernel of $\eta$ is
trivial.

\bigskip
\noindent
{\bf 3.13. Remark} (cf. Remark 1.18). Denote by $F$ the $N\times N$-matrix
formed by the elements $F_{ij}$ and set
$$
F(u):=1+(u\pm {1\over 2})^{-1}\cdot F. \leqno(1)
$$
Then the previous statements may be formulated as follows:
the fact that $\Cal S(u)$ satisfies the quaternary relation (3.6.2)
and the symmetry relation (3.6.3) is equivalent to the fact that the
elements
$F_{ij}$ satisfy relations (3.11.2) and (3.11.6).
Note that the summand $\pm {1\over 2}$ in (1) is essential.
In contrast to the case of the Yangian $\Y(N)$, the mapping
$S(u)\mapsto 1+Fu^{-1}$ does not define a morphism of algebras.

\bigskip
\noindent
{\bf 3.14. Remark.} There is an analogue of Theorem 1.22 for the twisted
Yangian. Namely, it follows from the proof of Theorem 3.8 that the
elements
$$
\bar s_{ij}^{(2k)},\quad i+j\leq 0;\qquad  \bar s_{ij}^{(2k-1)},\quad
i+j<0;\qquad k=1,2,\dots,
$$
in the case of $\Y^{+}(N)$, and the elements
$$
\bar s_{ij}^{(2k)},\quad i+j< 0;\qquad  \bar s_{ij}^{(2k-1)},\quad
i+j\leq 0;\qquad k=1,2,\dots,
$$
in the case of $\Y^{-}(N)$,
constitute a system of algebraically independent generators
of the algebra $\gr_1\Y^{\pm}(N)$. This fact may be regarded as
the Poincar\'e--Birkhoff--Witt theorem for the algebra
$\Y^{\pm}(N)$.

\bigskip
\noindent
{\bf 3.15.} We will now establish an analogue of Theorem 1.26 for the
twisted
Yangian.
Let us introduce the involutive automorphism $\sigma$ of the
polynomial current Lie algebra $\gl(N)[x]$:
$$
(\sigma(f))(x)=-(f(-x))^t,\qquad f\in\gl(N)[x].
$$
This involution determines a Lie subalgebra in $\gl(N)[x]$; denote it
by $\gl(N)[x]^{\sigma}$. It will be called the {\it twisted}
polynomial current Lie algebra corresponding to the orthogonal
Lie algebra $\oa(N)\subset\gl(N)$ or to the symplectic Lie algebra
$\spa(N)\subset\gl(N)$. If
$$
f=a_0+a_1x+\dots+a_kx^k
$$
is an element of $\gl(N)[x]^{\sigma}$, then the coefficients $a_{2i}$ lie
in the subalgebra $\oa(N)$ or $\spa(N)$ of $\gl(N)$, while the coefficients
$a_{2i-1}$ lie in the complement to that subalgebra, defined by the
restriction of $\sigma$ to $\gl(N)$.

The second filtration of $\Y(N)$ (see Definition 1.20) defines a filtration
of the subalgebra $\Y^{\pm}(N)\subset \Y(N)$. Denote by $\gr_2\Y^{\pm}(N)$
the
corresponding graded algebra.

\proclaim
{\bf Theorem} The graded algebra $\gr_2\Y^{\pm}(N)$ is isomorphic to the
universal enveloping algebra $\U(\gl(N)[x]^{\sigma})$.
\endproclaim

\Proof Let us consider the isomorphism
$
\U(\gl(N)[x])\ra \gr_2 \Y(N)
$
constructed in the proof of Theorem 1.26 and find the image of
the subalgebra $\U(\gl(N)[x]^{\sigma})$ under this isomorphism.
The Lie algebra $\gl(N)[x]^{\sigma}$ is the linear span of
the elements
$$
(E_{ij}+(-1)^M\theta_{ij}E_{-j,-i})x^{M-1},\quad -n\leq i,j\leq n;\
M=1,2,\dots.
$$
Their images in $\gr_2 \Y(N)$ have the form
$
\tilde t_{ij}^{\,(M)}+(-1)^M\theta_{ij}\tilde t_{-j,-i}^{\,(M)}.
$
Formula (3.5.4) implies that they are precisely the images of the
generators $s_{ij}^{(M)}$ in the $(M-1)$-th component of $\gr_2 \Y(N)$,
which proves the theorem.

\bigskip
\noindent
{\bf 3.16. Remark} (cf. Remark 1.27). The algebra $\Y^{\pm}(N)$ may be
considered as a flat deformation of the algebra $\U(\gl(N)[x]^{\sigma})$.
To see this we introduce new generators in $\Y^{\pm}(N)$:
$$
\tilde s_{ij}^{(M)}=s_{ij}^{(M)}h^{M-1},
$$
where $h\in\C\setminus\{0\}$ is the deformation parameter. Set
$$
\tilde s_{ij}(u)=\sum_{M=1}^{\infty}\tilde s_{ij}^{(M)}u^{-M}.
$$
Relations (3.7.2) and (3.6.4) will then take the form
$$\align
&\qquad[\tilde s_{ij}(u), \tilde s_{kl}(v)]=\\
&-{1\over u-v}(\delta_{kj}(\tilde s_{il}(u)-\tilde s_{il}(v))
-\delta_{il}(\tilde s_{kj}(u)-\tilde s_{kj}(v)))\\
&+{1\over u+v}(\delta_{i,-k}(\theta_{i,-l}\tilde s_{-l,j}(u)
-\theta_{k,-j}\tilde s_{-j,l}(v))-
\delta_{-j,l}(\theta_{k,-j}\tilde s_{i,-k}(u)
-\theta_{i,-l}\tilde s_{k,-i}(v)))\\
&+{h\over u-v}(
\tilde s_{kj}(u)\tilde s_{il}(v)-\tilde s_{kj}(v)\tilde s_{il}(u))\\
&-{h\over u+v}
(\theta_{k,-j}\tilde s_{i,-k}(u)
\tilde s_{-j,l}(v)-\theta_{i,-l}\tilde s_{k,-i}(v)\tilde s_{-l,j}(u))\\
&-{h\over u^2-v^2}\theta_{i,-j}
(\delta_{k,-i}(\tilde s_{-j,l}(u)-\tilde s_{-j,l}(v))
-\delta_{-j,l}(\tilde s_{k,-i}(u)-\tilde s_{k,-i}(v)))\\
&+{h^2\over u^2-v^2}
\theta_{i,-j}(\tilde s_{k,-i}(u)
\tilde s_{-j,l}(v)-\tilde s_{k,-i}(v)\tilde s_{-j,l}(u))
\endalign
$$
and
$$
\theta_{ij}\tilde s_{-j,-i}(-u)=\tilde s_{ij}(u)
\pm h{\tilde s_{ij}(u)-\tilde s_{ij}(-u)\over 2u}.
$$
Denote by $\Y_h^{\pm}(N)$ the algebra with abstract generators
$\tilde s_{ij}^{(M)},\ -n\leq i,j\leq n;\ M=1,2,\dots$ and the above
relations. Then the algebras $\Y_h^{\pm}(N)$ with $h\neq 0$ are
isomorphic to $\Y^{\pm}(N)=\Y_1^{\pm}(N)$, while setting $h=0$ in the above
formulae we get the following relations:
$$\align
[\tilde s_{ij}^{(M)},\tilde s_{kl}^{(L)}]&=
\delta_{kj}\tilde s_{il}^{(M+L-1)}-\delta_{il}\tilde s_{kj}^{(M+L-1)}\\
&+(-1)^M(
\delta_{i,-k}\theta_{k,-j}\tilde s_{-j,l}^{(M+L-1)}-
\delta_{-l,j}\theta_{i,-l}\tilde s_{k,-i}^{(M+L-1)})
\endalign
$$
and
$$
(-1)^M\theta_{ij}\tilde s_{-j,-i}^{(M)}=\tilde s_{ij}^{(M)}.
$$
They coincide with the commutation relations of the Lie algebra
$\gl(N)[x]^{\sigma}$ in the generators $(E_{ij}+(-1)^M\theta_{ij}E_{-j,-i})
x^{M-1}$. The flatness of the deformation follows from the
Poincar\'e--Birkhoff--Witt theorem for
$\Y^{\pm}(N)$ (see Remark 3.14).

\bigskip
\noindent
{\bf 3.17. Comments.} In this section we have presented a detailed
exposition of some
of the results announced in Olshanski\u\i's paper [O2]. The aim of
[O2] was to apply the approach of the work [O1] to the orthogonal
and symplectic algebras.

\newpage

\heading {\bf 4. The Sklyanin determinant $\qddet S(u)$ and
the center of $\Y^{\pm}(N)$}\endheading

\noindent
In this section we establish several facts about
the structure of the algebras $\Y^{\pm}(N)$ introduced in Section 3.
We find a system of algebraically independent generators of
the center of the algebra $\Y^{\pm}(N)$. We introduce the special
twisted Yangian $\SY^{\pm}(N)$ and prove that the algebra $\Y^{\pm}(N)$ is
isomorphic to the tensor product of its center and the algebra
$\SY^{\pm}(N)$.
We will keep to the notation of Section 3 and use
the $R$-matrix formalism of Subsections 1.3\ts--\ts1.8 extensively.

\bigskip
\noindent {\bf 4.1.} Let $u_1,\dots,u_m$ be formal variables. As in
Subsection 2.1 we put
$$
R(u_1,\dots,u_m):=(R_{m-1,m})(R_{m-2,m}R_{m-2,m-1}) \cdots (R_{1m} \cdots
R_{12}),
$$
where $R_{ij}:=R_{ij}(u_i-u_j)$. We regard $R_{ij}$ as an element of the
algebra
$$
\Y^{\pm}(N)[[u_1^{-1},\dots,u_m^{-1}]]_{ext}\otimes\End\E^{\otimes m},
\leqno (1)
$$
where $\Y^{\pm}(N)[[u_1^{-1},\dots,u_m^{-1}]]_{ext}$ is the localization of
 $\Y^{\pm}(N)[[u_1^{-1},\dots,u_m^{-1}]]$ with respect to the
multiplicative family generated by the elements $u_k^{-1}-u_l^{-1}$ and
$u_k^{-1}+u_l^{-1}$, \ $k\ne l$ (cf. (1.7)). We also need the following
elements of the algebra (1):
$$
S_i:=S_i(u_i),\quad 1\leq i\leq m \qquad \text{and}\qquad
R'_{ij}=R'_{ji}:=R'_{ij}(-u_i-u_j),\quad 1\leq i<j \leq m
$$
(see Subsections 3.3 and 3.5).
For an arbitrary permutation $(p_1,\ldots,p_m)$ of the numbers
$1,\ldots,m$, we abbreviate
$$
\langle S_{p_1},\ldots,S_{p_m}\rangle=S_{p_1}(R'_{p_1p_2}\ldots
R'_{p_1p_m}) S_{p_2}(R'_{p_2p_3}\ldots R'_{p_2p_m})\ldots S_{p_m}.\leqno
(2)
$$

\bigskip
\proclaim {\bf 4.2. Proposition} We have the fundamental identity
{\rm (cf. (2.1.2))}
$$
R(u_1,\ldots,u_m)\langle S_1,\ldots,S_m\rangle=\langle
S_m,\ldots,S_1\rangle R(u_1,\ldots,u_m).\leqno (1)
$$
\endproclaim

\Proof We shall prove (1) in several steps.

{\sl Step} 1. We have the following equalities:
$$
R_{ij}S_iR'_{ij}S_j=S_jR'_{ji}S_iR_{ij}, \leqno(2)
$$
$$
R_{ij}R'_{ik}R'_{jk}=R'_{jk}R'_{ik}R_{ij}, \leqno(3)
$$
where $i,j,k$ are pairwise distinct. Indeed, the equality (2) coincides
with the quaternary relation (3.6.2). It follows from the ternary relation
that (3) is
equivalent to the following fact: the mapping $T(u)\mapsto R'(-u)$, i.e.
$$
t_{ij}(u)\mapsto \delta_{ij}+\theta_{ij}E_{-i,-j}u^{-1},
$$
defines a homomorphism of algebras $\Y(N)\ra \U(\gl(N))$.
However, that has
already been proved for the mapping
$$
t_{ij}(u)\mapsto \delta_{ij}+E_{ij}u^{-1}
$$
(see Proposition 1.16). It remains to note that the mapping
$E_{ij}\mapsto \theta_{ij}E_{-i,-j}$ defines an automorphism of
the algebra $\U(\gl(N))$.

{\sl Step} 2. Observe that (3) can be rewritten as
$$
R_{ij}R'_{ki}R'_{kj}=R'_{kj}R'_{ki}R_{ij}\leqno (4)
$$
since $R'_{ik}=R'_{ki}$ and $R'_{jk}=R'_{kj}$. We shall also need the
following generalization of (3):
$$
R_{ij}(R'_{ik_1}\ldots R'_{ik_r})(R'_{jk_1}\ldots R'_{jk_r})=
(R'_{jk_1}\ldots R'_{jk_r})(R'_{ik_1}\ldots R'_{ik_r})R_{ij},\leqno (5)
$$
provided $i,j,k_1,\ldots,k_r$ are pairwise distinct.\par
\indent To verify (5) we observe that $R'_{ik_a}$ and $R'_{jk_b}$
commute when $a\ne b$, so that (5) can be rewritten as
$$
R_{ij}\cdot\prod_{a=1}^r (R'_{ik_a}R'_{jk_a})=\prod_{a=1}^r
(R'_{jk_a}R'_{ik_a})\cdot R_{ij},
$$
but this equality is an immediate consequence of (3).

{\sl Step} 3. Let us prove
that for any $i=1,\ldots, m-1$ and any permutation $(p_1,\ldots,p_m)$ of
the
numbers $1,\ldots,m$ $$
R_{p_ip_{i+1}}\langle S_{p_1},\ldots,S_{p_m}\rangle=
\langle S_{p_1},\ldots,S_{p_{i-1}},S_{p_{i+1}},S_{p_i},S_{p_{i+2}},
\ldots,S_{p_m}\rangle R_{p_ip_{i+1}}.\leqno (6)
$$
First, we examine the fragment of the product $\langle
S_{p_1},\ldots,S_{p_m}\rangle$ (see (4.1.2)) which precedes $S_{p_i}$. All
the factors of this fragment commute with $R_{p_ip_{i+1}}$ except
$R'_{p_kp_i}$ and $R'_{p_kp_{i+1}}$, where $k=1,\ldots,i-1$. To permute
$R_{p_ip_{i+1}}$ with these factors we use the rule
$$
R_{p_ip_{i+1}}R'_{p_kp_i}R'_{p_kp_{i+1}}=
R'_{p_kp_{i+1}}R'_{p_kp_i}R_{p_ip_{i+1}}
$$
which is a special case of (4). After these transformations the fragment
under
consideration takes the same form as the corresponding fragment
in the right hand side of (6).
\par \indent Second, we examine the fragment
$$
S_{p_i}R'_{p_ip_{i+1}}(\prod_{k>i+1}R'_{p_ip_k})S_{p_{i+1}}
(\prod_{k>i+1}R'_{p_{i+1}p_k}).\leqno (7)
$$
Since $R'_{p_ip_k}$ and $S_{p_{i+1}}$ commute for any $k>i+1$, we may
rewrite (7) as
$$
S_{p_i}R'_{p_ip_{i+1}}S_{p_{i+1}}(\prod_{k>i+1}R'_{p_ip_k})
(\prod_{k>i+1}R'_{p_{i+1}p_k}).\leqno (8)
$$
To permute $R_{p_ip_{i+1}}$ with (8), we use the identities
$$\gather
R_{p_ip_{i+1}}S_{p_i}R'_{p_ip_{i+1}}S_{p_{i+1}}=
S_{p_{i+1}}R'_{p_ip_{i+1}}S_{p_i}R_{p_ip_{i+1}},\\
R_{p_ip_{i+1}}(\prod_{k>i+1}R'_{p_ip_k})
(\prod_{k>i+1}R'_{p_{i+1}p_k})=(\prod_{k>i+1}R'_{p_{i+1}p_k})
(\prod_{k>i+1}R'_{p_ip_k})R_{p_ip_{i+1}}
\endgather
$$
which are special cases of (2) and (5), respectively. Then we rewrite
$R'_{p_ip_{i+1}}$ as $R'_{p_{i+1}p_i}$ and permute $S_{p_i}$ with the
product
$$
\prod_{k>i+1}R'_{p_{i+1}p_k}.
$$
Again after these transformations our fragment takes the same form as the
corresponding fragment in the right hand side of (6).\par
\indent Third, we look at the remaining fragment, which is just
$\langle S_{p_{i+2}},\ldots,S_{p_m}\rangle$. All the factors of this
fragment commute with $R_{p_ip_{i+1}}$; on the other hand, this fragment
appears in the right hand side of (6) in the same form.\par
\indent Thus the proof of (6) has been completed.

{\sl Step} 4. Finally, we
observe that (1) can be deduced from (6). In fact, using (6) repeatedly, we
permute $R_{12}$ with $\langle S_1,\ldots,S_m\rangle$, then we permute
$R_{13}$
with $\langle S_2,S_1,S_3,\ldots,S_m\rangle$ etc. The total effect of the
permutation with all the
factors $R_{ij}$ occuring in $R(u_1,\ldots,u_m)$ clearly amounts to
rearranging the factors $S_i$ into reverse order, just as they appear in
the
right hand side of (1).

\bigskip
\noindent {\bf 4.3.} Note that in the case of the twisted Yangian
the morphisms like (2.2.1) (see Remark 2.2) may be used as well.
So, the fundamental identity (4.2.1) remains true when the variables
$u_1,\dots,u_m$ are subjected to certain relations.

\proclaim {\bf Proposition} The following identity holds
in the algebra {\rm (4.1.1)} with m=N
$$
A_NS_1R'_{12}\dots R'_{1N}S_2\dots S_{N-1}R'_{N-1,N}S_N
=S_NR'_{N,N-1}\dots R'_{N1}S_{N-1}\dots S_2R'_{21}S_1A_N, \leqno(1)
$$
where $S_i=S_i(u-i+1)$,\ \ $R'_{ij}=R'_{ij}(-2u+i+j-2)$ and $u$ is
a formal variable.
\endproclaim

\Proof In the fundamental identity (4.2.1) set $m=N$ and $u_i=u-i+1$
for $i=1,\dots,N$. Then using Proposition 2.3, we get the equality (1).

\bigskip
\proclaim {\bf 4.4. Proposition} There exists a formal series
$$
\qddet S(u):=1+c_1u^{-1}+c_2u^{-2}+\dots \in \Y^{\pm}(N)[[u^{-1}]]
$$
such that both sides of {\rm (4.3.1)} are equal to
$\qddet S(u)A_N$.
\endproclaim

\Proof The proof is similar to that of Proposition 2.5. The required
statement
follows  from the fact that $A_N$ is a one-dimensional projection in
$\E^{\otimes N}$ and each of the elements $S_i(u-i+1),\ \ 1\leq i\leq N$
and
$R'_{ij}(-2u+i+j-2),\ \  1\leq i,j\leq N,\ i\ne j$ is a formal series in
$u^{-1}$
which begins with $1$.

\bigskip
\noindent {\bf 4.5. Definition.} The series $\qddet S(u)$ is called
the {\it Sklyanin determinant} of the matrix $S(u)$.

\medskip
For example, if $N=2$, then $\qddet S(u)$ is equal to
$$
\gather
s_{-1,-1}(u)s_{1,1}(u-1)-s_{1,-1}(u)s_{-1,1}(u-1)
+\frac{s_{-1,-1}(u)\mp s_{1,1}(u)}{2u-1}s_{1,1}(u-1)=
\\
s_{-1,-1}(u-1)s_{1,1}(u)-s_{-1,1}(u-1)s_{1,-1}(u)
+s_{-1,-1}(u-1)\frac{s_{1,1}(u)\mp s_{-1,-1}(u)}{2u-1}.
\endgather
$$
Using the symmetry relation (3.6.4), we can rewrite this as
follows:
$$
\qddet S(u)=
{2u+1\over 2u\pm 1}(s_{-1,-1}(u-1)s_{-1,-1}(-u)\mp
s_{-1,1}(u-1)s_{1,-1}(-u))
$$
$$
={2u+1\over 2u\pm 1}(s_{1,1}(-u)s_{1,1}(u-1)\mp s_{1,-1}(-u)s_{-1,1}(u-1)).
$$
An analogue of the latter expression for $\qddet S(u)$ with general $N$
is given in [M3].

\bigskip
\noindent {\bf 4.6. Remark.} Let $p$ be an arbitrary element of
the symmetric group
$\Sym_N$. Replace the factors $S_i(u-i+1)$ and
$R'_{ij}(-2u+i+j-2)$ in identity (4.3.1) with $S_{p(i)}(u-i+1)$ and
$R'_{p(i),p(j)}(-2u+i+j-2)$ respectively. Then Proposition 4.4
holds for the same series $\qddet S(u)$. This follows immediately from
the equalities:
$$
PA_NP^{-1}=A_N,\ \ PS_iP^{-1}=S_{p(i)},\ \ PR'_{ij}P^{-1}=R'_{p(i),p(j)},
$$
where $P$ denotes the image of $p$ in $\End\E^{\otimes N}$.

\bigskip
\proclaim {\bf 4.7. Theorem} We have
$$
\qddet S(u)=\gamma_N(u)\ts\qdet T(u)\ts\qdet T(-u+N-1),
$$
where $\gamma_N(u)\equiv1$ for $\Y^+(N)$ and $\dsize\gamma_N(u)=
\frac{2u+1}{2u-N+1}$ for $\Y^-(N)$.
\endproclaim

\Proof We use identity (4.3.1).

{\sl Step} 1. Observe that $S_i=T_iT_i^{\sigma}$,
where $T_i=T_i(u-i+1)$ and $\sigma$ denotes the involutive automorphism
of $\Y(N)$ (see Corollary 3.4): $T^{\sigma}(u)=T^t(-u)$. Therefore, the
left
hand side of (4.3.1) takes the form
$$
A_NT_1T_1^{\sigma}R'_{12}\dots R'_{1N}T_2T_2^{\sigma}R'_{23}\dots R'_{2N}
T_3T_3^{\sigma}\dots T_{N-1}T_{N-1}^{\sigma}R'_{N-1,N}T_NT_N^{\sigma},
\leqno(1)
$$
where $R'_{ij}=R'_{ij}(-2u+i+j-2)$. The equality (3.3.6) implies that
for $1\leq i,j\leq N,\ \ i\ne j$
$$
T_i^{\sigma}R'_{ij}T_j=T_jR'_{ji}T_i^{\sigma}. \leqno(2)
$$
Since the elements $T_i$ and $T_i^{\sigma}$ commute with $R'_{jk}$
for $i\ne j,k$, we can rewrite (1) in the following way:
$$
A_NT_1(T_1^{\sigma}R'_{12}T_2)R'_{13}\dots R'_{1N}(T_2^{\sigma}R'_{23}T_3)
\dots (T_{N-1}^{\sigma}R'_{N-1,N}T_N)T_N^{\sigma}.
$$
Applying (2) to the products enclosed in brackets, we obtain the expression
$$
A_NT_1T_2R'_{12}(T_1^{\sigma}R'_{13}T_3)R'_{14}\dots R'_{1N}R'_{23}
(T_2^{\sigma}R'_{24}T_4)\dots (T_{N-2}^{\sigma}R'_{N-2,N}T_N)
R'_{N-1,N}T_{N-1}^{\sigma}T_N^{\sigma}.
$$
Further applying (2) repeatedly, we bring (1) to the form
$$
A_NT_1\dots T_NR'_{12}\dots R'_{1N}R'_{23}\dots R'_{2N}\dots
R'_{N-1,N}T_1^{\sigma}\dots T_N^{\sigma}.
$$
Replacing here $A_N$ by $A_N^2$ and using Proposition 2.4, we transform
this expression into
$$
A_NT_N\dots T_1A_NR'_{12}\dots R'_{1N}R'_{23}\dots R'_{2N}\dots
R'_{N-1,N}T_1^{\sigma}\dots T_N^{\sigma}. \leqno(3)
$$

Further we will consider the algebras
$\Y^+(N)$ and $\Y^-(N)$ separately.

{\sl Step} 2. Let us show first that in the case of $\Y^+(N)$
$$
A_NR'_{12}\dots R'_{1N}R'_{23}\dots R'_{2N}\dots R'_{N-1,N}=A_N. \leqno(4)
$$
Indeed, $A_N={1\over 2}A_N(1-P_{ij})$ for all $i\neq j$. However,
$$
(1-P_{ij})R'_{ij}=(1-P_{ij})(1+{1\over 2u-i-j+2}Q_{ij})=1-P_{ij},
$$
since $P_{ij}Q_{ij}=Q_{ij}$ by Proposition 3.2.
Therefore, $A_NR'_{ij}=A_N$ and (4) is proved.
Hence, (3) takes the form
$$
A_NT_N\dots T_1A_NT_1^{\sigma}\dots T_N^{\sigma}.
$$
Since $\sigma$ is an automorphism of $\Y(N)$, this is equal to
$$
A_NT_N\dots T_1(A_NT_1\dots T_N)^{\sigma}=A_N \qdet T(u)\ts\sigma(\qdet
T(u))
$$
by Proposition 2.5. Finally, applying Proposition 4.4,
we conclude that
$$
\qddet S(u)=\qdet T(u)\ts\sigma(\qdet T(u)).
$$

{\sl Step} 3. In the case of $\Y^-(N),\ \ N=2n$, we verify that
$$
A_NR'_{12}\dots R'_{1N}R'_{23}\dots R'_{2N}\dots R'_{N-1,N}=
{2u+1\over 2u-N+1}A_N. \leqno(5)
$$
First we note that the fundamental identity (4.2.1) implies the relation
$$
A_NR'_{12}\dots R'_{1N}\dots R'_{N-1,N}=
R'_{N-1,N}\dots R'_{N,1}\dots R'_{21}A_N. \leqno(6)
$$
To prove this, we consider the trivial homomorphism $\Y^-(N)\ra\C$,
\ $s_{ij}(u)\mapsto \delta_{ij}$,\ \ $-n\leq i,j\leq n$, and put
$u_i=u-i+1$ for $i=1,\dots,N$. Then identity (4.2.1) becomes (6).
Thus,
$$
A_NR'_{12}\dots R'_{1N}\dots R'_{N-1,N}=\gamma_N(u)A_N \leqno(7)
$$
for a certain scalar function $\gamma_N(u)$.
To calculate it, we apply the left hand side of (7) to the
vector
$$
v=e_{-n}\otimes\dots\otimes e_{-1}\otimes e_1\otimes\dots\otimes e_n
\in\E^{\otimes N}.
$$
It is clear that $R'_{ij}v=v$ for $n+1\leq i<j\leq N$. Let
$0\leq a\leq n$. Using induction on $a$ we shall prove that
$$
A_NR'_{12}\dots R'_{1N}\dots R'_{a,a+1}\dots R'_{a,N}v=
{2u+1\over 2u-2a+1}A_Nv. \leqno(8)
$$
Then in the case $a=n$ we shall obtain the required equality
$\dsize\gamma_N(u)=\frac{2u+1}{2u-N+1}$.
Let $A_m^{(i)}$ denote the normalized antisymmetrizer over the
indices $\{i,i+1,\dots,m\}$ (see Subsection 2.3), so that
$A_N^{(1)}$ coincides with $A_N$. Then $A_N=A_NA_N^{(2)}$ and
$$
A_N^{(2)}R'_{12}\dots R'_{1N}=R'_{1N}\dots R'_{12}A_N^{(2)}. \leqno (9)
$$
Indeed, by Proposition 2.3 we can write
$$
A_N^{(2)}={1\over (N-1)!}R_{N-1,N}R_{N-2,N}R_{N-2,N-1}\dots R_{2N}\dots
R_{23}. $$
Using relation (4.2.3), we obtain
$$
R_{2N}\dots R_{23}R'_{12}R'_{13}\dots R'_{1N}=
R'_{13}\dots R'_{1N}R'_{12}R_{2N}\dots R_{23}.
$$
An easy induction argument gives (9). Analogously, $A_N^{(m+1)}=
A_N^{(m+1)}A_N^{(m+2)}$ for $m=1,\dots,N-2$ so that the left
hand side of (8) can be rewritten as
$$
A_N^{(1)}R'_{1N}\dots R'_{12}A_N^{(2)}R'_{2N}\dots R'_{23}A_N^{(3)}\dots
A_N^{(a)}R'_{aN}\dots R'_{a,a+1}A_N^{(a+1)}v.
$$
Therefore, to verify (8), we may replace $v$ by any vector of the form
$$
e_{-n}\otimes\dots\otimes e_{-(n-a+1)}\otimes e_{p_1}\otimes\dots\otimes
e_{p_{N-a}},
$$
where $(p_1,\dots,p_{N-a})$ is a permutation of the indices
$(-(n-a),-(n-a)+1,\dots,n)$. Let us fix such a vector $v'$, for which
$p_1=n-a+1$. Then $R'_{am}v'=v'$ for $m=a+2,\dots,N$ while
$R'_{a,a+1}v'$ equals
$$
v'+{1\over 2u-2a+1}\sum_k\theta_{k,-(n-a+1)}
e_{-n}\otimes\dots\otimes e_{-(n-a+2)}\otimes e_k\otimes
e_{-k}\otimes e_{p_2}\otimes\dots\otimes e_{p_{N-a}}.
$$
Let $w$ denote the right hand side of the latter equality. Note that
$$
A_N^{(a)}e_{-n}\otimes\dots\otimes e_{-(n-a+2)}\otimes e_k\otimes
e_{-k}\otimes e_{p_2}\otimes\dots\otimes e_{p_{N-a}}=0
$$
unless $k=\pm (n-a+1)$. Hence, by the induction hypothesis
$$
A_NR'_{12}\dots R'_{1N}\dots R'_{a-1,a}\dots R'_{a-1,N}w=
{2u+1\over 2u-2a+3}A_Nw,
$$
while $\dsize A_Nw=\frac{2u-2a+3}{2u-2a+1}A_Nv'$.
Thus, (8) and hence (5) are proved.

Repeating the same arguments as in the case of $\Y^+(N)$, we obtain from
(3)
and (5) that
$$
\qddet S(u)={2u+1\over 2u-N+1}\qdet T(u)\ts\sigma(\qdet T(u)).
$$

{\sl Step} 4. It remains to verify that
$$
\sigma(\qdet T(u))=\qdet T(-u+N-1).
$$
We shall do this simultaneously for both algebras $\Y^+(N)$ and $\Y^-(N)$.
By Proposition 2.7 we have
$$
\qdet T(u)=\sum_{p\in \Sym_N}\sgn(p)t_{-n,p(-n)}(u-N+1)\dots t_{n,p(n)}(u),
$$
hence
$$
\sigma(\qdet T(u))=\sum_{p\in \Sym_N}\sgn(p)\theta_{-n,p(-n)}\dots
\theta_{n,p(n)}t_{-p(-n),n}(-u+N-1)\dots t_{-p(n),-n}(-u).
$$
Let $q$ be the permutation of the indices $(-n,-n+1,\dots,n)$
such that $q(i)=-i$. Then the permutation $p'=(-p(-n),\dots,-p(n))$
equals $qp$. Since $$\theta_{-n,p(-n)}\dots\theta_{n,p(n)}=1,$$
by using Remark 2.8 we obtain
$$
\sigma(\qdet T(u))=\sum_{p'\in \Sym_N}\sgn(qp')t_{p'(-n),n}(-u+N-1)
\dots t_{p'(n),-n}(-u)
$$
$$
=\sgn(q)\sum_{p'\in \Sym_N}\sgn(p')t_{p'(-n),q(-n)}(-u+N-1)
\dots t_{p'(n),q(n)}(-u)=\qdet T(-u+N-1),
$$
which proves the theorem.

\bigskip
\proclaim {\bf 4.8. Theorem} $\qddet S(u)$ lies in the center
of $\Y^{\pm}(N)$, i.e. all its coefficients are central elements.
\endproclaim

\Proof As in the proof of Theorem 2.10 we consider the tensor space
$\E^{\otimes(N+1)}$ where the copies of $\E$ are enumerated by the
indices $0,1,\dots,N$. Set
$$
S_0:=S_0(v),\ \  S_i:=S_i(u-i+1),\quad i=1,\dots,N. \leqno(1)
$$
Then the statement of the theorem follows from the equality
$$
S_0(v)\qddet S(u)A_N=\qddet S(u)S_0(v)A_N. \leqno(2)
$$
To prove (2), we will use the fundamental identity (4.2.1). We have
$$
R(v,u,u-1,\dots,u-N+1)S_0R'_{01}\dots R'_{0N}S_1R'_{12}\dots R'_{1N}S_2
\dots S_{N-1}R'_{N-1,N}S_N
$$
$$
=S_NR'_{N,N-1}\dots R'_{N0}S_{N-1}\dots
S_1R'_{10}S_0R(v,u,u-1,\dots,u-N+1).
\leqno(3) $$
It was proved in Subsection 2.10 that
$$
R(v,u,u-1,\dots,u-N+1)=N!f(u,v)A_N,
$$
where $f(u,v)$ is defined by (2.10.8). Since $S_0$ and $A_N$ commute,
the left hand side of (3) takes the form
$$
N!f(u,v)S_0A_NR'_{01}\dots R'_{0N}
S_1R'_{12}\dots R'_{1N}S_2\dots S_N. \leqno(4)
$$
However, (4.7.9) implies that
$$
A_NR'_{01}\dots R'_{0N}=A_N^2R'_{01}\dots R'_{0N}=A_NR'_{0N}\dots
R'_{01}A_N.
$$
By Remark 1.15,
$$
A_NR'_{0N}\dots R'_{01}A_N=A_N(\tilde R_{01}\dots \tilde R_{0N})^{t_0}A_N
=(A_N\tilde R_{01}\dots \tilde R_{0N}A_N)^{t_0},
$$
where $\tilde R_{0i}=R_{0i}(-v-u+i-1)$.
Repeating the arguments of the proof of the equality (2.10.8),
we obtain that
$$
A_N\tilde R_{01}\dots \tilde R_{0N}A_N=g(u,v)A_N,
$$
where $g(u,v)$ is a non-zero element of the algebra
$\C[u][[v^{-1}]]$. Thus, (4) takes the form
$$
N!f(u,v)g(u,v)S_0(v)\qddet S(u)A_N.
$$
Similar transformations allow us to rewrite
the right hand side of (3) as
$$
N!f(u,v)g(u,v)\qddet S(u)S_0(v)A_N,
$$
which proves the theorem.

\bigskip
\noindent
{\bf 4.9. Remark.} One could prove Theorem 4.8 by using the inclusion
$\Y^{\pm}(N)\hra \Y(N)$. For this, note that
Theorem 4.7 implies that all the
coefficients of $\qddet S(u)$ belong to the center of the algebra
$\Y(N)$ and, therefore, to the center of its subalgebra $\Y^{\pm}(N)$.

\bigskip
\noindent
{\bf 4.10.} The following generalization of Proposition 2.12 will be used
in the proof of Theorem 4.11.
Let $\frak a$ be a Lie algebra and $\sigma$ be an involutive
automorphism of $\frak a$. Denote by $\frak a_0$ (resp. $\frak a_1$)
the set of elements $a\in \frak a$ such that $\sigma (a)=a$ (resp.
$\sigma (a)=-a$). Let $\frak a[x]^{\sigma}$ denote the corresponding
twisted
polynomial current Lie algebra:
$$
\frak a[x]^{\sigma}=\frak a_0\oplus\frak a_1x\oplus\frak a_0x^2\oplus
\frak a_1x^3\oplus\cdots.
$$

\proclaim {\bf Proposition} Suppose that the center of the Lie algebra
$\frak a_0$ is trivial and the $\frak a_0$-module $\frak a_1$ has no
nontrivial
invariant elements. Then the center of the universal enveloping
algebra $\U(\frak a[x]^{\sigma})$ is trivial.
\endproclaim

\Proof Let $\{f_1,\dots,f_r\}$ and $\{e_1,\dots,e_n\}$ be bases of $\frak
a_0$
and $\frak a_1$ respectively. Then for $1\leq i\leq r$ and $1\leq j\leq n$
$$
[f_i,e_j]=\sum_{k=1}^nc_{ij}^ke_k,
$$
where $c_{ij}^k$ are structure constants. As in the proof
of Proposition 2.12 it suffices to verify
that if an element $A\in \SS(\frak a[x]^{\sigma})$ is invariant under the
adjoint action of $\frak a[x]^{\sigma}$, then $A=0$. Let $m$ be the maximum
integer such that the element $e_ix^m$ occurs in $A$ for some
$i\in\{1,\dots,n\}$. Then $A$ may be written in the form $$
A=\sum_d A_d (e_1\,x^m)^{d_1}\dots(e_n\,x^m)^{d_n},
$$
where $d=(d_1,\dots,d_n),\ \ d_1\geq 0,\dots,d_n\geq 0$ and $A_d$ is a
polynomial in  the variables $e_ix^s,\ s<m$ with coefficients from the
subalgebra $\SS(\frak a_0[x^2])$. Just as in the proof of Proposition 2.12
we
deduce  the equalities (2.12.2) and (2.12.3) from
the relations
$
\text{ad}(f_ix)(A)=0 \quad \text{for}\quad i=1,\dots,r.
$
Repeating again the arguments of the proof of Proposition 2.12, we conclude
that $A_d=0$ for all $d\ne 0$, i.e. $A$ belongs to the subalgebra
$\SS(\frak
a_0[x^2])$. To complete the proof, it remains to apply Proposition 2.12 to
the
Lie algebra $\frak a_0$.

\bigskip
\noindent{\bf 4.11.} It follows from Theorem 4.7 that the Sklyanin
determinant of the matrix $S(u)$ satisfies the relation
$$
\gamma_N(u)\ts\qddet S(-u+N-1)=\gamma_N(-u+N-1)\ts\qddet S(u).
$$
Therefore, in contrast to the case of $\Y(N)$, the coefficients
$c_1,c_2,\dots$
of $\qddet S(u)$ are not algebraically independent.
The following statement takes the place of Theorem 2.13.

\proclaim {\bf Theorem} The coefficients $c_2,c_4,c_6,\dots$ of the series
$\qddet S(u)$
are algebraically independent and generate the center of $\Y^{\pm}(N)$.
In particular, $c_1,c_3,\dots$ can be expressed in terms of
$c_2,c_4,\dots$\,.
\endproclaim

\Proof We use the same idea as in the proof of Theorem 2.13. It consists of
reducing the assertion to its analogue for the algebra $\gr_2\Y^{\pm}(N)$,
which
is isomorphic to $\U(\gl(N)[x]^{\sigma})$ by Theorem 3.15.
As in Subsection 2.13 we set
$$
Z=E_{-n,-n}+E_{-n+1,-n+1}+\dots+E_{n,n}.
$$

{\sl Step} 1. For any $m=1,2,\dots$ the coefficient $c_{2m}$ of
$\qddet S(u)$ has the
degree $2m-1$ relative to $\text{deg}_2$ and its image in
the $(2m-1)$-th component of
$\gr_2\Y^{\pm}(N)$ coincides with $Zx^{2m-1}$.
Indeed, Proposition 4.4 implies that
$$
\multline
\qddet S(u)A_N=A_N(S_1(u)\dots S_N(u-N+1)
+
\\{1\over 2u-1}S_1(u)Q_{12}S_2(u-1)\dots
S_N(u-N+1)+\dots+
\\
{1\over 2u-2N+3}S_1(u)\dots S_{N-1}(u-N+2)Q_{N-1,N}S_N(u-N+1)+\dots+
\\
\prod_{1\leq i<j\leq N}{1\over 2u-i-j+1}S_1(u)Q_{12}\dots
Q_{1N}\times
\\
S_2(u-1)Q_{23}
\dots Q_{2N}S_3(u-2)\dots S_N(u-N+1)).
\endmultline
\tag{1}
$$

Let us apply both sides of (1) to the vector
$$
v=e_{-n}\otimes e_{-n+1}\otimes \dots\otimes e_n\in \E^{\otimes N}.
$$
Comparing the coefficients of $u^{-M}A_Nv$ in the left and
right hand sides, we find that $c_M$ is a linear combination of
monomials of the form $$
s_{i_1j_1}^{(M_1)}\dots s_{i_Nj_N}^{(M_N)},\qquad\text{where}
\quad M_1+\dots+M_N\leq M. \leqno(2)
$$
Moreover, the monomials (2) with $M_1+\dots+M_N=M$ arise only from
the first summand of the right hand side of (1). It is clear
that such monomials have the form
$$
s_{p(-n),-n}^{(M_1)}\dots s_{p(n),n}^{(M_N)},\qquad p\in \Sym_N.
$$
{}From Definition 1.20 of $\text{deg}_2$ we obtain the following formula:
$$
c_M=s_{-n,-n}^{(M)}+\dots+s_{n,n}^{(M)}+(\text{terms of degree}\ < M-1),
$$
which is an analogue of (2.13.3) for the algebra $\Y^{\pm}(N)$.
This proves the assertion and the fact that the elements $c_2,c_4,\dots$
are algebraically independent.

{\sl Step} 2. Now the theorem will follow from the fact that
the center of the algebra $\U(\gl(N)[x]^{\sigma})$ is generated
by $Zx,Zx^3,Zx^5,\dots$\,. To see this, we note that
$$
\U(\gl(N)[x]^{\sigma})=\C[Zx,Zx^3,\dots]\otimes\U(\ssl(N)[x]^{\sigma}).
$$
It remains, therefore, to prove that the center of
$\U(\ssl(N)[x]^{\sigma})$
is trivial. But this follows from Proposition 4.10 applied to
the Lie algebra $\frak a=\ssl(N)$ and the involution $\sigma$:
$$
\sigma(E_{ij})=-\theta_{ij}E_{-j,-i}.
$$
The theorem is proved.

\bigskip
\noindent
{\bf 4.12. Remark.} It follows from Theorems 4.7 and 4.11 that
the center of the algebra $\Y^{\pm}(N)$ is contained in the
center of the algebra $\Y(N)$. Furthermore, if $N$ is even then
the centers of $\Y^+(N)$ and $\Y^-(N)$ coincide with each other
as subalgebras in $\Y(N)$.

\bigskip
\noindent
{\bf 4.13. Definition.} Set
$$
\SY^{\pm}(N):=\SY(N)\cap \Y^{\pm}(N).
$$
This algebra is called the {\it special twisted Yangian}.
In other words, $\SY^{\pm}(N)$ can be regarded as the subalgebra
of $\Y^{\pm}(N)$ consisting of the elements which are stable
under all of the automorphisms of the form $\nu_g$ (see Subsection 3.10).

\bigskip
\proclaim {\bf 4.14. Proposition} The algebra $\Y^{\pm}(N)$ is isomorphic
to
the tensor product of its center $\Z^{\pm}(N)$ and the subalgebra
$\SY^{\pm}(N)${\rm :}
$$
\Y^{\pm}(N)=\Z^{\pm}(N)\otimes \SY^{\pm}(N). \leqno(1)
$$
In particular, the center of $\SY^{\pm}(N)$ is trivial.
\endproclaim

\Proof First we prove that
$$
\Y^{\pm}(N)=\Z^{\pm}(N)\, \SY^{\pm}(N). \leqno(2)
$$
We shall use the notations of Subsection 2.16. Consider
the series
$$
\sigma_{ij}(u)=(\tilde d(u)\tilde d(-u))^{-1}s_{ij}(u),\qquad -n\leq
i,j\leq n;
$$
by (3.5.2) it coincides with
$$
\sum_a\theta_{ja}\tau_{ia}(u)\tau_{-j,-a}(-u).
$$
Let us verify that all the coefficients of the series $\tilde d(u)\tilde
d(-u)$
belong to $\Z^{\pm}(N)$. By Theorem 4.7,
$$
\gamma_N(u)^{-1}\qddet S(u)=\qdet T(u)\qdet T(-u+N-1)
$$
$$
=(\tilde d(u)\tilde d(-u))(\tilde d(u-1)\tilde d(-u+1))\dots
(\tilde d(u-N+1)\tilde d(-u+N-1)).
$$
Proposition 2.15 implies that all the coefficients of the series
$\tilde d(u)\tilde d(-u)$ may be expressed as polynomials
in the coefficients of the series $\qddet S(u)$. By Theorem 4.8,
this proves that $\tilde d(u)\tilde d(-u)\in \Z^{\pm}(N)[[u^{-1}]]$.
Note that $\sigma_{ij}(u)\in \SY^{\pm}(N)[[u^{-1}]]$, since these series
are stable under all the automorphisms $\nu_g$ (see (2.16.1) and (3.10.1)).
Now (2) follows from the decomposition
$$
s_{ij}(u)=\tilde d(u)\tilde d(-u)\sigma_{ij}(u),\qquad -n\leq i,j\leq n.
$$
Finally, the decomposition (1) is a consequence of Proposition 2.16 and
the fact that $\Z^{\pm}(N)\subset \Z(N)$ and $\SY^{\pm}(N)\subset \SY(N)$.

\bigskip
\proclaim {\bf 4.15. Corollary} The subalgebra $\SY^{\pm}(N)\subset \SY(N)$
is
generated by all the coefficients of the series $\sigma_{ij}(u), \quad
-n\leq
i,j\leq n$. \endproclaim

\Proof The proof is the same as that of Corollary 2.17.

\bigskip
\proclaim {\bf 4.16. Corollary} $\SY^{\pm}(N)$ is isomorphic to the
factor-algebra $$\Y^{\pm}(N)/(\qddet S(u)=1).$$
\endproclaim

\Proof Proposition 4.14 implies that
$
\Y^{\pm}(N)=\I^{\pm}\oplus \SY^{\pm}(N),
$
where $\I^{\pm}$ is the ideal of $\Y^{\pm}(N)$ generated by all of the
coefficients of the series $\qddet S(u)$. This proves the assertion.

\bigskip
\noindent
{\bf 4.17.} The twisted Yangian $\Y^{\pm}(N)$ seems not to possess any
natural
Hopf algebra structure. Nevertheless, the following proposition holds.

\proclaim {\bf Proposition} $\Y^{\pm}(N)$ is a left coideal of the Hopf
algebra
$\Y(N)$, i.e.,
$$
\Delta(\Y^{\pm}(N))\subset \Y(N)\otimes \Y^{\pm}(N).
$$
Moreover,
$$
\Delta(s_{ij}(u))=\sum_{k,l}\theta_{lj}t_{ik}(u)t_{-j,-l}(-u)\otimes
s_{kl}(u),
\leqno(1)
$$
where $-n\leq i,j\leq n$.
\endproclaim

\Proof It is enough to prove (1). We use the notation of Subsection 1.28.
It is clear that
$$
\Delta(T^t(-u))=T^t_{[2]}(-u)T^t_{[1]}(-u).
$$
Therefore,
$$
\Delta(S(u))=\Delta(T(u)T^t(-u))
$$
$$
=T_{[1]}(u)T_{[2]}(u)T_{[2]}^t(-u)T_{[1]}^t(-u)
=T_{[1]}(u)S_{[2]}(u)T_{[1]}^t(-u).
$$
Rewriting this in terms of the matrix elements, we obtain (1).

\bigskip
\proclaim {\bf 4.18. Corollary} $\SY^{\pm}(N)$ is a left coideal of the
Hopf
algebra  $\SY(N)$.
\endproclaim

\Proof We have to verify that
$$
\Delta(\SY^{\pm}(N))\subset \SY(N)\otimes \SY^{\pm}(N).
$$
Using Corollary 2.20 and Proposition 4.17, we obtain
$$
\Delta(\sigma_{ij}(u))=\Delta((\tilde d(u)\tilde d(-u))^{-1}s_{ij}(u))
$$
$$
=((\tilde d(u)\tilde d(-u))^{-1}\otimes (\tilde d(u)\tilde d(-u))^{-1})
\sum_{k,l}\theta_{lj}t_{ik}(u)t_{-j,-l}(-u)\otimes s_{kl}(u)
$$
$$
=\sum_{k,l}\theta_{lj}\tau_{ik}(u)\tau_{-j,-l}(-u)\otimes \sigma_{kl}(u).
$$
The assertion then follows from Corollary 4.15.

\bigskip
\noindent
{\bf 4.19. Comments.} The results of this section, as those of Section 3,
were announced in Olshanski\u\i\ [O2]. In [O2], the Sklyanin
determinant $\qddet S(u)$ was called `the double quantum determinant' and
was denoted by ddet\ts$S(u)$. We think that the new terminology adopted in
the
present work is more emphatic. It is motivated by the fact that
E.K.Sklyanin was the first to define the new type of determinant involving
intermediate factors between matrix coefficients, see [S2]. One of the
differences between the algebras studied in [S2] and the twisted
Yangians is that here we must use, as the intermediate factors, the
$R'$-matrices instead of Sklyanin's $R^{-1}$.

\newpage

\heading {\bf 5. The quantum contraction
and the quantum Liouville formula for the Yangian}\endheading

\bigskip
\noindent
Here we develop another approach to the investigation
of the structure of the Yangian. This approach is based upon the use
of a one-dimensional projection $Q$ different from $A_N$.
We construct a series $z(u)$ (called quantum contraction of
the matrix $T(u)$), whose coefficient
belong to the center of $\Y(N)$ and generate the center.
We
establish a link between the quantum contraction and the quantum
determinant of the matrix $T(u)$ (the quantum Liouville formula).
Then we calculate the square of the antipodal map $\SS$.

\bigskip
\noindent
{\bf 5.1.} We will use here the notation of Sections 1 and 2.
Here $t$ will denote the usual transposition,
for which $(E_{ij})^t=E_{ji}$. Denote
$$
\hat T(u)=(T^t(u))^{-1},\qquad
\hat R(u)=(R'(u))^{-1} \leqno(1)
$$
where
$
R'(u):=R^{t_1}(u)=R^{t_2}(u)=1-Qu^{-1},
$
and
$$
Q:=P^{t_1}=P^{t_2}=\sum_{i,j}E_{ij}\otimes E_{ij}.
$$
A simple calculation (cf. Proposition 3.2) shows that
$$
Q^2=NQ\qquad\text{and}\qquad
Q\E^{\otimes2}=\C\eta\leqno(2)
$$
where $\eta=e_1\otimes e_1+\dots+e_N\otimes e_N$, so that $N^{-1}Q$ is
a one-dimensional projection in $\E^{\otimes 2}$. It follows from the first
equality in (3) that
$$
(1-Qu^{-1})(1+Q(u-N)^{-1})=1.
$$
Therefore we have $\hat R(u)=1+Q(u-N)^{-1}$.

\bigskip
\proclaim {\bf 5.2. Proposition} The following identity holds:
$$
Q\hat T_1(u)T_2(u-N)=T_2(u-N)\hat T_1(u)Q. \leqno(1)
$$
\endproclaim

\Proof We start with the ternary relation (1.8.1):
$$
R_{12}T_1T_2=T_2T_1R_{12}, \leqno(2)
$$
where $R_{12}=R_{12}(u_1-u_2),\  T_1=T_1(u_1),\  T_2=T_2(u_2)$.
Further, we apply the transposition $t_1$ to both sides of (2).
By Remark 1.15, we get
$$
T_1^{t_1}R'_{12}T_2=T_2R'_{12}T_1^{t_1}.
$$
After multiplying both sides of the last equality by $(T_1^{t_1})^{-1}$ and
$(R'_{12})^{-1}$, we obtain that
$$
T_2\hat T_1\hat R_{12}=\hat R_{12}\hat T_1T_2. \leqno(3)
$$
Now we multiply (3) by $u_1-u_2-N$ and put $u_1=u,\ \ u_2=u-N$. Then (3)
turns into (1). (In other words, we use the fact that the rational function
$\hat R(u)$ in the variable $u$ with values in $\End\E^{\otimes2}$ has
a simple pole at the point $u=N$ and
$\underset {u=N}\to{\text{res}}\ts\hat R(u)=Q$).

\bigskip
\proclaim {\bf 5.3. Proposition} There exists a formal series
$$
z(u)=1+z_1u^{-1}+z_2u^{-2}+\dots\in \Y(N)[[u^{-1}]]
$$
such that {\rm (5.2.1)} equals $z(u)Q$.
\endproclaim

\Proof It follows from the second equality in (5.1.2) that (5.2.1)
equals $Q$ times a formal
series
$z(u)$ in $u^{-1}$ with coefficients in $\Y(N)$. Since the coefficients of
$u^0$
in the series $\hat T_1(u)$ and $T_2(u-N)$ are equal to $1$, the same is
true
for $z(u)$.

We will call the series $z(u)$ the {\it quantum contraction} (of the
matrix
$T(u)$).

\bigskip
\noindent
{\bf 5.4.} Let $t'_{ij}(u),\  1\leq i,j\leq N$, denote the image of
the series $t_{ij}(u)$ under
the antipodal map $\SS: T(u)\mapsto T^{-1}(u)$, i.e., $t'_{ij}(u)$ is
the matrix element of the matrix $T^{-1}(u)$.

\proclaim {\bf Proposition} For any $i=1,\dots,N$
$$
\eqalignno{
z^{-1}(u)&=\sum_{a=1}^Nt_{ai}(u)t'_{ia}(u-N)&(1)\cr
&=\sum_{a=1}^Nt'_{ai}(u-N)t_{ia}(u),&(2)\cr}
$$
and hence
$$
z^{-1}(u)=\frac{1}{N}\operatorname{tr}
\left( T(u)\ts T^{-1}(u-N)\right)
=\frac{1}{N}\operatorname{tr}
\left(T^{-1}(u-N)\ts T(u) \right).
$$

\endproclaim

\Proof Observe that
$$
1=\hat T_1(u)T_2(u-N)T_2^{-1}(u-N)T^t_1(u)=
T^t_1(u)T_2^{-1}(u-N)T_2(u-N)\hat T_1(u).
$$
Hence, by Proposition 5.3,
$$
Q=Q\hat T_1(u)T_2(u-N)T_2^{-1}(u-N)T^t_1(u)=z(u)QT_2^{-1}(u-N)T^t_1(u).
$$
Similarly,
$$
T^t_1(u)T_2^{-1}(u-N)Qz(u)=Q.
$$
Thus,
$$
QT_2^{-1}(u-N)T^t_1(u)=T^t_1(u)T_2^{-1}(u-N)Q=Qz^{-1}(u)=z^{-1}(u)Q.
\leqno(3)
$$
On the other hand, we have
$$
T^t_1(u)T_2^{-1}(u-N)Q(e_1\otimes
e_1)=T^t_1(u)T_2^{-1}(u-N)\sum_ae_a\otimes e_a
$$
$$
=\sum_{a,i,j}t_{ai}(u)t'_{ja}(u-N)(e_i\otimes e_j).
$$
Therefore, by (3),
$$
\delta_{ij}z^{-1}(u)=\sum_{a}t_{ai}(u)t'_{ja}(u-N), \leqno(4)
$$
which is a slight generalization of (1) and will be used later on.

To prove (2), we apply the relation
$$
z^{-1}(u)Q=QT^{-1}_2(u-N)T^t_1(u)
$$
to the vector $e_i\otimes e_i$. Then a calculation similar to that
performed
above shows that the coefficients of the vector $\eta$ (see (5.1.4)) in
the left and right hand sides coincide with those of (2).

\bigskip
\proclaim {\bf 5.5. Theorem} All the coefficients of the series $z(u)$
belong to the
center of $\Y(N)$.
\endproclaim

\Proof Consider the tensor space $\E^{\otimes3}$ where the copies of $\E$
are
labelled by the indices $0,1,2$ and put $T_i=T_i(u_i),\  i=0,1,2$ for
formal variables $u_0,u_1,u_2$. The statement of
the theorem will follow from the identity
$$
T_0z(u_1)Q_{12}=z(u_1)T_0Q_{12}. \leqno(1)
$$

{\sl Step} 1. We prove the auxiliary identity
$$
R_{20}\hat R_{10}\hat R_{12}=\hat R_{12}\hat R_{10}R_{20}, \leqno(2)
$$
where $R_{ij}=R_{ij}(u_i-u_j)$. Applying the transposition $t_1$ to both
sides of the
Yang--Baxter identity
$$
R_{12}R_{10}R_{20}=R_{20}R_{10}R_{12},
$$
we obtain (see Remark 1.15) that
$$
R_{10}^{t_1}R_{12}^{t_1}R_{20}=R_{20}R_{12}^{t_1}R_{10}^{t_1}.
$$
To get (2), it is sufficient to multiply both sides of
this identity by each of $(R_{10}^{t_1})^{-1}$
and $(R_{12}^{t_1})^{-1}$ from the left and from the right.

We will also need another identity
$$
R_{20}\hat R_{10}Q_{12}=Q_{12}\hat
R_{10}R_{20}=Q_{12}(1-(u_0-u_2)^{-2}),\qquad
u_1-u_2=N. \leqno(3)
$$
To prove the first equality in (3), we take the residue of both sides of
(2) at
$u_1-u_2=N$. By (5.1.4), in order to verify the second equality in (3) it
suffices to
apply $Q_{12}\hat R_{10}R_{20}$ to the basis vectors $e_i\otimes e_1\otimes
e_1,
\quad 1\leq i\leq N$. We have
$$
Q_{12}\hat R_{10}R_{20}(e_i\otimes e_1\otimes e_1)=Q_{12}\hat R_{10}
(e_i\otimes e_1\otimes e_1-{1\over u_2-u_0}e_1\otimes e_1\otimes e_i)
$$
$$
=Q_{12}(e_i\otimes e_1\otimes e_1-{1\over u_2-u_0}e_1\otimes e_1\otimes e_i
+{\delta_{i1}\over u_1-u_0-N}\sum_je_j\otimes e_j\otimes e_1
$$
$$
-{1\over (u_2-u_0)(u_1-u_0-N)}\sum_je_j\otimes e_j\otimes e_i)
$$
$$
=e_i\otimes\eta-{\delta_{i1}\over u_2-u_0}e_1\otimes\eta
+{\delta_{i1}\over u_1-u_0-N}e_1\otimes\eta
-{1\over (u_2-u_0)(u_1-u_0-N)}e_i\otimes\eta.
$$
Since $u_1-N=u_2$, this proves (3).

{\sl Step} 2. Let us verify that for arbitrary variables $u_0,u_1,u_2$ the
following identity holds:
$$
R_{20}\hat R_{10}T_2\hat T_1\hat R_{12}T_0=T_0T_2\hat T_1\hat R_{12}\hat
R_{10}R_{20}.
\leqno(4)
$$
Note that if $i\ne j,k$, then $T_i$ and $\hat T_i$ commute with $R_{jk}$
and
$\hat R_{jk}$. Therefore, the left hand side of (4) can be transformed in
the following way:
$$\align
R_{20}\hat R_{10}T_2\hat T_1\hat R_{12}T_0&=R_{20}T_2(\hat R_{10}\hat
T_1T_0)\hat R_{12}\\
&=(R_{20}T_2T_0)\hat T_1\hat R_{10}\hat R_{12}\qquad \text{by \ \
(5.2.3)}\\
&=T_0T_2R_{20}\hat T_1\hat R_{10}\hat R_{12}\ \ \qquad \text{by \ \
(1.8.1)}\\
&=T_0T_2\hat T_1(R_{20}\hat R_{10}\hat R_{12})\\
&=T_0T_2\hat T_1\hat R_{12}\hat R_{10}R_{20}\ \ \qquad \text{by (2),}
\endalign$$
which coincides with the right hand side of (4).

{\sl Step} 3. Let us take the residue of both sides of (4) at $u_1-u_2=N$.
We obtain:
$$
R_{20}\hat R_{10}T_2\hat T_1Q_{12}T_0=T_0T_2\hat T_1Q_{12}\hat
R_{10}R_{20},
\qquad u_1-u_2=N.
$$
By Proposition 5.3, we can rewrite this as follows:
$$
R_{20}\hat R_{10}z(u_1)Q_{12}T_0=T_0z(u_1)Q_{12}\hat R_{10}R_{20}.
$$
By (3), the left hand side is
$$
z(u_1)T_0Q_{12}(1-(u_0-u_2)^{-2}),
$$
while the right hand side is
$$
T_0z(u_1)Q_{12}(1-(u_0-u_2)^{-2}).
$$
Thus, the equality (1) is established and Theorem 5.5 is proved.

\bigskip
\noindent
{\bf 5.6.} Let $\Cal A$ be an arbitrary associative algebra. For any
$p=1,2,\dots,m$
we introduce the $p$-th partial trace as the map
$$
\tra_p: \Cal A\otimes\End\E^{\otimes m}\ra \Cal A\otimes\End\E^{\otimes
(m-1)}
$$
such that
$$
\tra_p: E_{i_1j_1}\otimes\dots\otimes E_{i_pj_p}\otimes\dots\otimes
E_{i_mj_m}
\mapsto E_{i_1j_1}\otimes\dots\otimes\delta_{i_pj_p}\otimes\dots\otimes
E_{i_mj_m}.
$$
Furthermore, for any subset $\{p_1,\dots,p_k\}\subset\{1,\dots,m\}$
one defines the map
$$
\tra_{p_1,\dots,p_k}: \Cal A\otimes\End\E^{\otimes m}\mapsto
\Cal A\otimes\End\E^{\otimes (m-k)}
$$
as the composition of $\tra_{p_1},\dots,\tra_{p_k}$.
We will simply write $\tra$ instead of $\tra_{1,\dots,m}$.
Let
$$
A_r=\sum_{i_1,\dots,i_m,j_1,\dots,j_m}a^{(r)}_{i_1j_1\dots i_mj_m}\otimes
E_{i_1j_1}\otimes\dots\otimes E_{i_mj_m}\ts;\qquad r=1,2
$$
be two arbitrary elements of $\Cal A\otimes\End\E^{\otimes m}$.
It follows immediately from the definition of the trace, that
if the elements $a^{(1)}_{i_1j_1\dots i_mj_m}$ and
$a^{(2)}_{j_1i_1\dots j_mi_m}$ commute for any sets of indices
$\{i_1,\dots,i_m\}$ and $\{j_1,\dots,j_m\}$, then
$$
\tra (A_1A_2)=\tra (A_2A_1). \leqno(1)
$$

\bigskip
\proclaim {\bf 5.7. Theorem} We have
$$
z(u)={\qdet T(u-1)\over\qdet T(u)}. \leqno(1)
$$
\endproclaim

\Proof Consider the auxiliary algebra
$$
\Y(N)[[u^{-1}]]\otimes\End\E^{\otimes(N+1)}, \leqno(2)
$$
where the copies of $\E$ are enumerated by the indices
$0,1,\dots,N$. As in Subsection 4.7, $A^{(i)}_m$ denotes the normalized
antisymmetrizer over the indices $\{i,i+1,\dots,m\}$.

{\sl Step} 1. We prove the identity
$$
P_{0N}A^{(1)}_N\qdet T(u-1)T^{-1}_N(u-N)T_0(u)
$$
$$
=A^{(0)}_{N-1}P_{0N}T_{N-1}(u-N+1)\dots T_0(u)A^{(1)}_{N-1}.
\leqno(3)
$$
By Proposition 2.5,
$$
A^{(1)}_N\qdet T(u-1)=A^{(1)}_NT_1(u-1)\dots T_N(u-N).
$$
Hence, the left hand side of (3) can be rewritten as
$$
P_{0N}A^{(1)}_NT_1(u-1)\dots T_{N-1}(u-N+1)T_0(u). \leqno (4)
$$
Proposition 2.3 and the fundamental identity (2.1.2) imply that
$$
A^{(1)}_{N-1}T_1(u-1)\dots T_{N-1}(u-N+1)=T_{N-1}(u-N+1)
\dots T_1(u-1)A^{(1)}_{N-1}.
\leqno(5)
$$
It is clear that $A^{(1)}_N=A^{(1)}_NA^{(1)}_{N-1}$, so, making use of (5),
we rewrite (4) as follows
$$
P_{0N}A^{(1)}_NT_{N-1}(u-N+1)\dots T_1(u-1)A^{(1)}_{N-1}T_0(u).
$$
Moving $A^{(1)}_{N-1}$ to the right and using the obvious relation
$P_{0N}A^{(1)}_N=A^{(0)}_{N-1}P_{0N}$, we obtain the right hand side of
(3).

{\sl Step} 2. Now we calculate the trace of both sides of (3). To do this,
we apply each side of (3) to the vector
$$
v=e_{i_0}\otimes e_{i_1}\otimes\dots\otimes e_{i_{N}}\in\E^{\otimes(N+1)}
$$
and decompose the image with respect to the canonical basis in the space
$\E^{\otimes(N+1)}$. We are interested in the coefficient of $v$
in this decomposition. It is clear that the trace is equal to the sum of
these coefficients over all the vectors $v$. For the left hand side we
have:
$$\align
\tr(P_{0N}A^{(1)}_N\qdet T(u-1)T^{-1}_N(u-N)T_0(u))&\\
=\qdet &T(u-1)\tr(P_{0N}A^{(1)}_NT^{-1}_N(u-N)T_0(u)),
\endalign$$
which by (5.6.1) equals
$$
\qdet T(u-1)\tr(A^{(1)}_NT^{-1}_N(u-N)T_0(u)P_{0N}).
$$
Furthermore,
$$
A^{(1)}_NT^{-1}_N(u-N)T_0(u)P_{0N}v
=A^{(1)}_NT^{-1}_N(u-N)T_0(u)
(e_{i_N}\otimes e_{i_1}\otimes\dots\otimes e_{i_{N-1}}\otimes e_{i_0})
$$
$$
=\sum_{a,b}A^{(1)}_N t'_{bi_0}(u-N)t_{ai_N}(u)
(e_a\otimes e_{i_1}\otimes\dots\otimes e_{i_{N-1}}\otimes e_b).
$$
Hence, the coefficient of $v$ is zero, unless
$(i_1,\dots,i_N)$ is a permutation of the indices $(1,\dots,N)$.
In the latter case, the coefficient is equal to
$$
{1\over N!} t'_{i_Ni_0}(u-N)t_{i_0i_N}(u). \leqno(6)
$$
Using (5.4.2), we obtain that the sum of the elements (6) over all the
vectors
$v$ equals $z^{-1}(u)$.
Thus, the trace of the left hand side of (3) is $\qdet T(u-1)z^{-1}(u)$.

Now, again using (5.6.1), for the right hand side of (3) we have:
$$\align
\tr(A^{(0)}_{N-1}P_{0N}T_{N-1}(u-N+1)\dots T_0(u)A^{(1)}_{N-1})&\\
=\tr(T_{N-1}&(u-N+1)\dots T_0(u)A^{(1)}_{N-1}A^{(0)}_{N-1}P_{0N}).
\endalign
$$
Since $A^{(1)}_{N-1}A^{(0)}_{N-1}=A^{(0)}_{N-1}$, we transform it as
follows:
$$
\tr(T_{N-1}(u-N+1)\dots T_0(u)A^{(0)}_{N-1}P_{0N})
=\tr(\qdet T(u)A^{(0)}_{N-1}P_{0N})
$$
$$
=\qdet T(u)\tr(A^{(0)}_{N-1}P_{0N}).
$$
Here we used Proposition 2.5. Further,
$$
A^{(0)}_{N-1}P_{0N}v=A^{(0)}_{N-1}
(e_{i_N}\otimes e_{i_1}\otimes\dots\otimes e_{i_{N-1}}\otimes e_{i_0}).
$$
The coefficient of $v$ in this decomposition is zero, unless
$i_0=i_N$ and
$(i_0,\dots,i_{N-1})$ is a permutation of the indices $(1,\dots,N)$.
In the latter case it equals $(N!)^{-1}$. Taking the sum over all the
vectors $v$,
we find that the trace of the right hand side of (3) is
$\qdet T(u)$, which proves the theorem.

\bigskip
\noindent
{\bf 5.8. Remark.} Relation (5.7.1) may be regarded as a `quantum
analogue' of the classical Liouville formula for the derivative of the
determinant of a matrix-valued function. To see this, for each $h\in{\Bbb
C}\setminus\{0\}$ consider the algebra $\Y(N,h)$ introduced in Subsection
1.25. Define the generating series $t_{ij}(u)$ for the elements
$t_{ij}^{(1)}, t_{ij}^{(2)},\ldots\in \Y(N,h)$ in the same way as it was
done in Subsection 1.6 for the algebra $\Y(N)$; then form the matrix
$T(u)$. The quantum determinant and the quantum contraction for the new
algebra are given by
$$
\qdet T(u)=\sum_{p\in \Sym_N} \sgn(p)\ts t_{p(1),1}(u)\ts
t_{p(2),2}(u-h)\ldots
t_{p(N),N}(u-Nh+h),
\leqno (1)
$$
$$
z(u)=\left(\frac{1}{N}\operatorname{tr}
\left( T(u)\ts T^{-1}(u-Nh)\right)\right)^{-1}.
\leqno (2)
$$
Arguments similar to those used in Subsections 2.10 and 5.5 show that
$\qdet T(u)$ and $z(u)$ are central in $\Y(N,h)$. The equality (5.7.1) is
then generalized to
$$
z(u)=\frac{\qdet T(u-h)}{\qdet T(u)}.
$$
Due to the definition (2) this equality can be rewritten as
$$
\operatorname{tr}\left( T^{-1}(u-Nh)\cdot\frac{T(u)-T(u-Nh)}{Nh}\right)
$$
$$
=\frac{1}{\qdet T(u-h)}\cdot\frac{\qdet T(u)-\qdet T(u-h)}{h}.\leqno (3)
$$
In the limit $h\to 0$ the entries of the matrix $T(u)$ become commutative
while the quantum determinant (1) tends to the usual $\det T(u)$. In
this limit we obtain from (3) the equality
$$
\operatorname{tr}\left( T^{-1}(u)\frac{d}{du}T(u)\right)=\frac{1}{\det
T(u)}\cdot\frac{d}{du}\det T(u)
$$
which is the Liouville formula. For this reason we shall refer to
relation (5.7.1) as the {\it quantum Liouville formula} for the
$T$-matrix.

Note also that the proof of Theorem 5.7 does not use the fact that
$z(u)$ is central (Theorem 5.5). Thus, Theorem 5.5 could also be derived
from Theorems 2.10 and 5.7.

\bigskip
\proclaim {\bf 5.9. Corollary} The coefficients $z_2,z_3,\dots$ of the
series
$z(u)$ are algebraically independent and generate the whole center
of the algebra $\Y(N)$.
\endproclaim

\Proof By Theorem 5.7
$$
(1+z_1u^{-1}+z_2u^{-2}+\dots)(1+d_1u^{-1}+d_2u^{-2}+\dots)=
1+d_1(u-1)^{-1}+d_2(u-1)^{-2}+\dots\,.
$$
As
$$
(u-1)^{-k}=\sum_{i=0}^{\infty}\binom{k+i-1}iu^{-k-i},
$$
we obtain that $z_1=0$ and
$$
z_k+d_k+\sum_{i=1}^{k-1}z_{k-i}d_i=
d_k+\sum_{j=1}^{k-1}\binom{k-1}jd_{k-j}
$$
for $k\geq 2$. An easy induction shows that for every $n\geq1$ the
coefficient
$d_n$ is a polynomial in the variables $z_2,\dots,z_{n+1}$ and the
coefficient $z_{n+1}$ may be expressed as
$
z_{n+1}=nd_n+(\dots),
$
where $(\dots)$ stands for a polynomial in the variables
$d_1,\dots,d_{n-1}$. By Theorem 2.13 this proves the assertion.

\bigskip
\noindent
{\bf 5.10. Remark.} All the arguments and results of Subsections 5.1-5.9,
in
particular the construction of the quantum contraction $z(u)$ and Theorem
5.7,
remain valid when the transposition $t$ is changed to the transposition
with respect to the forms $<\cdot,\cdot>_+$ and $<\cdot,\cdot>_-$ on
the space $\E$ introduced in Subsection 3.1.

\bigskip
\noindent
{\bf 5.11.} We conclude this section with a theorem which provides
a link between the antipode $\SS$ and the quantum contraction $z(u)$.

\proclaim {\bf Theorem} We have
$$
\SS^2=\sigma_N\circ\mu_{z(u)},
$$
where  $\sigma_N$ and $\mu_{z(u)}$ are the automorphisms of $\Y(N)$
defined by the formulas {\rm (1.12.1)} and {\rm (1.12.2)} respectively.
\endproclaim

\Proof The equality $T(u)T^{-1}(u)=1$ implies that
$$
\sum_{a=1}^Nt_{ia}(u)t'_{aj}(u)=\delta_{ij}. \leqno(1)
$$
By applying the antiautomorphism $\SS$ to both sides of (1), we get
$$
\sum_{a=1}^Nt''_{aj}(u)t'_{ia}(u)=\delta_{ij}, \leqno(2)
$$
where $t''_{ij}(u)$ denotes the image of $t_{ij}(u)$ under
the automorphism $\SS^2$. On the other hand, by relation (5.4.4),
$$
\sum_{a=1}^Nt_{aj}(u+N)t'_{ia}(u)=\delta_{ij}z^{-1}(u+N). \leqno(3)
$$
Comparing (2) and (3), we find that
$$
t''_{ij}(u)=t_{ij}(u+N)z(u+N),
$$
which proves the theorem.

\bigskip
\noindent
{\bf 5.12. Comments.} The approach to the description of the center of the
Yangian $\Y(N)$ presented in this chapter was proposed by the second
author in [N1]. Theorems 5.5, 5.7, 5.11 and sketches of their proofs
are contained in [N1] (in fact they are stated there in a greater
generality -- for the Yangian of the Lie superalgebra $\gl(N| M)$).

\newpage

\heading {\bf 6. The quantum contraction
and the quantum Liouville formula for the twisted Yangian}\endheading

\bigskip
\noindent
In this section, we extend the results of
Section 5 to the twisted Yangian $\Y^{\pm}(N)$. We start with the
introduction of
a `covering' algebra $\tilde \Y^{\pm}(N)$ by removing the symmetry relation
(3.6.3)
from the definition of the twisted Yangian $\Y^{\pm}(N)$. Then we construct
certain series $\delta(u)$ whose coefficients are central elements of the
`covering' algebra and prove that the symmetry condition on the $S$-matrix
can be expressed as the equality $\delta(u)=1$ (Theorem 6.4). By using
the series $\delta(u)$
we define an analogue of the quantum contraction $z(u)$ for the
twisted Yangian. This is a series $y(u)$ whose coefficients form
a new system of generators for the center. We describe the
relationship between $y(u)$, $z(u)$  and the Sklyanin
determinant $\qddet(u)$; see Theorems 6.7, 6.8. The latter theorem is an
analogue of
the quantum Liouville formula for the twisted Yangian.

\bigskip
\noindent
{\bf 6.1.} Let us denote by $\tilde \Y^{\pm}(N)$ the complex associative
algebra with generators $\td s_{ij}^{(1)},\td s_{ij}^{(2)},\dots$
where $-n\leq i,j\leq n$; subject to the following relations.
Introduce the generating series $\td s_{ij}(u)$ and form the matrix $\td
S(u)$
in the same fashion as it was done in (3.5.3) and (3.5.2) for the
generators
$s_{ij}^{(1)}, s_{ij}^{(2)},\dots$. We impose on the matrix $\td S(u)$
the quaternary relation but not the symmetry relation (see Subsection 3.6):
$$
R(u-v)\td S_1(u) R^\prime(-u-v)\td S_2(v)=
\td S_2(v)R^\prime(-u-v)\td S_1(u) R(u-v).
\tag{1}
$$
In the next few subsections we will establish several facts
about the structure of the algebra $\tilde \Y^{\pm}(N)$.

\bigskip
\proclaim {\bf 6.2. Proposition} There exists a formal series
$$
\delta(u)=1+\delta_1u^{-1}+\delta_2u^{-2}+\dots\in\tilde
\Y^{\pm}(N)[[u^{-1}]]
$$
such that
$$
Q\td S_1(u)R(2u)\td S^{-1}_2(-u)=\td S^{-1}_2(-u)R(2u)\td S_1(u)Q
=(1\mp {1\over 2u})\delta(u)Q. \leqno(1)
$$
\endproclaim

\Proof By multiplying both sides of the quaternary relation (6.1.1)
by $\td S^{-1}_2(v)$ we obtain the identity
$$
\td S^{-1}_2(v)R(u-v)\td S_1(u)R'(-u-v)=R'(-u-v)\td S_1(u)R(u-v)\td
S^{-1}_2(v).
\leqno(2)
$$
Note that the rational function $R'(-u)=1+Qu^{-1}$ has a simple
pole at $u=0$ and
$\underset {u=0}\to{\text{res}}\ts R'(-u)=Q$.
Taking the residue
of both sides of (2) at $u+v=0$, we get
the first equality in (1).
Now, by Proposition 3.2, the assertion follows from the fact that
the coefficients of $u^0$ in the series $\td S_1(u)$, $R(2u)$, $\td
S^{-1}_2(-u)$
and $(1\mp (2u)^{-1})$ are equal to 1.

\bigskip
\proclaim {\bf 6.3. Theorem} All the coefficients of the series
$\delta(u)$ belong to the center of the algebra $\tilde \Y^{\pm}(N)$.
\endproclaim

\Proof The proof is quite similar to that of Theorem 5.5. Consider the
tensor
space $\E^{\otimes3}$, where the copies of $\E$ are enumerated
by the indices 0,1,2 and set
$$
\gather
\td S_i=\td S_i(u_i);\qquad i=0,1,2;
\\
R_{ij}=R_{ij}(u_i-u_j),\quad R'_{ij}=R'_{ij}(-u_i-u_j);\qquad
0\leq i<j\leq 2
\endgather
$$
where $u_0,u_1,u_2$ are formal variables.
We shall prove the identity
$$
\td S_0\delta(u_1)Q_{12}=\delta(u_1)\td S_0Q_{12}, \leqno(1)
$$
which implies the statement of the theorem.

{\sl Step} 1. We verify that the following auxiliary
identities hold provided that $u_1+u_2=0$:
$$
\gather
Q_{12}R'_{01}R_{02}=R_{02}R'_{01}Q_{12}=Q_{12}(1-(u_0+u_1)^{-2}),
\tag2
\\
Q_{12}R'_{02}R_{01}=R_{01}R'_{02}Q_{12}=Q_{12}(1-(u_0-u_1)^{-2}).
\tag3
\endgather
$$
Indeed, by (4.2.3)
$$
R'_{12}R'_{01}R_{02}=R_{02}R'_{01}R'_{12}.
$$
Taking the residue at $u_1+u_2=0$, we obtain the first equality in (2).
By Proposition 3.2, to verify the second equality in (2), it suffices
to apply $Q_{12}R'_{01}R_{02}$ to the basis vectors $e_i\otimes e_{-1}
\otimes e_1,\ i=-n,-n+1,\dots,n$. We have
$$
\gather
Q_{12}R'_{01}R_{02}(e_i\otimes e_{-1}\otimes e_1)=Q_{12}R'_{01}
(e_i\otimes e_{-1}\otimes e_1-{1\over u_0-u_2}e_1\otimes e_{-1}\otimes e_i)
\\
=Q_{12}(e_i\otimes e_{-1}\otimes e_1+
{\delta_{i1}\over u_0+u_1}\sum_j\theta_{ji}e_j\otimes e_{-j}\otimes e_1
-{1\over u_0-u_2}e_1\otimes e_{-1}\otimes e_i
\\
-{1\over (u_0-u_2)(u_0+u_1)}\sum_j\theta_{j1}e_j\otimes e_{-j}\otimes e_i)
\\
=e_i\otimes\xi+{\delta_{i1}\theta_{i1}\over u_0+u_1}e_1\otimes\xi
-{\delta_{i1}\over u_0-u_2}e_1\otimes\xi
-{1\over (u_0-u_2)(u_0+u_1)}e_i\otimes\xi.
\endgather
$$
Since $u_2=-u_1$, this implies (2). The proof of (3) is quite similar.

{\sl Step} 2. We prove that for arbitrary formal variables $u_0,u_1,u_2$
the following identity holds:
$$
R_{01}R'_{02}\td S_0R_{02}R'_{01}\td S^{-1}_2R_{12}\td S_1R'_{12}
=\td S^{-1}_2R_{12}\td S_1R'_{12}R'_{01}R_{02}\td S_0R'_{02}R_{01}.
\leqno(4)
$$
We use the fact that $\td S_i$ and $\td S^{-1}_i$ commute with $R_{jk}$ and
$R'_{jk}$, if $i\ne j,k$. Let us transform the left hand side of (4)
in the following way:
$$\align
R_{01}(R'_{02}\td S_0R_{02}\td S^{-1}_2)R'_{01}R_{12}\td S_1R'_{12}
&=R_{01}\td S^{-1}_2R_{02}\td S_0(R'_{02}R'_{01}R_{12})\td
S_1R'_{12}\qquad\text{by
(6.2.2)}\\
&=R_{01}\td S^{-1}_2R_{02}\td S_0R_{12}R'_{01}R'_{02}\td S_1R'_{12}\ \
\qquad\text{by
(4.2.3)}\\ &=\td S^{-1}_2(R_{01}R_{02}R_{12})\td S_0R'_{01}\td
S_1R'_{02}R'_{12}\\
&=\td S^{-1}_2R_{12}R_{02}(R_{01}\td S_0R'_{01}\td
S_1)R'_{02}R'_{12}\qquad\text{by
(1.5.2)}\\
&=\td S^{-1}_2R_{12}R_{02}\td S_1R'_{01}\td
S_0(R_{01}R'_{02}R'_{12})\qquad\text{by
(6.1.1)}\\
&=\td S^{-1}_2R_{12}R_{02}\td S_1R'_{01}\td S_0R'_{12}R'_{02}R_{01}\ \
\qquad\text{by
(4.2.3)}\\ &=\td S^{-1}_2R_{12}\td S_1(R_{02}R'_{01}R'_{12})\td
S_0R'_{02}R_{01},
\endalign
$$ which coincides with the right hand side of (4) by (4.2.3).

{\sl Step} 3. Let us take the residues of both sides of (4) at $u_1+u_2=0$.
We obtain the equality
$$
R_{01}R'_{02}\td S_0R_{02}R'_{01}(\td S^{-1}_2R_{12}\td S_1Q_{12})
=(\td S^{-1}_2R_{12}\td S_1Q_{12})R'_{01}R_{02}\td S_0R'_{02}R_{01},
$$
provided $u_1+u_2=0$. By Proposition 6.2, this may be rewritten as follows:
$$
R_{01}R'_{02}\td S_0R_{02}R'_{01}Q_{12}\delta(u_1)
=\delta(u_1)Q_{12}R'_{01}R_{02}\td S_0R'_{02}R_{01}. \leqno(5)
$$
By (2) and (3), the left hand side equals
$$
(1-(u_0+u_1)^{-2})R_{01}R'_{02}Q_{12}\td S_0\delta(u_1)
=(1-(u_0+u_1)^{-2})(1-(u_0-u_1)^{-2})\td S_0\delta(u_1)Q_{12}.
$$
Similarly, the right hand side of (5) is
$$
(1-(u_0+u_1)^{-2})(1-(u_0-u_1)^{-2})\delta(u_1)\td S_0Q_{12},
$$
which proves (1) and the theorem.

\bigskip
\proclaim {\bf 6.4. Theorem} The relation $\delta(u)=1$ is equivalent to
the
symmetry relation
$$
\td S(-u)=\td S(u)\pm\frac{\td S(u)-\td S(-u)}{2u}.
\tag{1}
$$
\endproclaim

\Proof We will make use of Proposition 6.2. Let us apply both sides of the
equality
$$
(1\mp {1\over 2u})\delta(u)Q=Q\td S_1(u)R(2u)\td S^{-1}_2(-u)
$$
to the vector $e_{-i}\otimes e_j\in\E^{\otimes2}$. By Proposition 3.2, we
have
$$
(1\mp {1\over 2u})\delta(u)Q(e_{-i}\otimes e_j)=\delta_{ij}(1\mp {1\over
2u})\theta_{i1}
\delta(u)\xi.
$$
Denote by $s'_{ij}(u),\ -n\leq i,j\leq n$ the matrix elements of the matrix
$\td S^{-1}(u)$.
Then
$$
Q\td S_1(u)R(2u)\td S^{-1}_2(-u)(e_{-i}\otimes e_j)=
Q\td S_1(u)R(2u)\sum_ks'_{kj}(-u)(e_{-i}\otimes e_k)
$$
$$
=Q\td S_1(u)\sum_ks'_{kj}(-u)(e_{-i}\otimes e_k-{1\over 2u}e_k\otimes
e_{-i})
$$
$$
=Q(\sum_{k,l}\td s_{l,-i}(u)s'_{kj}(-u)(e_l\otimes e_k)
-{1\over 2u}\sum_{k,l}\td s_{kl}(u)s'_{kj}(-u)(e_l\otimes e_{-i}))
$$
$$
=(\sum_k\theta_{k1}\td s_{-k,-i}(u)s'_{kj}(-u)
-{1\over 2u}\theta_{-i,1}\sum_k\td s_{ik}(u)s'_{kj}(-u))\xi.
$$
Thus,
$$
\delta_{ij}(1\mp {1\over 2u})\delta(u)=\sum_k(\theta_{ik}\td
s_{-k,-i}(u)\mp
{1\over 2u}\td s_{ik}(u))s'_{kj}(-u).
$$
On the other hand,
$$
\delta_{ij}=\sum_k\td s_{ik}(-u)s'_{kj}(-u).
$$
By comparing the last two equalities, we obtain that
$$
(1\mp {1\over 2u})\td s_{ij}(-u)\delta(u)=\theta_{ij}\td s_{-j,-i}(u)\mp
{1\over 2u}\td s_{ij}(u).
\tag2
$$
Hence, the relation $\delta(u)=1$ implies that
$$
\theta_{ij}\td s_{-j,-i}(u)=\td s_{ij}(-u)\pm {\td s_{ij}(u)-\td
s_{ij}(-u)\over 2u},
$$
which essentially coincides with (6.4.1); see the relation (3.6.4).

Conversely, if the symmetry relation (6.4.1) is valid, then (2)
for $i=j=-1$ becomes
$$
(1\mp {1\over 2u})\td s_{-1,-1}(-u)\delta(u)=\td s_{1,1}(u)\mp
{1\over 2u}\td s_{-1,-1}(u)=(1\mp {1\over 2u})\td s_{-1,-1}(-u).
$$
As $\td s_{-1,-1}(-u)$ is invertible, this implies that $\delta(u)=1$.
The theorem is proved.

\bigskip
\proclaim {\bf 6.5. Propoition} The mapping
$$
\inv: \td S(u)\mapsto \td S^{-1}(-u-{N\over2})
$$
defines an involutive automorphism of the algebra $\tilde \Y^{\pm}(N)$.
\endproclaim

\Proof By inverting the left and right hand sides of the
quaternary relation (6.1.1), we obtain
$$
\multline
\td S^{-1}_2(v)(R'(-u-v))^{-1}\td S^{-1}_1(u)R^{-1}(u-v)
\\
=R^{-1}(u-v)\td S^{-1}_1(u)(R'(-u-v))^{-1}\td S^{-1}_2(v).
\endmultline
\tag1
$$
Furthermore, we observe that
$$
R^{-1}(u-v)={(u-v)^2\over (u-v)^2-1}R(v-u),
$$
$$
(R'(-u-v))^{-1}=R'(u+v-N).
$$
Therefore, replacing $(u,v)$ by $(-u-N/2,-v-N/2)$, we transform
(1) into the quaternary relation for the matrix $\td S^{-1}(-u-N/2)$.

It remains to verify that $\inva\circ\inva=1$. Let us apply
$\inva$ to both sides of the relation
$$
(\inv \td S(u))\td S(-u-{N\over2})=1.
$$
We obtain that
$$(\inva\circ\inva)(\td S(u))\td S^{-1}(u)=1,\quad \text{i.e.,}\quad
(\inva\circ\inva)(\td S(u))=\td S(u).
$$

\medskip
\noindent
{\bf 6.6.} Let us consider the image $\inva(\delta(u))$ of the element
$\delta(u)$ under the automorphism $\inva$ of the algebra $\tilde \Y^{\pm}
(N)$. Denote by $ y(u)$ the image of $\inva(\delta(u-N/2))$ under
the factorization map $\tilde \Y^{\pm}(N)\ra \Y^{\pm}(N)$. By
Theorem 6.3, all the coefficients of the series $\inva(\delta(u))$ belong
to the center of $\tilde \Y^{\pm}(N)$, hence, all the coefficients of
$ y(u)$ are central elements in the algebra $\Y^{\pm}(N)$.
We shall call the series $ y(u)$ the {\it quantum
contraction} (of the matrix $S(u)$).

\proclaim {\bf Proposition} The following identities hold in the
algebra $\Y^{\pm}(N)$:
$$
\multline
QS^{-1}_1(-u)R(2u-N)S_2(u-N)
\\
=S_2(u-N)R(2u-N)S^{-1}_1(-u)Q
=(1\mp{1\over 2u-N}) y(u)Q.
\endmultline
\tag1
$$
\endproclaim

\Proof It suffices to apply the automorphism $\inva$ to each of the parts
of identity (6.2.1), replace $u$ by $u-N/2$, and take their
images in the algebra $\Y^{\pm}(N)$.

\bigskip
\noindent
{\bf 6.7.} Let us define the quantum contraction $z(u)$
corresponding to the transposition $t$ that
was used in Section 3: $(E_{ij})^t=\theta_{ij}E_{-j,-i}$; see Proposition
5.3 and
Remark 5.10. Then we obtain the following

\proclaim {\bf Theorem} We have the equality
$
y(u)=z(u)z^{-1}(-u+N).
$
\endproclaim

\Proof Recall that $S(u)=T(u)T^t(-u)$. It follows from Proposition 6.6
that
$$
(1\mp{1\over2u-N}) y(u)Q=Q(T_1(-u)T^t_1(u))^{-1}R(2u-N)
T_2(u-N)T^t_2(-u+N)
$$
$$
=Q\hat T_1(u)T^{-1}_1(-u)R(2u-N)T_2(u-N)T^t_2(-u+N), \leqno(1)
$$
where $\hat T(u)$ denotes the matrix $(T^t(u))^{-1}$.
However,
$$
T^{-1}_1(-u)R(2u-N)T_2(u-N)=T_2(u-N)R(2u-N)T^{-1}_1(-u),
$$
which is an immediate consequence of the ternary relation
written in the form
$$
R(u-v)T_2(u)T_1(v)=T_1(v)T_2(u)R(u-v).
$$
Indeed, it suffices to multiply both sides of the latter relation
by $T^{-1}_1(v)$ from the left and from the right
and to replace $(u,v)$ by $(u-N,-u)$. Therefore, the right
hand side of (1) takes the form:
$$
Q\hat T_1(u)T_2(u-N)R(2u-N)T^{-1}_1(-u)T^t_2(-u+N).
$$
By Proposition 5.3 (see Remark 5.10) the last expression
equals
$$
Qz(u)R(2u-N)T^{-1}_1(-u)T^t_2(-u+N)
=z(u)QR(2u-N)T^{-1}_1(-u)T^t_2(-u+N).
$$
Now, using (3.2.4), we obtain
$$
QR(2u-N)=Q(1-{P\over2u-N})=(1\mp {1\over 2u-N})Q.
$$
Hence, by (5.4.3), we have
$$
(1\mp {1\over 2u-N}) y(u)Q=(1\mp {1\over 2u-N})z(u)Q
T^{-1}_1(-u)T^t_2(-u+N)
$$
$$
=(1\mp {1\over 2u-N})z(u)z^{-1}(-u+N)Q,
$$
which proves the assertion.

\bigskip
\noindent
{\bf 6.8.} Now we prove an analogue of the quantum Liouville
formula for the twisted Yangian.

\proclaim {\bf Theorem} In the algebra $\Y^{\pm}(N)$ we have
$$
 y(u)=\varepsilon_N(u){\qddet S(u-1)\over\qddet S(u)}, \leqno(1)
$$
where $\varepsilon_N(u)\equiv 1$ for $\Y^+(N)$ and
$\dsize\varepsilon_N(u)={(2u+1)(2u-N-1)\over(2u-1)(2u-N+1)}$ for
$\Y^-(N)$.
\endproclaim

\Proof It follows from Theorems 6.7 and 5.7 that
$$
 y(u)={\qdet T(u-1)\over\qdet T(u)}
{\qdet T(-u+N)\over\qdet T(-u+N-1)}.
$$
Furthermore, by Theorem 4.7 we get
$$
 y(u)={\gamma_N(u)\over\gamma_N(u-1)}{\qddet S(u-1)\over\qddet S(u)}
=\varepsilon_N(u){\qddet S(u-1)\over\qddet S(u)},
$$
and the theorem is proved.

\bigskip
\proclaim {\bf 6.9. Corollary} The coefficients $ y_1, y_2,\dots$
of the quantum contraction $ y(u)$ generate the center of the algebra
$\Y^{\pm}(N)$.
\endproclaim

\Proof By using Theorem 4.11 and repeating the arguments of the proof of
Corollary 5.9, we obtain that the coefficients of the series
$\qddet S(u-1)(\qddet S(u))^{-1}$ generate the center of $\Y^{\pm}(N)$.
Since $\varepsilon_N(u)$ has the form $1+a_1u^{-1}+a_2u^{-2}+\dots,\ \
a_i\in\C$, the same is true for the series $ y(u)$.

\bigskip
\noindent
{\bf 6.10. Comments.} Theorem 6.4 was announced in Olshanski\u\i\ [O2]; it
has some similarity to Theorem 6 in Drinfeld [D1].

\newpage

\heading {\bf 7. The quantum determinant and the Sklyanin
determinant of block matrices}\endheading
\

Here we will gather several results related to dividing the $T$- and
$S$-matrices into rectangular blocks.

\bigskip
\noindent
{\bf 7.1.} Let us fix a partition of the number $N$ into a sum of
two nonnegative integers,
$
N=r+s.
$
For any $N\times N$-matrix $A$ we will denote by
$^{11}A,\ ^{12}A,\ ^{21}A,\ ^{22}A$ the
blocks of the matrix $A$ with respect to this partition so that
$$
A=\binom{^{11}A\quad ^{12}A}{^{21}A\quad ^{22}A}.
$$

\bigskip
\proclaim {\bf 7.2. Proposition} The matrix elements of the matrices
$^{11}T(u)$ and $^{22}(T^{-1}(v))$ commute with each other.
\endproclaim

\Proof Multiplying both sides of the ternary relation (1.8.1)
by $T^{-1}_2(v)$ from the left and from the right, we obtain
$$
T^{-1}_2(v)R(u-v)T_1(u)=T_1(u)R(u-v)T^{-1}_2(v).
$$
Since $R(u-v)=1-P(u-v)^{-1}$, this may be expressed as
$$
[T_1(u),T^{-1}_2(v)]={1\over u-v}(T_1(u)PT^{-1}_2(v)-
T^{-1}_2(v)PT_1(u)).
$$
Rewriting this in terms of matrix elements (see the proof of
Proposition 1.8), we obtain that
$$
[t_{ij}(u), t'_{kl}(v)]={1\over u-v}(\delta_{kj}\sum_a
t_{ia}(u)t'_{al}(v)-\delta_{il}\sum_at'_{ka}(v)t_{aj}(u)),
$$
where, as before, $t'_{ij}(u)$ denotes the matrix element of
the matrix $T^{-1}(u)$.
Thus, if $1\leq i,j\leq r$ and $r<k,l\leq N$, then
$
[t_{ij}(u), t'_{kl}(v)]=0,
$
which proves the proposition.

\bigskip
\noindent
{\bf 7.3.} For an invertible $N\times N$-matrix $A$ over $\C$
one has the following formula for the determinant of $A$:
$$
\text{det\ }A\ \text{det\ } ^{11}(A^{-1})=\text{det\ } ^{22}A.
$$
Here we prove an analogue of this formula for the quantum determinant of
the $T$-matrix. Denote by $t^*_{ij}(u),\ 1\leq i,j\leq N$, the image
of the series $t_{ij}(u)$ under the automorphism $\inva$ of the algebra
$\Y(N)$. That is, $t^*_{ij}(u)$ is
the matrix element of the matrix
$T^*(u):=T^{-1}(-u)$. Since $T^*(u)$ satisfies the ternary relation
(see Subsection 1.12), we can repeat the construction of the quantum
determinant from Section 2 for the matrix $T^*(u)$. In particular,
analogues of formulae (2.8.1) and (2.8.2) hold for $\qdet T^*(u)$.
We shall use the following one below: for $q\in \Sym_N$
$$
\qdet T^*(u)=\sgn(q)\sum_{p\in \Sym_N}\sgn(p)t^*_{p(1),q(1)}(u)\dots
t^*_{p(N),q(N)}(u-N+1). \leqno(1)
$$

\proclaim {\bf Theorem} We have
$$
\qdet T(u)\ \qdet ^{11}(T^*)(-u+N-1)=\qdet ^{22}T(u).
$$
\endproclaim

\Proof By Proposition 2.5,
$$
\qdet T(u)A_N=A_NT_1\dots T_N, \leqno(2)
$$
where $T_i=T_i(u-i+1)$ for $i=1,\dots,N$. Let us multiply both
sides of (2) by $T_N^{-1}\dots T_{s+1}^{-1}$ from the right. Then (2)
takes the form
$$
\qdet T(u)A_NT_N^{-1}\dots T_{s+1}^{-1}=A_NT_1\dots T_s. \leqno(3)
$$
Now we apply both the sides of (3) to the basis vector
$$
v_0=e_{r+1}\otimes\dots\otimes e_N\otimes e_1\otimes\dots\otimes e_r
\in \E^{\otimes N}.
$$
For the right hand side of (3) we have
$$
A_NT_1\dots T_sv_0=A_N\sum_{i_1,\dots,i_s}t_{i_1,r+1}(u)\dots
t_{i_s,N}(u-s+1)e_{i_1}\otimes\dots\otimes e_{i_s}
\otimes e_1\otimes\dots\otimes e_r.
$$
The coefficient of $v_0$ in this decomposition equals
$$
{1\over N!}\sum_p\sgn(p)t_{p(r+1),r+1}(u)\dots t_{p(N),N}(u-s+1),
$$
where $p$ runs over all the permutations of the set of indices
$\{r+1,\dots,N\}$. By (2.7.1), this is nothing else but
$(N!)^{-1}\qdet ^{22}T(u)$.

Similarly, for the left hand side of (3) we obtain
$$\align
A_N&T_N^{-1}\dots T_{s+1}^{-1}v_0\\
&=A_N\sum_{i_{s+1},\dots,i_N}t'_{i_N,r}(u-N+1)
\dots t'_{i_{s+1},1}(u-s)e_{r+1}\otimes\dots\otimes e_N
\otimes e_{i_{s+1}}\otimes\dots\otimes e_{i_N}.
\endalign$$
It is clear that the coefficient of $v_0$ in this expression equals
$$
\multline
{1\over N!}\sum_{p\in \Sym_r}\sgn(p)t'_{p(r),r}(u-N+1)\dots
t'_{p(1),1}(u-s)
\\
={1\over N!}\sum_{p\in \Sym_r}\sgn(p)t^*_{p(r),r}(-u+N-1)\dots
t^*_{p(1),1}(-u+s).
\endmultline
\tag4
$$
It follows immediately from (1) that (4) coincides with
$(N!)^{-1}\qdet ^{11}(T^*)(-u+N-1)$ and the theorem is proved.

\bigskip
\noindent
{\bf 7.4.} Now we shall prove analogues of Proposition 7.2 and Theorem 7.3
for the twisted Yangian $\Y^{\pm}(N)$. We will keep using the notation of
Sections 3, 4 and 6.

Let us fix a nonnegative integer $M\leq N$ such that $N-M$ is even.
Put $m=[M/2]$. For a $N\times N$-matrix $A$ denote by $^{11}A$ and
$^{22}A$ the submatrices of $A$ whose rows and columns are enumerated by
the
indices $\{-m,-m+1,\dots,m\}$ and $\{-n,-n+1,\dots,-m-1,m+1,\dots,n\}$
respectively.

\bigskip
\proclaim {\bf 7.5. Proposition} The matrix elements of the matrices
$^{11}S(u)$ and $^{22}(S^{-1}(v))$ commute with each other.
\endproclaim

\Proof Multiplying both sides of the quaternary relation (3.6.2) by
$S_2^{-1}(v)$ from the left and from the right, we obtain the relation
$$
S_2^{-1}(v)R(u-v)S_1(u)R'(-u-v)=R'(-u-v)S_1(u)R(u-v)S_2^{-1}(v).
$$
Using the equalities
$$
R(u-v)=1-{P\over u-v}\qquad\text{and}\qquad R'(-u-v)=1+{Q\over u+v}
$$
we rewrite the last relation as follows:
$$
\align
[S_1(u), S_2^{-1}(v)]=&{1\over u-v}(S_1(u)PS_2^{-1}(v)-S_2^{-1}(v)PS_1(u))
\\
&-{1\over u+v}(QS_1(u)S_2^{-1}(v)-S_2^{-1}(v)S_1(u)Q)
\\
&+{1\over u^2-v^2}(QS_1(u)PS_2^{-1}(v)-S_2^{-1}(v)PS_1(u)Q),
\endalign
$$
or in terms of the matrix elements:
$$
\align
[s_{ij}(u), s'_{kl}(v)]=&{1\over u-v}(\delta_{kj}\sum_as_{ia}(u)s'_{al}(v)
-\delta_{il}\sum_as'_{ka}(v)s_{aj}(u))
\\
&-{1\over u+v}(\delta_{i,-k}\sum_a\theta_{ak}s_{-a,j}(u)s'_{al}(v)
-\delta_{j,-l}\sum_a\theta_{al}s'_{ka}(v)s_{i,-a}(u))
\\
&+{1\over u^2-v^2}(\delta_{i,-k}\theta_{kj}\sum_as_{-j,a}(u)s'_{al}(v)
-\delta_{j,-l}\theta_{il}\sum_as'_{ka}(v)s_{a,-i}(u)).
\endalign
$$
Thus, if $-m\leq i,j\leq m$ and $m<|k|,|l|\leq n$, then
$[s_{ij}(u), s'_{kl}(v)]=0$, and the proposition is proved.

\bigskip
\noindent
{\bf 7.6.} Set $S^*(u):=\inva(\td S(u))=\td S^{-1}(-u-N/2)$ and denote by
$s^*_{ij}(u)$
the matrix elements of the matrix $S^*(u)$. Note that in the proof of the
fundamental identity (4.2.1) and hence in the construction of the Sklyanin
determinant of the $S$-matrix (Propositions 4.3 and 4.4) we have only used
the quaternary relation (3.6.2) and have never used the symmetry relation
(3.6.3).
Therefore, these results remain valid for the algebra $\tilde \Y^{\pm}(N)$;
see Subsection 6.1.
In particular, all of the coefficients of $\qddet \td S(u)$ belong to the
center
of $\tilde \Y^{\pm}(N)$. Since inv is an automorphism of the algebra
$\tilde \Y^{\pm}(N)$,
we can repeat the construction of the Sklyanin determinant for the matrix
$S^*(u)$ and obtain the analogues of
Propositions 4.3 and 4.4 for this matrix. By applying the factorization map
$\tilde \Y^{\pm}(N)\ra\Y^{\pm}(N)$ to the equality (1) below,
we will obtain an analogue of
Theorem 7.3 for the twisted Yangian.

\proclaim {\bf Theorem} In the algebra $\td\Y^{\pm}(N)$ we have
$$
\qddet\td S(u)\ \qddet ^{22}(S^*)(-u+{N\over 2}-1)=\qddet ^{11}\td S(u).
\leqno(1)
$$
\endproclaim

\Proof
By the analogue of Proposition 4.3 for the matris $\td S(u)$ we have
$$
\qddet \td S(u)A_N=A_N\td S_1R'_{12}\dots R'_{1N}\td S_2\dots
\td S_{N-1}R'_{N-1,N}\td S_N,
\leqno(2)
$$
where $\td S_i=\td S_i(u-i+1)$ for $1\leq i\leq N$ and
$R'_{ij}=R'_{ij}(-2u+i+j-2)$
for $1\leq i,j\leq N,\ i\ne j$. Using the fact that $\td S_i$ and $R'_{jk}$
commute
provided $i\ne j,k$, we rewrite the right hand side of (2) in the form
$$
A_N\td S_1R'_{12}\dots R'_{1M}\td S_2\dots \td S_{M-1}R'_{M-1,M}\td
S_MR'_{1,M+1}
\dots
R'_{1N}
\dots R'_{M,M+1}\dots R'_{MN}
$$
$$
\td S_{M+1}R'_{M+1,M+2}\dots R'_{M+1,N}\td S_{M+2}\dots \td
S_{N-1}R'_{N-1,N}
\td S_N.
$$
Since all of the matrices $\td S_i$ and $R'_{ij}$ are invertible, relation
(2) is equivalent to the following one:
$$
\qddet \td S(u)A_N\td S^{-1}_N(R'_{N-1,N})^{-1}
\td S^{-1}_{N-1}\dots \td S^{-1}_{M+2}(R'_{M+1,N})^{-1}
\dots(R'_{M+1,M+2})^{-1}\td S^{-1}_{M+1}\leqno(3)
$$
$$
=A_N\td S_1R'_{12}\dots R'_{1M}\td S_2\dots \td S_{M-1}R'_{M-1,M}\td
S_MR'_{1,M+1}
\dots
R'_{1N}
\dots R'_{M,M+1}\dots R'_{MN}.
$$
Now we compare the diagonal elements of the matrices in the
left and right hand sides of (3) corresponding to the vector
$$
v_0=e_{-m}\otimes e_{-m+1}\otimes\dots
\otimes e_m\otimes e_{-n}\otimes\dots\otimes e_{-m-1}\otimes
e_{m+1}\otimes\dots\otimes e_n\in\E^{\otimes N}.
$$
It is clear that $R'_{ij}v_0=v_0$, if $i\leq M$ and $j>M$.
Thus, for the right hand side of (3) we have:
$$
A_N\td S_1R'_{12}\dots R'_{1M}\td S_2\dots \td S_{M-1}R'_{M-1,M}\td
S_MR'_{1,M+1}
\dots
R'_{1N}
\dots R'_{M,M+1}\dots R'_{MN}v_0\leqno(4)
$$
$$
=A_N\td S_1R'_{12}\dots R'_{1M}\td S_2\dots \td S_{M-1}R'_{M-1,M}\td S_Mv_0
$$
$$
=A_N\sum_{i_1,\dots,i_M}a_{i_1,\dots,i_M}(u)e_{i_1}\otimes\dots
\otimes e_{i_M}\otimes e_{-n}\otimes\dots\otimes e_{-m-1}\otimes
e_{m+1}\otimes\dots\otimes e_n,
$$
where $a_{i_1,\dots,i_M}(u)$ are certain elements of
$\tilde \Y^{\pm}(N)[[u^{-1}]]$. To calculate the coefficient of $v_0$
in this decomposition we may take into account only those
summands for which the sequence $(i_1,\dots,i_M)$ is obtained by
a permutation of the indices $(-m,-m+1,\dots,m)$.
It is not difficult to see that this allows us to replace in (4)
the matrix $\td S$ by its submatrix $^{11}\td S$; $A_N$ by the
antisymmetrizer
$(N!)^{-1}M!A_M$ in the space $(\C^M)^{\otimes M}$ where $\C^M$ is spanned
by the basis vectors $e_{-m},e_{-m+1},\dots,e_m$; and $R'$ by its
restriction
to the space $\C^M\otimes\C^M$. Therefore, by Proposition 4.3, the
coefficient
of $v_0$ in (4) equals $(N!)^{-1}\qddet ^{11}\td S(u)$.

Furthermore, set $\dsize\tilde u_i=-u+i-{N\over 2}-1$ for $i=M+1,\dots,N$.
Since
$$
(R'_{ij})^{-1}=R'_{ij}(2u-i-j+N+2),
$$
we can express the
left hand side of (3) as follows:
$$
\qddet \td S(u)A_NS^*_NR^*_{N,N-1}\dots R^*_{N,M+1}S^*_{N-1}\dots
S^*_{M+2}R^*_{M+2,M+1}S^*_{M+1}, \leqno(5)
$$
where $S^*_i=S^*_i(\tilde u_i)$ and
$R^*_{ij}=R'_{ij}(-\tilde u_i-\tilde u_j)$. Repeating the same arguments as
for the right hand side of (3) and using Remark 4.6, we obtain that
the matrix element of the operator (5) on the basis vector $v_0$
equals $(N!)^{-1}\qddet \td S(u)\qddet ^{22}(S^*)(\tilde u_N)$,
which proves the theorem.

\bigskip
\noindent
{\bf 7.7. Comments.} Proposition 7.2 (which in fact holds
for even more general $R$-matrices) is due to Cherednik [C1, Theorem 2.4].
The use of quantum
determinants of certain submatrices of $T(u)$ was significant in
Drinfeld [D3], Nazarov--Tarasov [NT], Molev [M2].

\newpage
\heading
{\bf References}
\endheading
\bigskip

\itemitem{[B]}
{F. A. Berezin},
{\sl Introduction to Superanalysis},
D. Reidel,
Dordrecht-Boston,
1987.

\itemitem{[BG]}
{A. J. Bracken and H. S. Green},
{\sl Vector operators and a polynomial identity for SO(n)},
{J. Math. Phys.}
{\bf 12}
(1971),
2099--2106.

\itemitem{[BCC]}
{D. M. O'Brien, A. Cant and A. L. Carey},
{\sl On characteristic identities for Lie algebras},
{Ann. Inst. Henri Poincar\'e}
{\bf A26}
(1977),
405--429.

\itemitem{[BD1]}
{A. A. Belavin and V. G. Drinfeld},
{\sl Solutions of the classical Yang--Baxter equation for
simple Lie algebras},
{Functional. Anal. Appl.}
{\bf 16}
(1982),
159--180.

\itemitem{[BD2]}
{A. A. Belavin and V. G. Drinfeld},
{\sl Classical Yang--Baxter equation for simple Lie algebras},
{Functional Anal. Appl.}
{\bf 17}
(1983),
220--221.

\itemitem{[C1]}
{I. V. Cherednik},
{\sl Factorized particles on the half-line and root systems},
{Theor. Math. Phys.}
{\bf 61}
(1984),
no. 1,
35--44.

\itemitem{[C2]}
{I. V. Cherednik},
{\sl A new interpretation of Gelfand--Tzetlin bases}, {Duke Math. J.}
{\bf 54}
(1987),
563--577.

\itemitem{[C3]}
{I. V. Cherednik},
{\sl Quantum groups as hidden symmetries of classic
representation theory},
in \lq Differential Geometric Methods in Physics
(A. I. Solomon, Ed.)',
World Scientific,
Singapore,
1989,
pp. 47--54.

\itemitem{[Ca1]}
{A. Capelli},
{\sl \"Uber die Zur\"uckf\"uhrung der Cayley'schen Operation
$\Omega$ auf ge\-w\"ohn\-liche Polar-Operationen}, {Math. Ann.}
{\bf 29}
(1887),
331--338.

\itemitem{[Ca2]}
{A. Capelli},
{\sl Sur les op\'erations dans la th\'eorie des formes
alg\'ebriques},
{Math. Ann.}
{\bf 37}
(1890),
1--37.

\itemitem{[CL]}
{R. W. Carter and G. Lusztig},
{\sl On the modular representations of the general linear and
symmetric groups},
{Math. Z.}
{\bf 136}
(1974),
193--242.

\itemitem{[CP1]}
{V. Chari and A. Pressley},
{\sl Yangians and $R$-matrices},
{L'Enseign. Math.}
{\bf 36}
(1990),
267--302.

\itemitem{[CP2]}
{V. Chari and A. Pressley},
{\sl Fundamental representations of Yangians and
singularities of $R$-matrices},
{J. Reine Angew. Math.}
{\bf 417}
(1991),
87--128.

\itemitem{[D1]}
{V. G. Drinfeld},
{\sl Hopf algebras and the quantum Yang--Baxter equation}, {Soviet Math.
Dokl.}
{\bf 32}
(1985),
254--258.

\itemitem{[D2]}
{V. G. Drinfeld},
{\sl Degenerate affine Hecke algebras and Yangians}, {Functional Anal.
Appl.}
{\bf 20}
(1986),
56--58.

\itemitem{[D3]}
{V. G. Drinfeld},
{\sl A new realization of Yangians and quantized affine
algebras},
{Soviet Math. Dokl.}
{\bf 36}
(1988),
212--216.

\itemitem{[D4]}
{V. G. Drinfeld},
{\sl Quantum Groups},
in \lq Proc. Int. Congress Math., Berkeley, 1986', AMS,
Providence RI,
1987,
pp. 798--820.

\itemitem{[FT]}
{L. D. Faddeev and L. A. Takhtajan},
{\sl Spectrum and scattering of excitations in the
one-dimensional isotropic Heisenberg model}, {J. Soviet Math.}
{\bf 24}
(1984),
241--267.

\itemitem{[G]}
{M. D. Gould},
{\sl Characteristic identities for semi-simple Lie algebras},
{J. Austral. Math. Soc.}
{\bf B26}
(1985),
257--283.

\itemitem{[Gr]}
{H. S. Green},
{\sl Characteristic identities for generators of GL(n), O(n) and Sp(n)},
{J. Math. Phys.}
{\bf 12}
(1971),
2106--2113.

\itemitem{[H]}
{R. Howe},
{\sl Remarks on classical invariant theory}, {Trans. AMS}
{\bf 313}
(1989),
539--570.

\itemitem{[HU]}
{R. Howe and T. Umeda},
{\sl The Capelli identity, the double commutant theorem,
and multiplicity-free actions},
{Math. Ann.}
{\bf 290}
(1991),
569--619.

\itemitem{[IK]}
{A. G. Izergin and V. E. Korepin}
{\sl A lattice model related to the nonlinear Schrodinger equation},
{Sov. Phys. Dokl.}
{\bf 26}
(1981)
653--654.

\itemitem{[KR]}
{A. N. Kirillov and N. Yu. Reshetikhin},
{\sl Yangians, Bethe ansatz and combinatorics},
{Lett. Math. Phys.}
{\bf 12}
(1986),
199--208.

\itemitem{[K]}
{B.\ Kostant},
{\sl On the tensor product of a finite and an infinite
dimensional representations},
{J. Funct. Anal.}
{\bf 20}
(1975),
257--285.

\itemitem{[KuR]}
{P. P. Kulish and N. Yu. Reshetikhin},
{\sl $GL_3$-invariant solutions of the Yang-Baxter equation}, {J. Soviet
Math.}
{\bf 34}
(1986),
1948--1971.

\itemitem{[KRS]}
{P. P. Kulish, N. Yu. Reshetikhin and E. K. Sklyanin},
{\sl Yang--Baxter equation and representation theory}, {Lett. Math. Phys.}
{\bf 5}
(1981),
393--403.

\itemitem{[KS1]}
{P. P. Kulish and E. K. Sklyanin},
{\sl On the solutions of the Yang--Baxter equation}, {J. Soviet Math.}
{\bf 19}
(1982),
1596--1620.

\itemitem{[KS2]}
{P. P. Kulish and E. K. Sklyanin},
{\sl Quantum spectral transform method: recent developments},
in \lq Integrable Quantum Field Theories', {Lecture Notes in Phys.}
{\bf 151}
Springer,
Berlin-Heidelberg,
1982,
pp. 61--119.

\itemitem{[KS3]}
{P. P. Kulish and E. K. Sklyanin},
{\sl Algebraic structures related to reflection equations}, {J. Phys.}
{\bf A25}
(1992),
5963--5975.

\itemitem{[KSS]}
{P. P. Kulish, R. Sasaki and G. Schwiebert},
{\sl Constant solutions of reflection equations and quantum
groups },
{J. Math. Phys.}
{\bf 34}
(1993),
286--304.

\itemitem{[L1]}
{S. Z. Levendorski\u\i},
{\sl On PBW bases for Yangians},
{Lett. Math. Phys.}
{\bf 27}
(1993),
37--42.

\itemitem{[L2]}
{S. Z. Levendorski\u\i},
{\sl On generators and defining relations of Yangians},
{J. Geom. Phys.}
{\bf 12}
(1993),
1--11.

\itemitem{[M1]}
{A. I. Molev},
{\sl Representations of twisted Yangians}, {Lett. Math. Phys.}
{\bf 26}
(1992),
211--218.

\itemitem{[M2]}
{A. I. Molev},
{\sl Gelfand--Tsetlin bases for representations of Yangians},
Preprint CMA-MR19-93, Canberra, 1993.

\itemitem{[M3]}
{A. I. Molev},
{\sl Yangians and classical Lie algebras, Part II. Sklyanin
determinant, Laplace operators and characteristic identities},
Preprint CMA MRR 024-94, Canberra, 1994.

\itemitem{[N1]}
{M. L. Nazarov},
{\sl Quantum Berezinian and the classical Capelli identity},
{Lett. Math. Phys.}
{\bf 21}
(1991),
123--131.

\itemitem{[N2]}
{M. L. Nazarov},
{\sl Yangians of the\lq strange' Lie superalgebras},
in \lq Quantum Gr\-ou\-ps (P. P. Kulish, Ed.)', {Lecture Notes in Math.}
{\bf 1510},
Springer,
Berlin-Heidelberg,
1992,
pp. 90--97.

\itemitem{[No]}
{M. Noumi},
{\sl Macdonald's symmetric polynomials as zonal spherical
functions on some quantum homogeneous spaces},
{Preprint UTMS 93--42}, Tokyo, 1993.
June 1993.

\itemitem{[NT]}
{M. Nazarov and V. Tarasov},
{\sl Yangians and Gelfand--Zetlin bases}, Preprint RIMS, Kyoto Univ.,
February 1993.

\itemitem{[O1]}
{G. I. Olshanski\u\i},
{\sl Representations of infinite-dimensional classical
groups, limits of enveloping algebras, and Yangians},
in \lq Topics in Representation Theory (A. A. Kirillov, Ed.), {Advances in
Soviet Math.} {\bf 2},
AMS,
Providence RI,
1991,
pp. 1--66.

\itemitem{[O2]}
{G. I. Olshanski\u\i},
{\sl Twisted Yangians and infinite-dimensional classical Lie algebras},
in \lq Quantum Groups (P. P. Kulish, Ed.)', {Lecture Notes in Math.}
{\bf 1510},
Springer,
Berlin-Heidelberg,
1992,
pp. 103--120.

\itemitem{[PP]}
{A. M. Perelomov and V. S. Popov},
{\sl Casimir operators for semisimple Lie algebras}, {Isv. AN SSSR}
{\bf 32}
(1968),
1368--1390.

\itemitem{[R]}
{N. Yu. Reshetikhin},
{\sl Integrable models of quantum one-dimensional magnets
with $O(n)$ and $Sp(2k)$-symmetry},
{Theor. Math. Phys.}
{\bf 63}
(1985),
no. 3,
347--366.

\itemitem{[RS]}
{N. Yu. Reshetikhin and M. A. Semenov-Tian-Shansky},
{\sl Central extensions of quantum current groups}, {Lett. Math. Phys.},
{\bf 19}
(1990),
133--142.

\itemitem{[RTF]}
{N. Yu. Reshetikhin, L. A. Takhtajan and L. D. Faddeev},
{\sl Quantization of Lie Groups and Lie algebras}, {Leningrad Math. J.}
{\bf 1}
(1990),
193--225.

\itemitem{[S1]}
{E. K. Sklyanin},
{\sl Quantum version of the method of inverse scattering
problem},
{J. Soviet. Math.}
{\bf 19}
(1982),
1546--1596.

\itemitem{[S2]}
{E. K. Sklyanin},
{\sl Boundary conditions for integrable quantum systems}, {J. Phys.}
{\bf A21}
(1988),
2375--2389.

\itemitem{[T1]}
{V. O. Tarasov},
{\sl Structure of quantum $L$-operators for the
$R$-matrix of of the $XXZ$-model},
{Theor. Math. Phys.}
{\bf 61}
(1984),
1065--1071.

\itemitem{[T2]}
{V. O. Tarasov},
{\sl Irreducible monodromy matrices for the $R$-matrix of the
$XXZ$-model and lattice local quantum Hamiltonians}, {Theor. Math. Phys.}
{\bf 63}
(1985),
440--454.

\itemitem{[TF]}
{L. A. Takhtajan and L.D. Faddeev},
{\sl Quantum inverse scattering method and the Heisenberg
$XYZ$-model},
{Russian Math. Surv.}
{\bf 34}
(1979),
no. 5,
11--68.

\itemitem{[W]}
{H. Weyl},
{\sl Classical Groups, their Invariants and Representations},
{Princeton Univ. Press},
Princeton NJ,
1946.

\enddocument